\documentclass[12pt,a4paper,twoside]{book}

\usepackage{geometry}                               
\usepackage{graphicx,subfigure,rotating,xy}         
\usepackage{amsmath,amsxtra,amscd,amssymb,latexsym} 
\usepackage{bbm,array,theorem,enumerate,stmaryrd}   
\usepackage{amstext,amsfonts,mathrsfs,yfonts,bbold} 
\usepackage{setspace}                               
\usepackage{float}                                  
\usepackage{fancyhdr}                               
\usepackage{caption}                                
\usepackage{color}                                  
\usepackage{url}                                    
\usepackage{listings}                               
\usepackage{wrapfig}                                
\usepackage{makeidx}                                
\usepackage{lmodern}                                
\usepackage{eso-pic}                                

\usepackage{nicechapter}
\usepackage{nicetheme}

\newcolumntype{I}{!{\vrule width 1.5pt}}

\newlength{\savedwidth}

\fancypagestyle{headings}{           
\fancyhead{}                         
\fancyfoot[C]{\bfseries\thepage}
}

\makeindex

\newcommand{\pop}[1]{\frac{\partial}{\partial #1}}

\newcommand{\ca}[1]{{\cal #1}}
\newcommand{\SO}{{\rm SO}}
\newcommand{\vergule}{\quad\quad}
\newcommand{\blanc}{\quad\quad}
\newcommand{\ket}[1]{\left| #1 \right>}

\newcommand{\D}{{\cal D}}
\newcommand{\F}{{\cal F}}
\newcommand{\A}{{\cal A}}
\newcommand{\N}{{\cal N}}
\newcommand{\U}{{\rm U}}
\newcommand{\SU}{{\rm SU}}

\newcommand{\p}[1]{(\ref{#1})}
\newcommand{\CS}{{\rm CS}}
\newcommand{\lb}{\label}
\newcommand{\B}{{\cal B}}
\newcommand{\vp}{\varphi}
\newcommand{\ve}{\varepsilon}
\newcommand{\hc}{\mbox{h.c.}}
\newcommand{\diag}{{\rm diag}}
\newcommand{\nn}{\nonumber}

\newcommand{\dalpha}{{\dot\alpha}}
\newcommand{\dbeta}{{\dot\beta}}
\newcommand{\dgamma}{{\dot\gamma}}

\newcommand{\askip}{\hskip 3mm}

\newcommand{\rmL}{{\rm L}}
\newcommand{\rmR}{{\rm R}}

\begin{document}



\thispagestyle{empty}

\begin{spacing}{1}

  \vspace{-2cm}

  \begin{center}
    {\small
    UNIVERSIT\'E DE NANTES\\
    FACULT\'E DES SCIENCES ET TECHNIQUES \\
    ------------\\
    \'ECOLE DOCTORALE \\
    MAT\'ERIAUX MATI\`ERE MOL\'ECULE EN PAYS DE LOIRE\\}
  \end{center}

  \vspace{0.6cm}

  \begin{minipage}[l]{5cm}
    {\small Ann\'ee : 2012}
  \end{minipage}
  \hfill
  \begin{minipage}[r]{7cm}
    \begin{center}
      {\scriptsize N$^{\circ}$ attribu\'e par la biblioth\`eque } \\
      \vspace{0.1cm}
      \begin{tabular}[b]{|l|l|l|l|l|l|l|l|l|l|}
        \hbox to .1cm{}&\hbox to .1cm{}&\hbox to
        .1cm{}&\hbox to .1cm{}&\hbox to .1cm{}&\hbox to .1cm{}&\hbox to
        .1cm{}&\hbox to .1cm{}&\hbox to .1cm{}&\hbox to .1cm{}\\
        \hline
      \end{tabular}
    \end{center}
  \end{minipage}~~~~~~~~~~~~~~~~~~

\end{spacing}

\vspace{1.5cm}

\begin{spacing}{2}
  \begin{center}
    {\bf \huge Extended supersymmetry and its applications in quantum mechanical models associated with self-dual gauge fields}
  \end{center}
\end{spacing}

\begin{spacing}{1.2}
  \begin{center}
    TH\`ESE DE DOCTORAT \\
    Discipline : Physique \\
    Sp\'ecialit\'e : Physique Subatomique \\
  \end{center}

  \begin{center}
    \small{\textsl{Pr\'esent\'ee et soutenue publiquement par}\\}

    \vspace{0.4cm}
    \textbf{\large Maxim KONYUSHIKHIN}

    \vspace{0.2cm}
    \small{\textsl{Le 29 Mars 2012, devant le jury ci-dessous}}
    \vfill
  \end{center}
\end{spacing}

\vspace{0.4cm}

\begin{spacing}{0.25}
  {\small
  \begin{tabular}{ l  l }
    \textsl{Pr\'esident}  & M. AICHELIN Joerg, {\sl Professeur, Subatech, Universit\'e de Nantes }\\
                          &      \\
		    	          &      \\
                          &      \\
    \textsl{Rapporteurs}  & M. DELDUC Fran\c{c}ois, {\sl Directeur de recherche CNRS, ENS Lyon} \\
                          &      \\
			              & M. FEDORUK Sergey, {\sl Professeur, JINR, Dubna, Russie } \\
                          &      \\
    \textsl{Examinateur} & M. IVANOV Evgeny, {\sl Professeur, JINR, Dubna, Russie } \\
                          &      \\
  \end{tabular}}

  \vspace{0.4cm}

  \begin{description}
    {\small \item [\textsl{Directeur de th\`ese : }]Andrei SMILGA,
    {\sl Professeur, Subatech, Universit\'e de Nantes }}
  \end{description}

  \vspace{0.3cm}

  \begin{flushright}
    {\small N$^\circ$ ED 500~~~~~~~~~~~~~~~~~~}
  \end{flushright}
  {\small \par}
\end{spacing}



\cleardoublepage
\setcounter{page}{1}
\chapitresimple{Acknowledgment}

\mbox{\ }

Throughout the various stages of completion of this thesis and during my graduate education as a whole, I have greatly benefited from discussions with Andrei Smilga, my research advisor. I warmly thank him for sharing his insight, offering advice, maintaining interest and providing encouragement.

The study involved in this thesis was done in an amiable and pleasant collaboration with Evgeny Ivanov. I deeply appreciate his participation in my education, his help and his advices during the writing of the manuscript.

Let me also express my deep gratitude to my wife Olga Driga. I would not have completed my thesis without her unconditional support.

I am also indebted to L.~Alvarez-Gaume, F.~Delduc, S.~Fedoruk, A.~Gorsky, O.~Lechtenfeld, M.~Shifman and A.~Wipf  for their illuminating discussions.


  \setcounter{tocdepth}{2} 
  \tableofcontents
  \numberwithin{equation}{section} 




  \cleardoublepage
\chapter{Introduction}

Supersymmetric quantum mechanics (SQM) provides a proper venue for exploring and modeling salient features
of supersymmetric field theories in diverse dimensions \cite{W}. Some SQM models represent one-dimensional reductions
of higher-dimensional supersymmetric theories. At the same time, many interesting models of this kind
can be constructed directly in \mbox{(0+1)} dimensions, without any reference to the dimensional reduction procedure.
They exhibit some surprising properties related to peculiarities of one-dimensional supersymmetry. For any SQM model (like for any
supersymmetric field theory), it is desirable, besides the component Hamiltonian and Lagrangian description, to have
the appropriate superfield Lagrangian formulation. The latter makes supersymmetry manifest,  prompts
possible generalizations of the model and allows one to reveal relationships with other cognate theories.

This study is devoted to the Hamiltonian and the Lagrangian as well as the superfield Lagrangian formulation for a certain class of ${\cal N}=4$
SQM models~\footnote{Hereafter, in quantum mechanics, $\N$ counts the number of {\sl real} supercharges.} with self-dual or anti-self-dual Abelian or non-Abelian gauge field backgrounds \cite{KonSmi,Ivanov:2009tw,IvKon}. Surprisingly, such systems did not attracted much attention so far.
A natural framework for this formulation proves to be the harmonic superspace (HSS)
approach \cite{HSS} adapted to the one-dimensional case \cite{IvLecht}.

The models of supersymmetric quantum mechanics with background gauge fields
are of obvious interest for several reasons. One of them is a close relation of these systems to the Landau problem (motion of a charged particle in an external magnetic field) and its generalizations (see e.g.~\cite{Land}).
The Landau-type models constitute a basis of the theoretical description
of quantum Hall effect (QHE), and it is natural to expect that their supersymmetric
extensions, with extra fermionic variables added, may be relevant to
spin versions of QHE. Also, these systems can provide quantum-mechanical realizations
of various Hopf maps closely related to higher-dimensional QHE (see e.g. \cite{gknty} and references therein).

The first type of SQM models considered in this work represents a subclass of  well-known systems
which describe the motion of a fermion on an even-dimensional manifold with an arbitrary gauge background.
It was observed many years ago that one can treat these systems
as supersymmetric ones such that, e.g.,
the Atiyah-Singer index of the massless Dirac operator $\, /\!\!\!\!     \D $ can be interpreted as the
Witten index of a certain
supersymmetric Hamiltonian \cite{Gaume}.
The corresponding supercharges and the Hamiltonian are
\begin{equation*}
\label{QHchir}
 Q = /\!\!\!\!     \D (1 + \gamma_5) ,
\blanc
 \bar Q = /\!\!\!\!     \D (1 - \gamma_5)  ,
\blanc
 H =  /\!\!\!\!     \D^2  ,
\end{equation*}
where $\gamma_5$ is the appropriate ``fifth gamma matrix'' obeying $\gamma_5^2=1$ and anticommuting with the Dirac operator,
$\left\{\gamma_5,\, \, /\!\!\!\!     \D \right\} = 0$.
 Indeed, for any eigenstate $\Psi$ of the massless Dirac operator  $\, /\!\!\!\!     \D $
with a nonzero eigenvalue $\lambda$,
the state $\gamma^5 \Psi$ is also an eigenstate of  $\, /\!\!\!\!     \D $  with the eigenvalue
$-\lambda$. Thus, all excited states of $H$ are doubly degenerate.

It turns out that for a four-dimensional flat manifold and self-dual or anti-self-dual gauge field, Abelian or non-Abelian, the spectrum of $H$ is 4-fold
degenerate implying the extended ${\cal N} = 4$ supersymmetry.
For a flat Dirac operator in the instanton background, this can be traced back to Ref.~\cite{JR}.

${\cal N}=4$ SQM models with the background Abelian gauge fields were treated in the pioneer papers \cite{de Crombrugghe:1982un,Smilga:1986rb} and, more recently, e.g. in \cite{IKL,IvLecht,Kirchberg:2004za,KonSmi}.
In particular, in \cite{IvLecht} an off-shell Lagrangian superfield formulation
of the general models associated with
the multiplets $({\bf 4, 4, 0})$ and $({\bf 3, 4, 1})$ was given
in the ${\cal N}=4$, $d=1$ harmonic superspace~\footnote{The first
superfield formulation of general $({\bf 3, 4, 1})$ SQM
(without background gauge field couplings) was given in \cite{IvSmi}.}.
It was found that ${\cal N}=4$ supersymmetry
requires the gauge field to be (anti)self-dual in the four-dimensional $({\bf 4, 4, 0})$ case, or to obey a ``static'' version of the (anti)self-duality condition in  the three-dimensional $({\bf 3, 4, 1})$ case.
In the papers \cite{Kirchberg:2004za,KonSmi}, it was observed (in a Hamiltonian approach) that the Abelian $({\bf 4, 4, 0})$ ${\cal N}=4$ SQM
admits a simple generalization to arbitrary self-dual non-Abelian background. In \cite{Ivanov:2009tw}, an off-shell Lagrangian
formulation was shown to exist for a particular class of such non-Abelian ${\cal N}=4$ SQM models, with $\SU(2)$ gauge group
and 't~Hooft ansatz \cite{tHooft} for the self-dual $\SU(2)$ gauge field (see also \cite{KLS}). As in the Abelian case, it was the use of ${\cal N}=4$, $d=1$
harmonic superspace that allowed us to construct such an off-shell formulation.
A new non-trivial feature of the construction of \cite{Ivanov:2009tw} is the involvement of
an auxiliary ``semi-dynamical'' $({\bf 4, 4, 0})$ multiplet with the Wess-Zumino
type action possessing an extra  gauged $\U(1)$ symmetry. After quantization, the corresponding bosonic fields become
a sort of spin $\SU(2)$ variables to which the background gauge field naturally couples~\footnote{The use of such
auxiliary bosonic variables for setting up coupling of a particle to Yang-Mills fields  can be traced back to \cite{Bala}.
In the context of ${\cal N}=4$ SQM, they were employed in \cite{FIL0,FIL} and \cite{gknty,BKS}.}.

The second class of SQM models that we consider can be obtained at the component level by the Hamiltonian reduction of the systems discussed above from four to three dimensions. Their superfield description is nontrivial and consists in coupling the coordinate supermultiplet $({\bf 3, 4, 1})$ to an external non-Abelian gauge field through the introduction of the auxiliary $({\bf 4, 4, 0})$ superfield.
The off-shell ${\cal N}=4$ supersymmetry
restricts the external gauge field to be represented by a ``static'' version of the 't~Hooft ansatz for four-dimensional (anti)self-dual $\SU(2)$ gauge fields, i.e. to a particular solution of the general monopole Bogomolny equations \cite{Bog}~\footnote{Some BPS monopole backgrounds in the framework of ${\cal N}=2$ SQM were considered, e.g., in \cite{H1}.}.
A new feature of the three-dimensional case is the appearance of ``induced'' potential term in the action
as a result of eliminating the auxiliary field of the coordinate $({\bf 3, 4, 1})$ supermultiplet. This term is bilinear in the
$\SU(2)$ gauge group generators. As a particular ``spherically symmetric''  case of the construction (with the
exact $\SU(2)$ R-symmetry) we recover the ${\cal N}=4$ mechanics with Wu-Yang monopole \cite{WYa} (recently considered in \cite{BKS} with an essentially different treatment of the spin variables).

The chapter~\ref{chap2} is devoted to the introduction to supersymmetry in four-dimensional relativistic field theories. We discuss the motivation and the properties of supersymmetric theories as well as the their practical realization through the superfield approach. As an illustration, we consider the simplest example of the Wess-Zumino model -- a complex scalar field coupled to a Weyl spinor field.

In chapter~\ref{chap3}, we discuss supersymmetry in quantum mechanics. The ordinary superspace and harmonic superspace formalisms are given. In particular, the structure of the supermultiplets $({\bf 4, 4, 0})$ and $({\bf 3, 4, 1})$  is explained. Additionally, we introduce necessary notations which will be used in the chapter~\ref{chap4}.

The chapter~\ref{chap4} presents the original results of this study. We give the component and the superfield description of the four-dimensional and the three-dimensional models discussed above. In particular, the Hamiltonians and the corresponding supercharges are written.

  \cleardoublepage
\chapter{Supersymmetric extension of Poincar\'e symmetry}\label{chap2}

\vfill

\begin{intro}

This chapter is purely introductory and is devoted to supersymmetric field theories in four-dimensional Minkowski space.

\vskip 2mm

We explain {\em what} is supersymmetry and why it is the only possible nontrivial extension of the Poincar\'e symmetry. We discuss main properties of any supersymmetric field theory and  motivate {\em why} supersymmetric theories are interesting.

\vskip 2mm

Finally, we show how to work with such theories and explain the superfield formalism. As an illustration, we consider a simplest possible supersymmetric example -- the Wess-Zumino model which describes supersymmetric dynamics of a complex scalar field.

\end{intro}
\vfill
\clearpage

In a supersymmetric field theory, interactions between particles are fine-tuned in a special way, so that an additional continuous symmetry -- {\em supersymmetry} -- emerges. This symmetry mixes bosons and fermions (particles with different statistics) between each other.

Supersymmetry admits  natural resolution of certain inconsistency problems of field theories. For instance, a vacuum in a field theory usually has infinite energy density. In a theory with unbroken supersymmetry, however, the energy of a vacuum is exactly zero. This subject is discussed in details in Section~\ref{sec1.7.4}.

The infinite vacuum energy density in a field theory does not produce a problem by itself. Being coupled to gravity, however, such a theory becomes inconsistent. Thus, every known consistent field theory with gravity must be supersymmetric. In a similar manner, supersymmetry is included into every consistent string theory.

Another notable property of supersymmetric field theories is the equality in the number of bosonic and fermionic particles. It is not what we observe experimentally.
This does not mean, however, that the idea of supersymmetry is altogether unreasonable. Indeed, supersymmetry may describe particle interactions at very small distances and very high energies and be broken at our energy scale. Assuming that it is the case, the extra predicted particles may acquire large masses, which explains the fact that they have not been observed so far.
In addition, supersymmetric gauge theories provide the lightest supersymmetric particle as a natural candidate for dark matter.

All physical phenomena up to the TeV scale are well described by the Standard Model. Despite its success, however, it is conceivable that a new theory has to exist beyond the TeV scale. One reason is that we need a Higgs boson to break the electroweak symmetry. The radiative corrections to the mass of the Higgs boson are quadratically divergent and thus give an unacceptable large contribution if the cutoff scale is not of the TeV scale. This is called the naturalness problem. One way to address it is to consider a supersymmetric extension of the Standard Model. In the presence of supersymmetry, the mass of the Higgs boson is the same as the mass of its fermionic partner, while the fermion mass obtains only a logarithmic divergence due to the fact that an additional chiral symmetry appears in the absence of the mass term.

Of course, supersymmetry should be broken below the TeV scale to be able to describe our non-supersymmetric world. One can introduce supersymmetry breaking terms by hand. They should break supersymmetry softly in the sense that quadratic divergences should be absent. Alternatively, supersymmetry can be broken dynamically, so that the soft terms are generated in the a energy effective theory.

There is another aspect which makes supersymmetry attractive for a theorist.
In particle physics, symmetries restrict particle dynamics and allow one to make theoretical predictions on kinematical grounds without actually doing any concrete dynamical calculation.
The introduction of the extra symmetry on top of the Poincar\'e symmetry imposes more constraints on the amplitudes in a theory and makes it more accessible for theoretical studies.
In it even believed that in some cases supersymmetry makes a theory exactly solvable.
In fact, supersymmetry is a powerful instrument to study strong coupling dynamics and non-perturbative effects analytically.

In addition, theoretically appealing property of supersymmetry is that it offers the only ``loophole'' to the Coleman-Mandula theorem (see Section~\ref{sec2.4}) which prohibits any nontrivial extension of the Poincar\'e symmetry in a field theory.

It is possible to make a supersymmetric theory from almost any field theory. There exist en effective {\em superfield} technique for this. It is discussed in details in this chapter.

In a certain way, supersymmetric extension of a theory can be compared with the extension of real numbers $\mathbb R$ to complex numbers $\mathbb C$. Indeed, analytical functions on the complex plane are much more constrained than functions of real argument. It is well known that an analytical function of a complex argument, given in some region of the complex plane, can be uniquely analytically continued to other regions or even to the whole complex plane. It it also well known that an analytical function can be reconstructed from the knowledge of its zeros and poles, which is not the case for functions of real argument.

These central properties of complex functions are of great importance in supersymmetric theories: the introduction of supersymmetry renders  physical observables (e.g. amplitudes) depend on the parameters of the theory analytically. This allows one to calculate these quantities in one region of the theory (for example, in the region of the weak coupling, where the calculations can be carried out perturbatively) and then analytically continue the results to the strong coupling regime. As an example of such analysis, let us mention the paper by Seiberg and Witten \cite{SW} who studied the supersymmetric extension of quantum chromodynamics (QCD) and showed analytically that this theory is confining.

Supersymmetry, however, is reacher than just the complex analysis. Whereas the complex extension of real numbers is unique, several supersymmetric extensions of a quantum theory may be possible. A theory may have an ordinary or an extended supersymmetry. A quantum field theory in four space-time dimensions may have as much as four independent supersymmetries. The four-dimensional field theories with gravity may have as much as eight independent supersymmetries.
The number of independent supersymmetries is usually counted by the $\ca N$ symbol, e.g. $\ca N=2$ means that a field theory has two independent supersymmetries.

The more is the number of supersymmetries, the more a theory is constrained. This generally makes the theory more amenable for theoretical studies. The best known example is $\ca N =4$ super-Yang-Mills theory -- a theory similar to QCD, but extended to have four different supersymmetries. It is believed that this theory is exactly solvable. Still, this theory is very different from QCD, having zero $\beta$-function, no dimensional transmutation and no confinement. The simplest supersymmetric extension of QCD (with one supersymmetry) is still too complicated to be understood analytically.
The $\N=2$ super-Yang-Mills theory (which was studied by Seiberg and Witten) is intermediate between $\N=1$ and $\N=4$ theories.

Being invented as a form of mathematical construction, supersymmetry
produced the deepest impact on theoretical physics over the last several decades and became an essential part of modern high-energy physics.

\section{Basic notations in four-dimensional Minkowski space}

We denote the coordinates in four-dimensional Minkowski space as $x^\mu$ with the Lorentz indices taken from the middle of the Greek alphabet,
\begin{equation}
\mu,\, \nu,\, \rho,\,\dots = 0,1,2,3.
\end{equation}
Thereby, the coordinate  $x^0$ is associated with time. As for the three space components, $x^i$, we use the indices from the middle of the Latin alphabet,
\begin{equation}
 i,\, j,\, k,\, \dots = 1,2,3.
\end{equation}
Also, it is convenient to use the vectorial notation $\vec x$ for the space components of the four-vector $x^\mu$.

The Minkowski space metric tensor is
\begin{equation}
 g_{\mu\nu} = \left(
\begin{array}{cccc}
 1&0&0&0
\\
 0&-1&0&0
\\
 0&0&-1&0
\\
 0&0&0&-1
\end{array}
\right)
\equiv \diag \left(1,-1,-1,-1\right).
\end{equation}
As usual, it is used for raising and lowering Lorentz indices. For instance, one has for a tensor $A_{\mu\nu}$ with two Lorentz indices:
\begin{equation}
 A^{\mu}_{\,\,\nu} = g^{\mu\rho} A_{\rho\nu},
\quad\quad
 A_{\mu\nu} = g_{\mu\rho} A^{\rho}_{\,\,\nu},
\end{equation}
where $g^{\mu\nu}$ is the inverse metric tensor, which is equal to $g_{\mu\nu}$. As usual, summing over the repeated indices is assumed, as in the formulas above.

Supersymmetry involves fermions. Consequently, spinors and spinor notations are extensively exploited in all the chapters. The spinorial indices are denoted with undotted and dotted Greek letters from the beginning of the alphabet:
\begin{equation}
\alpha,\beta = 1,2
\quad\quad\mbox{and}\quad\quad
\dot \alpha, \dot\beta = 1,2.
\end{equation}
Throughout this chapter the following four-dimensional matrices are used:
\begin{equation}\label{sigma4d}
 \left(\sigma^\mu\right)_{\alpha\dot\alpha} = \left\{1,\, \vec \sigma\right\}_{\alpha\dot\alpha},
\quad\quad
 \left(\bar\sigma^\mu\right)^{\dot\alpha\alpha} = \left\{1,\, -\vec\sigma\right\}^{\dot\alpha\alpha},
\end{equation}
where $\vec \sigma$ are ordinary Pauli matrices. Note that starting from the next chapter, where a quantum-mechanical formalism is involved, the Euclidean version of these matrices will be used, see Eq.~(\ref{sigmas}).

\section{Poincar\'e group and Poincar\'e algebra}

The Poincar\'e group in Minkowski space parametrized by the coordinates $x^\mu$ can be realized by linear transformations
\begin{equation}\label{trans}
 x'^\mu = \Lambda^\mu_\nu x^\nu + c^\mu
\end{equation}
which preserve the space-time interval
\begin{equation}
 ds^2 = g_{\mu\nu} dx^\mu dx^\nu.
\end{equation}
The subgroup of homogeneous transformations (i.e. those with parameters $\Lambda^\mu_\nu$) form the Lorentz group ${\rm O}(1,3)$. The invariance of $ds^2$ implies
\begin{equation}\label{eq111}
 g_{\mu\nu}\Lambda^\mu_\rho \Lambda^\nu_\sigma = g_{\rho\sigma}
\end{equation}
and, as a consequence, ${\rm det} \,\Lambda = \pm 1$. Here we skip the consideration of the discrete Poincar\'e transformations (i.e. space-time reflections) some of which are related to the $\det \Lambda = -1$ branch of solutions of Eq.~(\ref{eq111}). Instead, we take the proper subgroup in the Lorentz group with ${\rm det}\, \Lambda = 1$. The infinitesimal (infinitely small) Lorentz transformation can be written as
\begin{equation}\label{eqomega}
 \Lambda^\mu_\nu = \delta^\mu_\nu + \omega^\mu_\nu,
 \quad\quad
  \omega_{\mu\nu} = - \omega_{\nu\mu},
\end{equation}
where, as usual, $\omega_{\mu\nu} = g_{\mu\rho}\omega^\rho_\nu$.
In this way, the infinitesimal form of the transformations (\ref{trans}) is given by
\begin{equation}
 \delta x^\mu = -i\left[c^\nu\hat P_\nu + \frac 12 \omega^{\nu\rho}\hat M_{\nu\rho}\right]x^\mu,
\end{equation}
where the differential operators
\begin{equation}\label{PM}
 \hat P_\mu = i\,\partial_\mu,
 \quad\quad
 \hat M_{\mu\nu} = -i\left(x_\mu\partial_\nu- x_\nu\partial_\mu\right)
\end{equation}
are the generators of the Poincar\'e algebra. The infinitesimal form for the action of the Poincar\'e group on functions $f(x)$ on the Minkowski space is
\begin{equation}\label{ftrans}
 f'\left(x\right) =f(x) - i\left[c^\nu\hat P_\nu + \frac 12 \omega^{\nu\rho}\hat M_{\nu\rho}\right] f(x).
\end{equation}

The Poincar\'e algebra generators -- four translations $P_\mu$ and six space-time rotations $M_{\mu\nu}$ -- form the Poincar\'e algebra~\footnote{
Note that here and below we omit ``hats'' on the operators $P_\mu$ and $M_{\mu\nu}$.}
\begin{equation}
\begin{array}{l}
 [ P_\mu\,,\, P_\nu] = 0,
\\[2mm]
 [ M_{\mu\nu}\,,\, P_\lambda ]= i\left( g_{\mu\lambda}\, P_\nu - g_{\nu\lambda}\, P_\mu \right),
\\[2mm]
[ M_{\mu\nu}\,,\, M_{\rho\sigma}]
= i\left( g_{\mu\rho}\,  M_{\nu\sigma}+g_{\nu\sigma}\,  M_{\mu\rho}
  -g_{\nu\rho}\,  M_{\mu\sigma} - g_{\mu\sigma}\,  M_{\nu\rho}
\right).
\end{array}
\label{eq1}
\end{equation}
Thus, the Poincar\'e algebra in Minkowski space in four dimensions has 10 independent generators: four space-time shifts, three space rotations and three boosts (transformations to other inertial reference frames).

\section{Two-component spinor notation}

Supersymmetry unifies bosons and fermions and thus extensively uses the spinorial formalism. Here we recall the basic properties of this formalism in four-dimensional Minkowski space.

Four-dimensional spinors realize irreducible representation of the Lorentz group (which has six generators: three spatial rotations and three Lorentz boosts). There are two types of spinors: left-handed and right-handed, which are marked by undotted and dotted indices, respectively, in the following way:
\begin{equation}
\begin{array}{lll}
 \mbox{left-handed:}\quad\quad& \xi_\alpha,\quad& \alpha = 1,2,
\\
 \mbox{right-handed:}\quad\quad& \bar\eta_{\dot\alpha},\quad& \dot\alpha = 1,2.
\end{array}
\end{equation}

It is possible to lower and raise spinor indices with the invariant Levi-Civita tensor {\em from the left}. For instance,
\begin{equation}
 \chi^\alpha = \ve^{\alpha\beta}\chi_\beta,
 \quad\quad
 \chi_\alpha = \ve_{\alpha\beta}\chi^\beta
\end{equation}
and similar for spinors with dotted indices. The two-index antisymmetric Lorentz-invariant Levi-Civita tensors $\ve^{\alpha\beta}$, $\ve^{\dot\alpha\dot\beta}$, $\ve_{\alpha\beta}$, and  $\ve_{\dot\alpha\dot\beta}$ are defined as
\begin{equation}
\begin{array}{l}
 \ve^{\alpha\beta} = -\ve^{\beta\alpha},
\quad
\ve_{\alpha\beta} = -\ve_{\beta\alpha},
 \quad \ve_{12}=-\ve^{12} = 1,
\\
 \ve^{\dot\alpha\dot\beta}=-\ve^{\dot\beta\dot\alpha},
\quad
\ve_{\dot\alpha\dot\beta}=-\ve_{\dot\beta\dot\alpha},
\quad \ve_{\dot 1\dot 2}=-\ve^{\dot 1\dot 2} = 1.
\end{array}
\end{equation}

The Lorentz transformation law for the undotted (left) spinors can be written as
\begin{equation}
 \xi'_\alpha = U_\alpha^\beta \xi_\beta,
\end{equation}
where the matrix $U$ has the form
\begin{equation}\label{eqU}
 U=\exp\left(-\frac i2 \omega_{\mu\nu} \sigma^{\mu\nu}\right)
\end{equation}
with $\omega_{\mu\nu}$ being the same as in Eqs.~(\ref{eqomega}), (\ref{ftrans}). The matrices $\sigma^{\mu\nu}$ give a particular {\em matrix} realization of the Lorentz rotations $M^{\mu\nu}$ and satisfy the last line in Eqs.~(\ref{eq1}). To be more specific,
\begin{equation}\label{sigmamunu}
 \sigma^{\mu\nu} =\frac i4\left(\sigma^\mu\bar\sigma^\nu - \sigma^\nu\bar\sigma^\mu\right)
\end{equation}
with the matrices $\sigma^\mu$ and $\bar\sigma^\mu$ being introduced in Eq.~(\ref{sigma4d}).

Let us consider a spatial rotation. The matrix from Eq.~(\ref{eqU}) takes the following form:
\begin{equation}
 U_{\rm rot} = \exp \left(-i\frac \theta 2 \vec n\vec \sigma\right),
 \quad\quad
 \theta\, n^i = \frac 12\ve^{ijk} \omega^{jk},
\end{equation}
where $\ve^{ijk}$ is antisymmetric Levi-Civita tensor ($\ve^{123}=1$), $\theta$ is the rotation angle and $\vec n$ -- the unit vector denoting the axis of rotation.
Analogously for a Lorentz boost, the matrix from Eq.~(\ref{eqU}) has the form
\begin{equation}
 U_{\rm boost} = \exp \left(\frac \phi 2 \vec n'\vec \sigma\right),
 \quad\quad
 \phi\, n^i = \omega^{oi}.
\end{equation}
Here $\tanh \phi = v$, where $v$ is the velocity in the units of speed of light of the first inertial reference frame with respect to the second, $\vec n'$ denotes the velocity direction. Note that in the case of the spatial rotation the matrix $U_{\rm rot}$ is unitary, $U^\dagger_{\rm rot} U_{\rm rot} = 1$, whereas in the case of the Lorentz boost the matrix $U_{\rm boost}$ is not. This reflects the fact that the Lorentz group is the noncompact ${\rm O}(1,3)$ group rather than the compact ${\rm O}(4)$ group.

Dotted spinors transform as complex conjugates of undotted spinors:
\begin{equation}
 \bar\eta_{\dot\alpha} \sim \left(\eta_\alpha\right)^*,
\end{equation}
where the sign $\sim$ means ``is transformed as''. Therefore, for dotted spinors the Lorentz transformation goes with the complex conjugated matrix,
\begin{equation}
 \bar\eta'_{\dot\alpha} = \left(U^*\right)_{\dot\alpha}^{\dot\beta} \bar\eta_{\dot\beta},
\end{equation}
where
\begin{equation}\label{eqUb}
 U^*=\exp\left(\frac i2 \omega_{\mu\nu} \bar\sigma^{\mu\nu}\right),
\end{equation}
and the matrices
\begin{equation}\label{barsigmamunu}
 \bar\sigma^{\mu\nu} =\frac i4\left(\bar\sigma^\mu\sigma^\nu - \bar\sigma^\nu\sigma^\mu\right)
\end{equation}
give another matrix realization of the Lorentz rotations $M^{\mu\nu}$ and satisfy the last equation in Eqs.~(\ref{eq1}).

Similarly for dotted spinors, one has for particular cases of spatial rotations and Lorentz boosts:
\begin{equation}
 \bar\eta'^{\dot\alpha}=\left\{
\begin{array}{ll}
 \left(U_{\rm rot}\right)_{\dot\beta}^{\dot\alpha}\, \bar\eta^{\dot\beta}, \quad& \mbox{for rotations},
 \\[2mm]
 \left(U_{\rm boost}^{-1}\right)_{\dot\beta}^{\dot\alpha}\, \bar\eta^{\dot\beta}, \quad& \mbox{for boosts},
\end{array}
\right.
\end{equation}
where for convenience the  index  for the spinor $\bar\eta_{\dot\alpha}$ is raised with the antisymmetric Levi-Civita tensor. Note that under spatial rotations the undotted spinor $\xi_\alpha$ and the dotted spinor $\bar\eta^{\dot\alpha}$ transform under one and the same matrix $U_{\rm rot}$.

The spinors $\xi_\alpha$ and $\bar\eta_{\dot\alpha}$ are referred to as {\em Weyl spinors}. In Minkowski space in four dimensions one undotted and one dotted Weyl spinor comprise one Dirac spinor (see, for example, the textbook \cite{LL4} for a more detailed description).

In order to be Lorentz-invariant, an equation which involves spinors must have the same number of undotted and dotted indices on each side, otherwise the equation becomes invalid under a change of reference frame. One should also remember, however, that complex conjugation implies the interchange of dotted and undotted indices. For instance, the relation
\begin{equation}
 \left(\xi_{\alpha\beta}\right)^* = \bar\eta_{\dot\alpha\dot\beta}
\end{equation}
is Lorentz-invariant.

Lorentz scalars can be built by convolution of either undotted or dotted spinor indices. For example, the products
\begin{equation}
 \chi^{\alpha}\xi_\alpha
\quad\quad\mbox{and}\quad\quad
\bar\psi_{\dot\beta}\bar\eta^{\dot\beta}
\end{equation}
are invariant under the Lorentz transformations.

\section{The Coleman-Mandula (no-go) theorem}\label{sec2.4}

The Poincar\'e algebra (\ref{eq1}) forms basis of geometric symmetries of a relativistic field theory. Other symmetries like flavour symmetry, isospin symmetry, {\em etc.} commute with the Poincar\'e group and are {\em internal} in the sense that they have nothing to do with Minkowski space. A natural question arises: is it possible to extend the Poincar\'e group with an additional symmetry which affects space-time coordinates? It was believed for a long time that this is not possible. In 1967 Coleman and Mandula
formulated a theorem which states that, in a dynamically nontrivial relativistic quantum field theories of space-time dimension $d\ge 3$ with interactions and with asymptotic states (particles), no geometric extension of the Poincar\'e group is possible~\cite{Colem}. In other words, besides already known conserved generators carrying Lorentz indices (the energy-momentum operator $P_\mu$ and the Lorentz transformations $M_{\mu\nu}$) no such new conserved charges (algebra generators) can appear. According to the theorem, the only allowed additional conserved charges must be Lorentz scalars, such as the electromagnetic charge. However, in 1970 Golfand and Likhtman
found a loophole in this theorem \cite{A1} which, together with the Coleman-Mandula theorem, singles out supersymmetry as the only possible geometric extension of the Poincar\'e invariance in a relativistic field theory. A reason for which this statement is not valid in one and in two space-time dimensions will become clear shortly.

\begin{wrapfigure}{r}{0.5\hsize}
 \centering
 \includegraphics[width=0.9\hsize,bb=1 0 340 209,keepaspectratio=true]{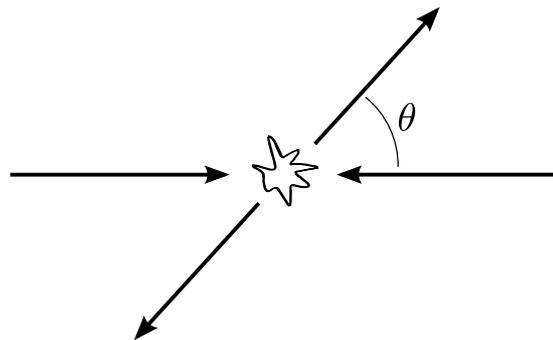}
 \caption{Two-particle scattering. Only the scattering angle $\theta$ is undefined from energy and momentum conservation laws.}
\label{222scattering}
\end{wrapfigure}

The essence of the proof of the Coleman-Mandula theorem is the following. Let us take an interacting field theory and consider two-particle scattering process. The energy and momentum conservation laws present in every Poincar\'e-invariant field theory. Particularly, in our case
\begin{equation}
  p_1^\mu + p_2^\mu =  p_3^\mu + p_4^\mu,
\quad\quad
 \mu = 0,1,2,3,
\end{equation}
where $p^\mu_{1}$, $p^\mu_2$ are the particle 4-momenta before the interaction while $p^\mu_{3}$, $p^\mu_{4}$ are the particle 4-momenta after the interaction.
The kinematic constraints above leave only one essential free parameter -- the scattering angle $\theta$, see Fig.~\ref{222scattering}. This angle cannot be determined on kinematical grounds and is defined by particular dynamics in the theory.
Imagine now that there is an additional symmetry generator of space-time with some Lorentz indices.
An additional exotic conservation law which have the same Lorentz indices  corresponds to this symmetry and involves the particles 4-momenta.
The presence of this conservation law would completely fix the scattering angle $\theta$ (or at most would leave only a discrete set of possible angles). Since the scattering amplitude is an analytic function of the angle, it then must vanish for all angles. In other words, the theory has trivial $S$-matrix, i.e. it is non-interacting.

Consequently, the Coleman-Mandula theorem is not applicable in one and in two space-time dimensions, where there is no scattering angle between the two particles.
The details of the proof and also its generalization to non-identical particles, particles with spin, {\em etc.} can be found in Refs.~\cite{Colem,sweinberg,wirend}.

Thus, no geometric extension of the Poincar\'e symmetry is possible on {\em asymptotic states} in a nontrivial field theory. Saying this differently, either the theory dynamics is trivial or the theory has no asymptotic states (particles), or extra geometric symmetries in the theory are broken on asymptotic states. The latter two statements can be illustrated on an example of a {\em conformal} field theory. The conformal symmetry adds scale invariance to a theory. For instance, any conformal field theory possesses an extra space-time symmetry
\begin{equation}
 x'^\mu = \lambda x^\mu,
\end{equation}
where $\lambda$ is an arbitrary positive number. In its infinitesimal form $\lambda=1+\epsilon$, $|\epsilon|\ll 1$ and
\begin{equation}
 x'^\mu = x^\mu + i\epsilon\hat D x^\mu,
\end{equation}
where $\hat D$ is the so called dilatational operator,
\begin{equation}
 \hat D = -i x^\mu\partial_\mu
\end{equation}
Together with certain additional conformal operators (special conformal transformations usually denoted as $\hat K_\mu$), the Poincar\'e algebra (\ref{eq1}) extends in a nontrivial way. However, the asymptotic states (particles) in the theory would break conformal symmetry down to the Poincar\'e symmetry, in full accordance with Coleman-Mandula theorem. The second possibility -- the theory has no asymptotic states at all. Such a theory is scale invariant so that the distance between two points in space is undetermined. For instance, $\N = 4$ super-Yang-Mills theory is of this kind.

\section{Supersymmetric extension of the Poincar\'e algebra}\label{suexpoal}

The Coleman-Mandula theorem assumes that all symmetry generators in a relativistic field theory are operators which possibly have some Lorentz indices (or, equivalently, even number of spinor indices) and thus are bosonic operators, i.e. they form a Lie algebra with certain {\em commutation relations}. Meanwhile, generators with odd number of spinor indices are of fermionic nature and are not considered in the proof of the theorem since they cannot participate in commutation relations.
The loophole in the theorem consists in the possibility to introduce the operators $Q_\alpha$ and ${\bar Q}_{\dot\alpha}$ with {\em spinor} indices.
The algebra which now includes such operators must involve not only commutators, e.g. $[B_1, \, B_2]$ and $[Q_\alpha,\, B_3]$ (with $B_{1,2,3}$ being the bosonic operators), but also anticommutators, e.g.
$\{Q_\alpha, \, \bar Q_\dalpha\}$.
Hence, the spinor operators $Q_\alpha$, $\bar Q_\dalpha$, due to their nature, cannot produce additional restrictions on particle momenta in scattering processes so that no new conservation laws appear. Nevertheless, they relate to each other various scattering amplitudes which greatly constrains the $S$-matrix in a quantum field theory.

The complex operators $Q_\alpha$ and ${\bar Q}_{\dot\alpha}$ are Hermitian conjugated,
\begin{equation}\label{c2Qbar}
{\bar Q}_{\dot\alpha} = \left(Q_\alpha\right)^\dagger, 
\end{equation}
and transform as ordinary Weyl spinors under the action of the Poincar\'e algebra, namely they satisfy the following commutation relations with the Poincar\'e algebra generators:
\begin{equation}
\begin{array}{l}
[P_\mu,\, Q_\alpha] = [P_\mu,\, {\bar Q}^{\dot\alpha}] =0,
\\[2mm]
[M^{\mu\nu},\, Q_\alpha] =  i \left(\sigma^{\mu\nu}\right)_\alpha^{\,\,\,\beta}\,Q_\beta,
\\[2mm]
[M^{\mu\nu},\, {\bar Q}^{\dot \alpha}] =  i \left(\bar\sigma^{\mu\nu}\right)^{\dot\alpha}_{\,\,\,\dot\beta}\, {\bar Q}^{\dot \beta},
\label{Qspinor}
\end{array}
\end{equation}
where the index for the supercharge $\bar Q_\dalpha$ is raised with the antisymmetric Levi-Civita tensor, $\bar Q^\dalpha=\ve^{\dalpha\dbeta}\bar Q_{\dbeta}$, and the matrices $\sigma^{\mu\nu}$ and $\bar\sigma^{\mu\nu}$ were introduced in Eqs.~(\ref{sigmamunu}) and~(\ref{barsigmamunu}).

The operators $Q_\alpha$ and ${\bar Q}_{\dot\alpha}$ are also referred to as {\em supercharges} or {\em supergenerators}. According to their indices, the minimum number of such supergenerators in four-dimensional Minkowski space is four. To close the algebra (\ref{eq1}), (\ref{Qspinor}), one needs to specify the anticommutators $\{Q_\alpha, {\bar Q}_{\dot\alpha}\}$, $\{Q_\alpha, Q_\beta\}$ and $\{\bar Q_\dalpha, \bar Q_\dbeta\}$. The first anticommutator can only be proportional to $P_\mu \left(\sigma^\mu\right)_{\alpha\dot\beta}$ since it is the only operator with the appropriate Lorentz indices. The standard normalization is
\begin{equation}
\{Q_\alpha, {\bar Q}_{\dot\alpha}\} = 2 P_\mu\left( \sigma^\mu\right)_{\alpha\dot\alpha}.
 \label{eq2}
\end{equation}
The simplest choice for the other two anticommutators allowed by the Jacobi identities is
\begin{equation}
\{Q_\alpha,  Q_{\beta}\} = \{{\bar Q}_{\dot\alpha},  {\bar Q}_{\dot\beta}\} = 0.
\label{eq3}
\end{equation}
Thereby, Eqs.~(\ref{eq1}), (\ref{Qspinor}), (\ref{eq2}), (\ref{eq3}) form the {\em super-Poincar\'e algebra}  first obtained by Golfand and Likhtman \cite{A1}.

This minimal super-Poincar\'e algebra with four supercharges can be further extended with additional supercharges. As was demonstrated in Ref.~\cite{HaagL}, one can construct {\em extended supersymmetries}, with up to sixteen supercharges in four dimensions.
The minimal supersymmetry is referred to as
${\mathcal N}=1$. Correspondingly, one can consider ${\mathcal N}=2$ (eight supercharges) or ${\mathcal N}=4$ (sixteen supercharges). The extended supersymmetry in a four-dimensional field theory is discussed in Section~\ref{c2sec2.7}.

As was also demonstrated in Ref.~\cite{HaagL}, it is possible to modify the super-Poincar\'e algebra by an introduction of central charges in it. Such superalgebras are referred to as {\em centrally extended}. A central charge is an element of the superalgebra which commutes with other generators. It acts as a number with numerical value being dependent on a sector of a theory under consideration. The presence of central charges reflect possible existence of conserved topological currents and topological charges \cite{WiOli}.
For instance, if a theory under consideration supports topologically stable domain walls, the right-hand side of (\ref{eq3}) can be modified in the following way:
\begin{equation}\label{eq4}
 \{Q_\alpha, Q_{\beta}\} = C_{\alpha\beta},
\quad\quad
 \{\bar Q_\dalpha, \bar Q_{\dbeta}\} = \left(C_{\alpha\beta}\right)^\dagger.
\end{equation}
Here $C_{\alpha\beta}=C_{\beta\alpha}$ are the central charges.
Let us remark that they have spinor indices and thus transform under Lorentz rotations. This is why $C_{\alpha\beta}$ are also called tensor central charges to distinguish them from ``standard'' central charges which commute with all superalgebra generators.
(See also the footnote in Section~\ref{c2sec2.7}, where it is shown how the central charges $Z^{IJ}$, which are Lorentz scalars, can be introduced for the case of extended supersymmetry.)

\section{Main properties of supersymmetric field theories}\label{sec2.6}

For any supersymmetric field theory, the following fundamental properties hold:
\begin{itemize}
 \item a state in a supersymmetric field theory cannot have negative energy;
 \item if supersymmetry is {\em unbroken}, the vacuum has exactly zero energy;
 \item if there is a boson with mass $m$, there must exist a fermion with exactly the same mass $m$, and {\em vice versa} (Bose-Fermi degeneracy);
 \item any supersymmetric field theory has equal number of bosonic and fermionic degrees of freedom in every {\em supermultiplet}.
\end{itemize}
Let us discuss these statements in details.

\subsection{Non-negative energy of an eigenstate and vanishing of the vacuum energy}

The first consequence which follows from the super-Poincar\'e algebra is the fact that a state in a quantum field theory cannot have negative energy. This straightforwardly follows from Eqs.~(\ref{c2Qbar}) and (\ref{eq2}) if one takes the sum
\begin{equation}
P^0 =
\frac{1}{4} \sum_{\alpha=1}^2 \left[Q_\alpha\left(Q_\alpha \right)^\dagger + \left(Q_\alpha \right)^\dagger Q_\alpha \right]  
\end{equation}
and calculates an average of the left and the right hand sides for a normalized eigenstate $\ket{\Psi}$ with the energy $E$. Indeed,
\begin{multline}\label{eq5}
\left<\Psi \left| P^0\right|\Psi\right> = E
=
\frac 14\sum\limits_{\alpha=1}^2\left< \Psi \left|  Q_\alpha\left(Q_\alpha
\right)^\dagger + \left(Q_\alpha \right)^\dagger Q_\alpha\right| \Psi\right>
\\[2mm]
=
\frac 14\sum\limits_{\alpha=1}^2
\big< \left(Q_\alpha\right)^\dagger \Psi \big|\left(Q_\alpha\right)^\dagger\Psi\big>^*
 +
\frac 14\sum\limits_{\alpha=1}^2\big<Q_\alpha\Psi \big| Q_\alpha\Psi\big>^*
\end{multline}
The second line in this equality is always non-negative. Thus, for any quantum eigenstate its energy $E \ge 0$.

The minimum $E=0$ is achieved on a {\em vacuum} state $\ket{0}$ which is annihilated by the supercharges,
\begin{equation}
Q_\alpha|0\rangle = \left(Q_\alpha \right)^\dagger |0\rangle =0.
\end{equation}
A field theory may have one or several vacua with zero energy. If a theory have no states with zero energy, i.e. $E_{\rm vac} > 0$, the supersymmetry is spontaneously broken. In fact, the vanishing of the vacuum energy is the necessary and sufficient condition for supersymmetry to be left unbroken.

\subsection{Bose-Fermi degeneracy}\label{s2.6.2}

In a supersymmetric theory, if there is a boson with the mass $m$, a fermion with the very same mass $m$ must exist too, and {\em vice versa}.

To elaborate more on this point, let us introduce a bosonic state $\ket{B}$ with the mass $m$ and associate with it one of the following fermionic states: $Q_\alpha\ket{B}$, $\left(Q_\alpha\right)^\dagger\ket{B}$, where $\alpha=1,2$. At least one of these four states is nonzero. Indeed, the sum of the norms of these four states in positive:
\begin{multline}
\sum\limits_{\alpha=1}^2
\big< \left(Q_\alpha\right)^\dagger B \big|\left(Q_\alpha\right)^\dagger B\big>
 +
\sum\limits_{\alpha=1}^2\big<Q_\alpha B \big| Q_\alpha B\big>
\\[2mm]
=
\sum\limits_{\alpha=1}^2\left< B \left|  Q_\alpha\left(Q_\alpha
\right)^\dagger + \left(Q_\alpha \right)^\dagger Q_\alpha\right| B\right>^*
=4\left<B \left| P^0\right|B\right>^* = 4 E_B,
\end{multline}
where $E_B > 0$ is the state $\ket{B}$ energy.

Note also that $P^2= P_\mu  P^\mu$ which is a Casimir operator of the Poincar\'e algebra (it commutes with all the Poincar\'e algebra generators) is also a Casimir operator of the super-Poincar\'e algebra, because
\begin{equation}
[P^2,\, Q_\alpha] = [P^2,\, {\bar Q}_{\dot\alpha}] = 0.
\end{equation}
Thus, from $P^2 \ket{B}=m^2 \ket{B}$ follows that
\begin{equation}
P^2 \big|Q_\alpha B\big> = m^2 \big|{Q_\alpha B}\big>
\quad\quad
 \mbox{and}
\quad\quad
 P^2\ket{\left(Q_\alpha\right)^\dagger B} = m^2 \ket{\left(Q_\alpha\right)^\dagger B}.
\end{equation}

Combining the two observations above, one arrives to the statement of this section. In a similar manner, one can prove the reverse statement:
for a fermion with the mass $m$ there exist a boson with the very same mass $m$.

\subsection{Supermultiplets}

The Poincar\'e group is not a compact group. That is why all its unitary representations (except for the trivial representation) are infinite-dimensional. This infinite dimensionality reveals itself in a widely known fact that particle states are labeled by the continuous parameters -- particle 4-momentum $p_\mu$.

The Poincar\'e algebra has two Casimir operators: $P^2 = P_\mu P^\mu$ and 
$W^2 = W_\mu W^\mu$, where $W^\mu$ is Pauli--Lubanski vector,
\begin{equation}
 W^\mu = \frac{1}{2}\, \varepsilon^{\mu\nu\rho\sigma}\, P_\nu\, M_{\rho\sigma}
\end{equation}
($\ve^{\mu\nu\rho\sigma}$ is antisymmetric Levi-Civita tensor).
The eigenvalues of the operator $P^2$ fix particle mass squared,
$p_\mu p^\mu = m^2$, while the eigenvalues of the operator $W^2$ are responsible for particle spin if the particle mass is not zero.

To understand the latter statement, let us boost to a reference frame where the particle is at rest: $p_\mu = (m,\,0,\,0,\,0)$.
One can check that in this reference frame
\begin{equation}
 W^2 =  W_\mu  W^\mu  = - m^2 s(s+1),
\end{equation}
where $s$ is the particle spin.

For massless particles $P^2=0$ and $W^2=0$. Then, instead of spin, one must consider particle helicity. One can boost to the reference frame where the particle 4-momentum is $p_\mu = (E,\,0,\,0,\,E)$ with $E$ being the particle energy. Then the eigenvalues of the operator $M_{12}$ are $\pm \lambda$ with $\lambda$ being the helicity.

Hence, besides the particle 4-momentum $p_\mu$, a  unitary irreducible representation of the Poincar\'e algebra is identified by the particle mass $m$ and the particle spin or helicity $s$, if the particle has zero mass.
In contrast with $P^2$, the operator $W^2$ does not commute with the supercharges, i.e.
$[W^2 ,\, Q_\alpha]\neq 0$, as follows from Eq.~(\ref{Qspinor}). The same is true for the operator $M_{12}$.
Thus, massive irreducible superalgebra representations must contain particles with different spins, while massless irreducible superalgebra representations must contain particles with different helicities.
Due to the property, $Q_\alpha^2 = \bar Q_\dalpha^2 = 0$, the supercharges may change the particle into another particle with different spin/helicity a finite number of times. The corresponding set of particles, all with the same mass $m$, but with different spins/helicities is called a {\em supermultiplet}. In the simplest case a supermultiplet consists of two particles with spins $s$, $s+1/2$ or helicities $\lambda$, $\lambda+1/2$. Further details on building the supermultiplets can be found e.g. in \cite{Wess,wirend,sweinberg,shifsusy}.

\subsection{Equal number of bosonic and fermionic degrees of freedom in every supermultiplet}\label{sec1.7.4}

We omit here a formal proof of the equality of the number of bosonic and fermionic states in a supermultiplet. It will be given in the case of supersymmetric quantum mechanics in Section~\ref{sec3.2.4}.
Instead, let us discuss how this fact follows from the vanishing of the vacuum energy.

Consider a free field theory. It is well known that bosons and fermions contribute to the vacuum energy due to zero-point oscillations. The bosonic contribution is
\begin{equation}
\sum_B \sum_{\vec p}\,\sqrt {m_B^2+\vec p^{\,2}},
\end{equation}
where the (divergent) sum runs over all bosonic degrees of freedom and over all spatial momenta.
The fermionic contribution is
\begin{equation}
- \sum_F \sum_{\vec p}\,\sqrt {m_F^2+\vec p^{\,2}},
\end{equation}
where the sum runs over all fermionic degrees of freedom. The extra minus sign
is due to $-1$ associated with the fermion loop in the corresponding Feynman diagram which describes the vacuum energy density. The vanishing of the vacuum energy density requires the cancellation of two contributions which is possible only if the following equations hold inside each supermultiplet:

\begin{equation}\label{eq7}
n_B=n_F
\quad\quad
\left[\begin{array}{c}
 \mbox{equal number of bosons and fermions}
\\
\mbox{in a supermultiplet}
\end{array}\right]
\end{equation}
and
\begin{equation}\label{eq777}
 m_B = m_F
\quad\quad
\left[\begin{array}{c}
 \mbox{equal masses of bosons and fermions}
\\
 \mbox{ in a supermultiplet}
\end{array}\right].
\end{equation}
Note that the latter property was already proven in Section~\ref{s2.6.2} from algebraic considerations.
Note also that by bosons and fermions we mean physical (positive norm) degrees of freedom. For instance, a photon has two degrees of freedom corresponding to the two transverse polarizations (the two helicities $\pm 1$).

\section{Superspace and superfields}

In a relativistic field theory, fields are functions (probably, with some vector or spinor indices) which locally depend on the space-time point $x^\mu$ and transform in a certain way under the action of the Poincar\'e group.
With introduction of supersymmetry which is the geometric extension of the Poincar\'e symmetry, it is very natural to expand the space-time by an addition of appropriate extra dimensions. By doing so, one expands the concept of space-time to the concept of {\em superspace}.
In the superspace, the supercharges are realized as differential operators which generate  {\em supertranslations} in a way similar to the energy-momentum operator which generates translations in four-dimensional space-time.
Due to the anticommuting nature of the supercharges, the extra dimensions in the superspace are described by coordinates of Grassmann (anticommuting) nature. Finally, the concept of fields is extended to the concept of {\em superfields} which are functions of the coordinates on the superspace.
This breakthrough idea was pioneered by Salam and Strathdee \cite{salams}.

The immediate advantage of this formalism is that it gives simple and explicit description of the action of supersymmetry on {\em component fields} (see below) and provides a very efficient method for constructing manifestly supersymmetric Lagrangians.

In an ordinary $\N=1$ supersymmetry, the superspace
\begin{equation}\label{d4superspace}
\{ x^\mu,\, \theta^\alpha,\,\bar\theta^{\dot\alpha}\},
\quad\quad
\bar\theta^{\dot\alpha} \equiv \left(\theta^{\alpha}\right)^*
\end{equation}
includes four complex Grassmann (anticommuting) variables $\theta^\alpha$ and $\bar\theta^{\dot\alpha}$ which represent ``quantum'' or ``fermionic'' dimensions of the superspace.
They are complex conjugated and anticommute between each other,
\begin{equation}
\{\theta^\alpha ,\, \theta^\beta\}
= \{ \bar\theta^{\dot\alpha} ,\, \bar\theta^{\dot\beta}\}
=\{\theta^\alpha,\, \bar\theta^{\dot\beta}\} = 0.
\end{equation}
Note also the peculiarity of the Leibniz rule for Grassmann derivatives, e.g.
\begin{equation}
\frac{\partial}{\partial \theta^\alpha} \left(\theta^\beta\theta^\gamma\right) =
\left(\frac{\partial}{\partial \theta^\alpha} \theta^\beta\right)\theta^\gamma
-\theta^\beta \left(\frac{\partial}{\partial \theta^\alpha}\theta^\gamma \right).
\end{equation}
In addition, Hermitian conjugation changes the order of anticommuting numbers:
\begin{equation}
 \left(\theta^1 \theta^2\right)^\dagger = (\theta^2)^\dagger (\theta^1)^\dagger = \bar \theta^{\dot 2}\bar\theta^{\dot 1}.
\end{equation}

A superfield is a function of the coordinates~(\ref{d4superspace}) \cite{salams,ferraw}.
One can expand it in power series of the Grassmann variables $\theta^\alpha$ and $\bar\theta^{\dot\alpha}$.
This expansion has finite number of terms since the square of a given Grassmann parameter vanishes. Thus, the highest term in this expansion is
$\theta^2\bar\theta^2$, where $\theta^2 = \theta^\alpha\theta_\alpha$,
$\bar\theta^2 = \bar\theta_{\dot\alpha}\bar\theta^{\dot\alpha}$.
The most general superfield with no external indices has the following form:
\begin{equation}
S(x, \theta, \bar\theta) = \phi + \theta^\alpha\psi_\alpha
+ \bar\theta_\dalpha\bar\chi^\dalpha
+ \theta^2 F + \bar\theta^2 G
+\theta^\alpha A_{\alpha\dot \alpha} \bar\theta^{\dot\alpha}
+\theta^2 (\bar\theta_\dalpha\bar\lambda^\dalpha )
+\bar\theta^2 (\theta^\alpha\rho_\alpha) +\theta^2 \bar\theta^2 D,
\label{spin98}
\end{equation}
where $\phi$, $\psi_\alpha$, $\bar\chi^\dalpha$, $\dots$, $D$ depend only on $x^\mu$ and are referred to as the component fields.

In what follows we will use shorthand notations for contraction of spinor indices:
\begin{equation}
 A B = A^\alpha B_\alpha,
\quad\quad
 \bar A \bar B = \bar A_{\dot\alpha} \bar B^{\dot\alpha}.
\end{equation}
In particular, if $A$ and $B$ are {\em anticommuting} variables,
\begin{equation}
A B = B A,
\quad\quad
\bar A \bar B = \bar B\bar A,
\quad\quad
 \left(A B\right)^\dagger = \bar A \bar B.
\end{equation}

\subsection{Supersymmetry transformations and differential operators on superspace}

The representation of the supercharges $Q_\alpha$ and $\bar Q_\dalpha$ as differential operators on the superspace~(\ref{d4superspace}) can be derived in the following standard way. Let us associate with each point of the superspace~(\ref{d4superspace}) an element of the group corresponding to the ${\mathcal N}=1$ superalgebra (\ref{eq1}), (\ref{Qspinor}), (\ref{eq2}), (\ref{eq3}) as
\begin{equation}
G(x^\mu, \theta ,\bar \theta) =e^{i\left(-x^\mu  P_\mu
  +\theta^\alpha  Q_\alpha +\bar\theta_{\dot\alpha}{\bar Q}^{\dot\alpha} 
\right)}.
\label{eq10}
\end{equation}
Then the product of two elements
$G(0, \epsilon,\bar \epsilon)$ and
$G(x^\mu, \theta ,\bar \theta)$
is~\footnote{
All the elements~(\ref{eq10}) form an invariant space under the action of the Poincar\'e group: the result of the action of any Poincar\'e group element on~(\ref{eq10}) is of the same type. In fact, this space is invariant under the action of the whole super-Poincar\'e group. This statement partially follows from the equality~(\ref{e2.7.6}).

Let us also remark that the superspace~(\ref{d4superspace}) can be obtained as the factor of the super-Poincar\'e group over the Lorentz group much in the same way this can be done for the four-dimensional Minkowski space:
\begin{equation*}
 \frac{\mbox{Poincar\'e group}}{\mbox{Lorentz group}}
\quad\quad\longrightarrow\quad\quad
 \frac{\mbox{super-Poincar\'e group}}{\mbox{Lorentz group}}.
\end{equation*}
The points in this factor space are orbits obtained by the action of the Lorentz group on the super-Poincar\'e group space. If we choose a certain point as the origin, then the superspace can be parametrized by (\ref{eq10}).
}
\begin{equation}\label{e2.7.6}
G(0, \epsilon,\bar{\epsilon}) \,\, G(x^\mu, \theta ,\bar \theta)
=G(x^\mu
 + i\theta^\alpha \sigma^\mu_{\alpha\dot\alpha} \bar\epsilon^{\dot\alpha}
 - i\epsilon^\alpha\sigma^\mu_{\alpha\dot\alpha} \bar\theta^{\dot\alpha}
 \,,\, \theta +\epsilon \,,\,\bar \theta +\bar{\epsilon}).
\end{equation}
This equality can be proven by using the Hausdorff formula
\begin{equation}
e^Ae^B =e^{A+B +\frac{1}{2} [A,B] + ...}
\end{equation}
(where the ellipsis corresponds to infinite series of multi-commutator terms) and taking into account the fact that the series on the right-hand side terminate at the first commutator for the algebra elements considered here. While doing this calculation, one should also remember that the parameters $\epsilon$, $\bar\epsilon$, $\theta$, $\bar\theta$ as well as the supercharges $Q$, $\bar Q$ all anticommute between each other.

Thereby, the action of the group element
$G(0, \epsilon,\bar \epsilon)$ on $G(x^\mu, \theta ,\bar \theta)$
induces the following motion in the parameter space~(\ref{d4superspace}):
\begin{equation}\label{susytr}
\begin{array}{l}
x^\mu\rightarrow x^\mu 
 + i\theta^\alpha \sigma^\mu_{\alpha\dot\alpha} \bar\epsilon^{\dot\alpha}
 - i\epsilon^\alpha\sigma^\mu_{\alpha\dot\alpha} \bar\theta^{\dot\alpha},
\\[2mm]
  \theta_\alpha\rightarrow \theta_\alpha + \epsilon_\alpha,
\\[2mm]
  \bar\theta_\dalpha \rightarrow \bar\theta_\dalpha + \bar\epsilon_\dalpha.
\end{array}
\end{equation}
This motion is generated by the operator
$i\left(\epsilon^\alpha Q_\alpha + \bar\epsilon_\dalpha \bar Q^\dalpha\right)$, where the Hermitian-conjugated supercharges are
\begin{equation}
 Q_\alpha= -i\frac{\partial}{\partial \theta^\alpha} -
  \sigma^\mu_{\alpha\dalpha}\,\bar\theta^{\dot \alpha}\,\partial_\mu,
\quad\quad
 {\bar Q}_{\dot\alpha}=i\frac{\partial}{\partial \bar\theta^{\dot\alpha}}
  +\theta^{\alpha}\sigma^\mu_{\alpha\dalpha}\,\partial_\mu .
\end{equation}
They satisfy the anticommutation relations
\begin{equation}
\begin{array}{lll}
\big\{Q_\alpha,\,Q_\beta\big\}
  &=& \left\{\bar Q_\dalpha,\, \bar Q_\dbeta\right\} \,\,=\,\, 0,
\\[2mm]
\left\{ Q_\alpha,\, {\bar Q}_{\dot\alpha}\right\}
  &=& 2i\sigma^\mu_{\alpha\dot\alpha}\,\partial_\mu
\end{array}
\end{equation}
and hence, together with $P_\mu = i\partial_\mu$ and an appropriate expression for $M_{\mu\nu}$~\footnote{
If $M_{\mu\nu}$ is acting on a scalar, its expression is given in Eq.~(\ref{PM}). One must add certain (matrix) extra terms in this expression, if $M_{\mu\nu}$ is acting on a field with some vector or spinor indices.}
give an explicit realization of the supersymmetry algebra,
Eqs.~(\ref{eq1}), (\ref{Qspinor}), (\ref{eq2}), (\ref{eq3}).

One could study right multiplication instead of left multiplication in~(\ref{e2.7.6}) and would found that the induced motion is generated by a different operator
$\epsilon^\alpha D_\alpha + \bar\epsilon_\dalpha \bar D^\dalpha$, with the
operators $D_\alpha$ and $\bar D_\dalpha$ defined as
\begin{equation}\label{e2.7.14}
D_\alpha=\frac{\partial}{\partial \theta^\alpha}
+ i \sigma^\mu_{\alpha\dalpha}\,\bar\theta^{\dot \alpha}\,\partial_\mu,
\quad\quad
\bar D_{\dot\alpha}=-\frac{\partial}{\partial \bar\theta^{\dot\alpha}}
 -i\theta^{\alpha}\sigma^\mu_{\alpha\dalpha}\,\partial_\mu .
\end{equation}
with $\bar D_\dalpha = -\left(D_\alpha\right)^\dagger$.
Note that we have used a different convention of multipliers in the operators above on purpose: this will be convenient in subsequent sections.

The operators $D_\alpha$ and $\bar D_\dalpha$ are called {\em superderivatives}.
By their very definition, they satisfy the following anticommutation relations:
\begin{equation} 
\big\{D_\alpha,\,D_\beta\big\}
  = \left\{\bar D_\dalpha,\, \bar D_\dbeta\right\} = 0,
\quad\quad
\left\{ D_\alpha,\, {\bar D}_{\dot\alpha}\right\}
  = -2i\sigma^\mu_{\alpha\dot\alpha}\,\partial_\mu.
\end{equation}
In addition, the superderivatives and the supercharges anticommute:
\begin{equation}
\big\{D_\alpha,\,Q_\beta\big\}
=\left\{D_\alpha,\,\bar Q_\dbeta\right\}
=\left\{\bar D_\dalpha,\,Q_\beta\right\}
=\left\{\bar D_\dalpha,\,\bar Q_\dbeta\right\} = 0.
\end{equation}
This will allow us to reduce the number of independent components in superfields by imposing covariant (consistent with supersymmetry) constraints on them (see Section~\ref{c2chirality}).

\section{Superfields}

Superfields form linear representations of superalgebra. In general, however, the representations are highly reducible. Extra components in superfields can be eliminated by imposing covariant constraints which (anti)commute with the supersymmetry algebra. One could say that superfield formalism shifts the problem of finding supersymmetry representations to that of finding appropriate constraints. Note that one should constrain superfields without restricting their $x^\mu$ dependence (i.e., for instance, by virtue of differential equations in the $x$ space).

\subsection{Chiral superfield}\label{c2chirality}

Let us remark first that the superspace~(\ref{d4superspace}) has two invariant subspaces in it:
\begin{equation}
 \{x^\mu_{\rm L},\,\theta^\alpha\},
\quad\quad
x_{\rm L}^\mu = {x}^\mu
+ i\theta^\alpha \sigma^\mu_{\alpha\dot\alpha} \bar\theta^{\dot\alpha},
\end{equation}
and
\begin{equation}
\{x^\mu_{\rm R},\,\bar\theta^{\dot\alpha}\},
\quad\quad
x_{\rm R}^\mu = {x}^\mu
- i\theta^\alpha \sigma^\mu_{\alpha\dot\alpha} \bar\theta^{\dot\alpha}.
\end{equation}
Indeed, the supertransformations~(\ref{susytr}) give the following supertransformations in the subspaces:
\begin{equation}\label{e2.8.6}
\begin{array}{l}
x_{\rm L}^\mu \rightarrow x_{\rm L}^\mu
  + 2i\theta^\alpha \sigma^\mu_{\alpha\dot\alpha} \bar\epsilon^{\dot\alpha},
\\[2mm]
\theta^\alpha \rightarrow \theta^\alpha + \epsilon^\alpha,
\end{array}
\end{equation}
and
\begin{equation}
\begin{array}{l}
x_{\rm R}^\mu \rightarrow x_{\rm R}^\mu
  - 2i\epsilon^\alpha \sigma^\mu_{\alpha\dot\alpha} \bar\theta^{\dot\alpha},
\\[2mm]
\bar\theta^\dalpha \rightarrow \bar\theta^\dalpha + \bar\epsilon^\dalpha.
\end{array}
\end{equation}
These two subspaces are referred to as {\em chiral} (or left) and {\em antichiral} (or right) respectively.
Each of them is spanned by half of the Grassmann coordinates. Due to this, a function which is defined on the chiral or the antichiral subspace have much shorter component expansion.

Consider, for instance, a {\em chiral superfield} $\Phi(x_{\rm L},\theta)$. Its component expansion
\begin{equation}
{\Phi (x_\rmL,\theta )} = \phi ({x}_\rmL) + \sqrt{2}\,\theta^\alpha
\psi_\alpha ({
x}_\rmL) +  \theta^2 F({x}_\rmL)
\label{chsup}
\end{equation}
includes one complex scalar field $\phi(x)$
(two bosonic states) and one complex Weyl spinor $\psi_\alpha
(x)$ (two fermionic states) as well as the auxiliary $F$ term which is non-propagating: as we will see shortly, this field will appear in Lagrangian without a kinetic term. Chiral superfields are used for constructing matter sectors of various theories.
Note that the chiral superfield $\Phi(x_\rmL, \theta)$, being expressed as a function of $x^\mu$, depends on $\theta^\alpha$ as well as on $\bar\theta^\dalpha$ variables.

In fact, the chiral superfield $\Phi$ or the antichiral superfield $\bar\Phi$ can be obtained as solutions of the covariant constraints \cite{wesszu}
\begin{equation}
\bar D_{\dot\alpha} \Phi = 0
\quad\quad {\rm and}\quad\quad
 D_{\alpha} \bar\Phi = 0
\label{spin104}
\end{equation}
which follow from
\begin{equation}
 \bar D_{\dot\alpha }\, x_\rmL^\mu =0
\quad\quad\mbox{and}\quad\quad
 D_{\alpha }\, x_\rmR^\mu =0.
 \label{spi107}
\end{equation}
Moreover, the covariant superderivatives~(\ref{e2.7.14}) in the chiral basis $\{x_\rmL^\mu,\, \theta^\alpha,\, \bar\theta^{\dot\alpha}\}$ are realized as
\begin{equation}\label{4.17}
D_\alpha=\frac{\partial}{\partial \theta^\alpha}
+ 2i \sigma^\mu_{\alpha\dalpha}\,\bar\theta^{\dot \alpha}
  \frac{\partial}{\partial x^\mu_\rmL},
\quad\quad
\bar D_{\dot\alpha}=-\frac{\partial}{\partial \bar\theta^{\dot\alpha}},
\end{equation}
while in the antichiral basis
$\{x_\rmR^\mu,\, \theta^\alpha,\, \bar\theta^{\dot\alpha}\}$ they are realized as
\begin{equation}\label{e2.8.12}
D_\alpha=\frac{\partial}{\partial \theta^\alpha},
\quad\quad
\bar D_{\dot\alpha}=-\frac{\partial}{\partial \bar\theta^{\dot\alpha}}
 -2i\theta^{\alpha}\sigma^\mu_{\alpha\dalpha}
  \frac{\partial}{\partial x^\mu_\rmR} .
\end{equation}
Thus, for instance, the chirality condition
$\bar D_\dalpha \Phi = 0$ translates itself into
\begin{equation}
 \frac{\partial}{\partial \bar\theta^\dalpha}\Phi = 0,
\end{equation}
from which the solution~(\ref{chsup}) is obvious.

Let us write also the induced transformations of the component fields in the chiral superfield~(\ref{chsup}) under the infinitesimal supertransformations~(\ref{e2.8.6}) or, equivalently,~(\ref{susytr}):
\begin{equation}\label{e2.8.14}
\begin{array}{l}
 \phi \rightarrow \phi + \sqrt{2}\, \epsilon^\alpha \psi_\alpha,
\\[2mm]
 \psi_\alpha \rightarrow \psi_\alpha
  + i\sqrt 2\,\partial_\mu \phi\, \sigma^\mu_{\alpha\dalpha}\,\bar\epsilon^\dalpha
  + \sqrt 2 F\epsilon_\alpha,
\\[2mm]
 F\rightarrow F -i\sqrt 2 \,\partial_\mu
  \left(\psi^\alpha\sigma^\mu_{\alpha\dalpha}\bar\epsilon^\dalpha\right) .
\end{array}
\end{equation}
Note that the $F$ term transforms through a total space-time derivative.

\subsection{Real superfield}\label{s2.8.1}

Let us inspect Eq.~(\ref{spin98}). It gives a reducible representation of the supersymmetry algebra. We can simply constrain it with the reality condition $S^\dagger = S$. This gives what is called vector superfield:
\begin{equation}
V(x,\theta,\bar\theta)= \phi+\theta\psi + \bar\theta \bar\psi
  + \theta^2 F + \bar\theta^2 {\bar F}
+\theta^\alpha A_{\alpha\dot \alpha} \bar\theta^{\dot\alpha}
+\theta^2 (\bar\theta\bar\lambda )
+\bar\theta^2 (\theta\lambda)
  + \theta^2\bar\theta^2 D.
\label{vecsf}
\end{equation}
The superfield $V$ is real, $V=V^\dagger$, implying that the bosonic fields
$\phi$, $D$ and $A^\mu=\frac 12\left(\bar\sigma^\mu\right)^{\dot\alpha\alpha}A_{\alpha\dot\alpha}$
are real. The fields $\psi_\alpha$, $\lambda_\alpha$, $F$ are complex, with $\bar\psi_\dalpha= (\psi_\alpha)^\dagger$, $\bar \lambda_\dalpha=(\lambda_\alpha)^\dagger$,
$\bar F = F^*$.

The real superfield $V$ is used in construction of supersymmetric gauge theories. In fact, the (super)gauge freedom (which for the Abelian gauge superfield is $V\rightarrow V+(\Lambda +\bar\Lambda)$, where $\Lambda$ is an arbitrary chiral superfield)
allows one to eliminate the unwanted components $\phi$, $\psi_\alpha$, $\bar\psi_\dalpha$, $F$, and $\bar F$, reducing the physical content of $V$ to
\begin{equation}
V\rightarrow
\theta^\alpha A_{\alpha\dalpha} \bar\theta^{\dot\alpha}
+\left\{\theta^2 (\bar\theta\bar\lambda )
+\bar\theta^2 (\theta\lambda)\right\}
  + \theta^2\bar\theta^2 D
\end{equation}
so that $\lambda_\alpha$ is a fermionic {\em superpartner} of the gauge field $A_{\alpha\dalpha}$, while $D$ is an auxiliary field which is not dynamical and can be excluded from a corresponding Lagrangian by algebraic equations.

The supersymmetry transformations~(\ref{susytr}) induce the transformations
of the component fields in $V$. Here, for later purposes, we quote only the corresponding transformation for the auxiliary field $D$:
\begin{equation}\label{c2VD}
 D\rightarrow D + \frac i2 \partial_\mu \left(
  \epsilon^\alpha \sigma^\mu_{\alpha\dalpha}\bar\lambda^\dalpha
  -\lambda^\alpha \sigma^\mu_{\alpha\dalpha} \bar\epsilon^\dalpha
 \right).
\end{equation}
Note that, like in the chiral superfield case, it transforms through a total space-time derivative.
This property is of a paramount importance for construction of supersymmetric theories.

\subsection{Properties of superfields}

Let us enumerate the main properties of superfields which are used in construction of supersymmetric Lagrangians.

\begin{itemize}
 \item 

Linear combinations of superfields as well as products of superfields are again superfields. In general, a function of superfields is a superfield.

 \item
  If $\Phi$ is a chiral superfield, then $\bar\Phi=\Phi^\dagger$ is an antichiral superfield and {\em vice versa}.

 \item

Given a superfield, one can use the space-time derivatives $\partial/\partial x^\mu$ to generate a new one.
At the same time, the Grassmann derivatives $\partial/\partial\theta^\alpha$ and  $\partial/\partial\bar\theta^{\dot\alpha}$, being
applied to a superfield, do not produce a superfield.
One can use the covariant superderivatives $D_\alpha$ and $\bar D_{\dot\alpha}$ instead since they anticommute with the supercharges.

 \item
The squares of the covariant superderivatives $\bar D^2$, $D^2$, being applied on  a generic superfield,
produce a chiral or antichiral superfield, respectively. This immediately follows from the expression for $\bar D_{\dot\alpha}$ in (\ref{4.17}) and from the expression for $D_\alpha$ in~(\ref{e2.8.12}).
For instance, $ \bar D^2 \,D_\alpha S$ (with arbitrary $S$) is a chiral superfield while
$D_\alpha\Phi$ (with $\Phi$ chiral) is not a chiral superfield.
Indeed,
\begin{equation}
\bar D_{\dot\alpha}\left(D_\alpha\Phi
\right) = \left\{\bar D_{\dot\alpha} , \, D_\alpha
\right\}\Phi = -2i \sigma^\mu_{\alpha\dot\alpha}\,\partial_\mu\Phi \neq 0.
\end{equation}
At the same time, $D^2\Phi$ is an antichiral superfield.

\end{itemize}

\section{Building supersymmetric Lagrangians}

With the extension of space-time to the superspace~(\ref{d4superspace}), the space-time integral $\int d^4x$ must also include the integration over the new coordinates $\theta^\alpha$ and $\bar\theta^\dalpha$. As it is shown below, one can covariantly integrate superfields in the following three ways:
\begin{equation}
 \int d^4x \,d^4\theta
\quad\mbox{or}\quad
 \int d^2\theta\, d^4x_\rmL,
\quad\mbox{or}\quad
 \int d^2\bar\theta\, d^4x_\rmR.
\end{equation}
Do do so, one needs to define the rules of Grassmann integration. After that, manifestly supersymmetric Lagrangians can be straightforwardly constructed.
We will consider the simplest example of a superfield action involving a single chiral superfield.

\subsection{Rules of the Grassmann integration}\label{rugrin}

The rules of integration over the Grassmann variables, also known as Berezin integrals \cite{berezinbook}, are the following. One-dimensional integrals are defined as
\begin{equation}
\int d\theta_\alpha = 0,
\qquad
\int \theta_\alpha\,d\theta_\beta  = \delta_{\alpha\beta}
\label{spin114}
\end{equation}
(and similarly for the Grassmann variables with dotted indices),
while multi-dimensional integrals involving two and more Grassmann variables are to be understood as product of one-dimensional integrals.

Note that the Grassmann variables $\theta$ and $\bar\theta$ have the dimension of $\mbox{[length]}^{1/2}$, while the differentials $d\theta$ and $d\bar\theta$ have the dimension of $\mbox{[length]}^{-1/2}$. If $c$ is a number, then $d(c\,\theta) =c^{-1} d\theta$. This follows from the right equation in (\ref{spin114}).

We normalize the integral over all four Grassmann dimensions of the superspace,
\begin{equation}
\int \, d^4\theta\equiv \int d^2\theta\, d^2\bar \theta \sim \int\,d\theta_1\, d\theta_2\,
d\bar\theta_{\dot 1}\, d\bar\theta_{\dot 2},
\end{equation}
in such a way that
\begin{equation}
 \int d^2\theta\, d^2\bar\theta\,\,\theta^2\,\bar\theta^2= 1.
\label{spin116}
\end{equation}
Respectively, the integrals over the chiral and the antichiral subspaces are normalized as
\begin{equation}
\int d^2\theta\,\,\theta^2 = 1,
\quad\quad
 \int d^2\bar\theta\,\,\bar\theta^2= 1.
\end{equation}

\subsection{Kinetic terms for matter fields}

Consider the integral
\begin{equation}
 \int\, d^4x\,d^4\theta \, V(\dots)
\end{equation}
with $V$ being the real superfield~(\ref{vecsf}) which can be a function of other superfields.
The rules of Grassmann integration~(\ref{spin114}), (\ref{spin116}) imply that $D=\int d^4\,\theta V$, there $D$ is the coefficient in front of $\theta^2\bar\theta^2$ in $V$.
Since the supertransformations change the $D$ term through a total space-time derivative, Eq.~(\ref{c2VD}), then the expression
\begin{equation}
\int\, d^4\theta\, V(\dots)
\end{equation}
is superinvariant up to a total derivative.

Let us exploit the above idea and construct kinetic term
\begin{equation}\label{c2skin}
S_{\rm kin} =\int d^4x\, d^4\theta \, \bar\Phi\Phi
\end{equation}
for a chiral superfield $\Phi(x_\rmL,\theta)$. The component expansion of $\Phi$ is given in Eq.~(\ref{chsup}).
The product $\Phi\bar\Phi$, where $\bar\Phi=\Phi^\dagger$, is a real superfield. The calculation of the highest term in $\theta^\alpha$ and $\bar\theta^\dalpha$ gives
\begin{multline}
\bar\Phi\,\Phi
=
\dots
+\theta^2\bar\theta^2
\left\{\frac{1}{2} \partial_\mu\bar\phi\,\partial^\mu \phi
 -\frac{1}{4}\bar\phi\,\partial^2\phi
 -\frac{1}{4}\phi\,\partial^2\bar\phi
\right.
\\[2mm]
\left.
-\frac{i}{2}\,\psi^{\alpha}\sigma^\mu_{\alpha\dot\alpha}\,\partial_\mu\bar\psi^{\dot\alpha}
+\frac{i}{2}\,\sigma^\mu_{\alpha\dot\alpha}\partial_\mu\psi^{\alpha}\,\bar\psi^{\dot\alpha}
+\bar FF
\right\},
\label{spin122}
\end{multline}
where all the component fields depend on the space-time point $x^\mu$, and
the ellipsis denotes all lower-order terms in $\theta^\alpha$ and $\bar\theta^\dalpha$.
Substituting the last expression into~(\ref{c2skin}), integrating over the Grassmann variables and integrating by parts reads
\begin{equation}
S_{\rm kin} 
 =
\int d^4x \left\{
  \partial_\mu\bar\phi\,\partial^\mu \phi
 - {i} \psi^\alpha \sigma^\mu_{\alpha\dalpha}\,\partial_\mu \bar\psi^\dalpha
 +\bar FF
\right\}.
\label{spin123}
\end{equation}
Thus, this expression presents the kinetic term for the complex field $\phi(x)$ and the kinetic term for the Weyl spinor $\psi_\alpha(x)$.
As one can see, the field $F(x)$ appears in the Lagrangian with no derivatives and does not represent any physical (propagating) degrees of freedom. It can be eliminated from the action by virtue of the equations of motion.

\subsection{Potential terms for matter fields}

Similarly to the previous section, the integral
\begin{equation}
 \int d^2\theta d^4x_\rmL \, K(\dots),
\end{equation}
where $K$ is a chiral superfield (see Eq.~(\ref{chsup}) for its component expansion), which can be a function of other superfields, is invariant under the action of the supersymmetry group. To prove this statement, note that
\begin{equation}\label{e2.9.12}
 \int d^2\theta \,K(\dots) = F_K(x)
  - \frac{i}{\sqrt{2}} \partial_\mu\left(\psi^\alpha_K(x)\,
\sigma^\mu_{\alpha\dalpha}\bar\theta^\dalpha\right),
\end{equation}
with $F_K$ being the coefficient in front of $\theta^2$, while $\psi_K^\alpha$ being the coefficient in front of $\sqrt 2\theta^\alpha$ in the component expansion of the chiral superfield $K$. In addition, the last term in~(\ref{e2.9.12}) is a full space-time derivative and thus vanishes if one integrates it over space-time. Thereby, we see that
\begin{equation}
 \int d^2\theta d^4x_\rmL \, K(\dots)
=
  \int d^4x\, d^2\theta \, K(\dots)
=
  \int d^4x \, F_K(x).
\end{equation}
Finally, according to Eq.~(\ref{e2.8.14}), the $F_K$ term transforms as a full space-time derivative under the action of the supersymmetry group, and thus the integral above is indeed superinvariant.
It is also clear now that the integral
\begin{equation}
 \int d^2\bar\theta d^4x_\rmR \, \bar K(\dots),
\end{equation}
where $\bar K$ is an antichiral superfield, is superinvariant.

Potential terms for component fields in the Lagrangian   are those which enter with no space-time derivatives. They can be produced for the chiral superfield $\Phi$ from the previous section by the following superfield action:
\begin{equation}\label{e2.9.15}
S_{\rm int} =
  \int d^2\theta\, d^4 x_\rmL\, {\mathcal W}(\Phi)
  +\int d^2\bar\theta\, d^4 x_\rmR\, \bar{\mathcal W}(\bar\Phi).
\end{equation}
Here ${\mathcal W} (\Phi)$ is a function of the chiral superfield termed {\em superpotential}. The second integral which involves $\bar{\mathcal W}(\bar\Phi) = \left({\mathcal W}(\Phi)\right)^\dagger$ ensures that the expression above is real.
After taking the integrals over the Grassmann variables, one obtains the following component action:
\begin{equation}
 S_{\rm int} = \int d^4x \left\{
  F\,{\cal W}'(\phi ) -  \frac{1}{2} {\cal W}''(\phi )\,\psi^2
  + \hc \right\},
\end{equation}
where $\psi^2=\psi^\alpha\psi_\alpha$, the prime denotes the derivative with respect to $\phi$ and $\hc$ means the Hermitian conjugated expression.

\subsection{The Wess--Zumino model}\label{subswz}

Let us combine the results of the previous two sections and write the full superfield action involving a single chiral superfield $\Phi$ and its Hermitian conjugated superfield $\bar\Phi$:
\begin{equation}
S_{\rm WZ} = \int\! { d}^4 x \,{ d}^4\theta  \,\Phi\bar{\Phi} +
\int { d}^2 \theta\,d^4 x_\rmL {\cal W}(\Phi) +
\int { d}^2 \bar\theta\,d^4x_\rmR \bar{\cal W}(\bar\Phi ) .
\label{lagrwz}
\end{equation}
This model of supersymmetric field theory was invented by Wess and Zumino \cite{wesszu} and bears their name.
Let us remark that the first term in the superfield action is the integral over the full superspace, while the second and the third terms run over the chiral and antichiral subspaces, respectively.

In components, the corresponding Lagrangian is
\begin{equation}
{\mathcal L}_{\rm WZ} =
  \partial_\mu\bar\phi\,\partial^\mu \phi
 - {i} \psi^\alpha \sigma^\mu_{\alpha\dalpha}\,\partial_\mu \bar\psi^\dalpha
 +\bar FF
  +\left(F\,{\cal W}'(\phi ) -  \frac{1}{2} {\cal W}''(\phi )\,\psi^2\right)
  + \hc 
\label{wzcomp}
\end{equation}
It is supersymmetry-invariant up to a total space-time derivative.
The auxiliary field $F$ is non-dynamical and can be eliminated by
virtue of its classical equation of motion,
\begin{equation}
\bar F = -{\mathcal W}'(\phi).
\label{feom}
\end{equation}
The final expression for the component Lagrangian is
\begin{equation}\label{WZ4d}
{\mathcal L}_{\rm WZ} =
  \partial_\mu\bar\phi\,\partial^\mu \phi
 - {i} \psi^\alpha \sigma^\mu_{\alpha\dalpha}\,\partial_\mu \bar\psi^\dalpha
  -\left|{\cal W}'(\phi )\right|^2
  - \frac{1}{2} {\cal W}''(\phi )\,\psi^\alpha\psi_\alpha
  - \frac{1}{2} \bar {\cal W}''(\bar\phi )\,\bar\psi_\dalpha\bar\psi^\dalpha .
\end{equation}
It contains the scalar potential $\left| {\cal W}'(\phi)\right|^2$ describing the self-interaction of the complex field $\phi$.
For a renormalizable field theory in four dimensions,
the superpotential ${\mathcal W} (\Phi)$ must be a polynomial function of $\Phi$ of power not higher than three. Then, it can be always reduced to the form
\begin{equation}
{\cal W}(\Phi) = \frac{m}{2} \,\Phi^2 - \frac{\lambda}{3}
\Phi^3 
\label{spotp}
\end{equation}
with two {\em complex} constants $m$ and $\lambda$. In fact, one can always choose the phases of  the constants $m$ and $\lambda$ at will.

As a simplest example, let us consider the case $\lambda=0$.
Then the last three terms in the Lagrangian~(\ref{WZ4d}),
\begin{equation}
 - |m|^2\phi\bar\phi
 -\frac{m}{2} \psi^2
 -\frac{\bar m}{2} \bar\psi^2,
\label{spin131}
\end{equation}
are the mass terms for the fields $\phi$ and $\psi_\alpha$.
As expected, the masses of scalar and spinor particles are equal and
are given by one and the same parameter $|m|$.

\section{R-symmetries}\label{aresym}

The Coleman--Mandula theorem states that all global bosonic symmetries must commute with the Poincar\'{e} group (i.e. be Lorentz scalars). However, it is not necessary for them to commute with all generators of the super-Poincar\'{e} group.
Indeed, the form of the super-Poincar\'e algebra~(\ref{eq1}), (\ref{Qspinor}), (\ref{eq2}), (\ref{eq3}) allows one to multiply the supercharges by a constant phase in such a way that the superalgebra itself stays unchanged:
\begin{equation}
 Q_\alpha \rightarrow e^{-i\vp} Q_\alpha,
\quad\quad
\bar Q_\dalpha \rightarrow e^{i\vp} \bar Q_\dalpha.
\end{equation}
The corresponding Hermitian $\U(1)$ generator $R$ has the following commutation relations with the supercharges,
\begin{equation}
[R,\,Q_\alpha] = -Q_\alpha,
\quad\quad
  [R,\, \bar Q_{\dot\alpha} ] = \,\bar Q_{\dot\alpha},
\label{aresymcom}
\end{equation}
and it can be realized as a differential operator in the superspace~(\ref{d4superspace}),
\begin{equation}
 R= \theta^\alpha \frac{\partial}{\partial \theta^\alpha}
    - \bar\theta^\dalpha \frac{\partial}{\partial \bar\theta^\dalpha}.
\end{equation}
This additional $\U(1)$ symmetry is called the {\em R-symmetry}. 
It transforms the Grassmann parameters,
\begin{equation}
\theta^\alpha \rightarrow e^{i\vp}\theta^\alpha,
\quad\quad
\bar\theta^\dalpha \rightarrow e^{-i\vp}\,\bar\theta^\dalpha ,
\label{spin174}
\end{equation}
and the measure of the Grassmann integration,
\begin{equation}
d^2\theta \rightarrow e^{-2i\vp}d^2\theta ,
\quad\quad
d^2\bar\theta \rightarrow e^{2i\vp}d^2\bar\theta ,
\quad\quad
d^4\theta\rightarrow d^4\theta .
\label{spin175}
\end{equation}
Thus, we assign the {\em R-charge} $+1$ to $\theta^\alpha$ ($R\,\theta^\alpha = +\theta^\alpha$) and the R-charge $-1$ to
$\bar\theta^\dalpha$ ($R\,\bar\theta^\dalpha=-\bar\theta^\dalpha$), while the R-charges of $d^2\theta$ and $d^2\bar\theta$ are $-2$ and $+2$, respectively. The R-charge of $d^4\theta$ is zero.

The R-symmetry can be a symmetry of a given system.
To see how it works, let us consider the Wess-Zumino model~(\ref{lagrwz}) of a chiral superfield $\Phi$ with the superpotential
\begin{equation}
{\mathcal W}= -\frac \lambda 3\Phi^3.
\label{spinwcube}
\end{equation}
Then, the action~(\ref{lagrwz}) is invariant under the R-symmetry, if the R-charge of the superfield $\Phi$ is 2/3,
\begin{equation}\label{e2.10.7}
 \Phi\rightarrow e^{2i\vp/3}\Phi.
\end{equation}
Indeed, the first term in the action, $\int d^4\theta \,\Phi\bar\Phi$, is invariant since the antichiral superfield $\bar\Phi$ transforms with the complex conjugated phase factor. The potential term is also invariant:
\begin{equation}
\int d^2\theta \,{\mathcal W}(\Phi)\longrightarrow
\int \left( e^{-2i\vp}\, d^2\theta
\right) \left( {\mathcal W}(\Phi) \, e^{2i\vp}\right)
=\int d^2\theta \,{\mathcal W}(\Phi).
\end{equation}

Since the R-symmetry does not commute with supersymmetry, the component fields~(\ref{chsup}) of the superfield $\Phi$ do not all carry the same R-charge. It follows from Eqs.~(\ref{spin174}) and~(\ref{e2.10.7}) that, under the R-symmetry, they are transformed as
\begin{equation}
\begin{array}{l}
\phi (x) \rightarrow e^{\frac 23 i\vp} \,\phi (x),
\\[2mm]
\psi (x) \rightarrow e^{\left(\frac{2}{3}-1\right)i\vp}\, \psi(x),
\\[2mm]
F(x) \rightarrow e^{\left(\frac{2}{3}-2\right)i\vp} \, F(x).
\end{array}
\end{equation}
As one can see, the R-charge of the lowest component $\phi(x)$ coincides with the R-charge of the superfield $\Phi$ itself.
The above formulas can be used to explicitly check that the component Lagrangian~(\ref{WZ4d}) of the Wess-Zumino model is R-symmetry invariant provided $m=0$.

The R-symmetries play very important role in harmonic superspace approach, which is discussed in the case of supersymmetric quantum mechanics in the next chapter.

\section{Extended supersymmetry in Minkowski space in four dimensions}\label{c2sec2.7}

As we already know, the Coleman-Mandula theorem can be circumvented by the introduction to the Poincar\'e algebra four complex supercharges $Q_\alpha$, $\bar Q_\dalpha$ of fermionic nature. The supersymmetry algebra~(\ref{eq1}), (\ref{Qspinor}), (\ref{eq2}) and (\ref{eq3}) is usually referred to as minimal supersymmetry or as ${\mathcal N}=1$ supersymmetry. In fact, it can be extended even more, with more generators of fermionic nature,
\begin{equation}
 Q_\alpha^I
\quad\quad\mbox{and}\quad\quad
 \bar Q_\dalpha^J,
\end{equation}
where new indices $I,J = 1, 2,\dots, \N$ numerate the ``flavours''  of supercharges.
Evidently, the defining relations~(\ref{eq2}) and~(\ref{eq3}) can be generalized with~\footnote{The relations~(\ref{spipp64}) do not include possible central charges.
One can, for instance, introduce them as
\begin{equation}
\{Q_\alpha^I,\, Q_{\beta}^J\} = \ve_{\alpha\beta} Z^{IJ}
\end{equation}
with $Z^{IJ}=-Z^{JI}$.
See also the discussion at the end of Section~\ref{suexpoal}.}
\begin{eqnarray}
\{Q_\alpha^I,\, \bar Q_{\dot\alpha}^J\}
&=&
 2 P_\mu\left( \sigma^\mu\right)_{\alpha\dot\alpha}
 \,\delta^{IJ}\,,
\\[3mm]
\{Q_\alpha^I,\, Q_{\beta}^J\} &=&
 \{Q_{\dot\alpha}^I,\, Q_{\dot\beta}^J\} \,\,=\,\, 0,
\label{spipp64}
\end{eqnarray}
while the relation~(\ref{Qspinor}) stays untouched since it defines how spinors transform under the Poincar\'e symmetry.

In $\N=1$ supersymmetric renormalizable four-dimensional field theory which do now include gravity, the supermultiplets involve particles with spins \mbox{(0, 1/2)} or \mbox{(1/2, 1)}, since the supercharges change a particle spin by 1/2. Similarly, with the introduction of {\em extended supersymmetry}, one can transform a particle with spin zero to a particle with spin one by using two supercharges of different flavours. Thus, such a theory must include particles with spins \mbox{(0, 1/2, 1)} in a supermultiplet, i.e. it must be a gauge theory.

Gauge theories of this type are ${\mathcal N}=2$ and ${\mathcal N}=4$
super-Yang--Mills theories.
They are obtained by dimensional reduction from the minimal super-Yang--Mills theories in six and ten dimensions, respectively.
These theories are unsuitable for phenomenology, because all fermion fields they contain are nonchiral.
Nevertheless, they have rich dynamics the study of which provides deep insights into a large number of problems in mathematical physics.

All the properties of a supersymmetric field theory such as vanishing of vacuum energy and equal number of bosons and fermions in a supermultiplet, which were discussed in Section~\ref{sec2.6}, remain intact.
Moreover, the Abelian $\U(1)$ R-symmetry from Eq.~(\ref{aresymcom}) is extended to be non-Abelian $\U(\N)$ symmetry. Simply speaking, the non-Abelian R-symmetry group just mixes the flavours of the supercharges (i.e. it mixes the supercharge flavour indices).

In addition, the superspace~(\ref{d4superspace}) is trivially extended to
\begin{equation}\label{d4exsuperspace}
\{ x^\mu,\, \theta^{I\,\alpha},\, \bar\theta^{J\,\dot\alpha}\}.
\end{equation}
Note, however, that the usefulness of this $\N =2$ or $\N =4$ superspace is limited due to the fact that there are no chiral subspaces in it (which span over half of Grassmann coordinates). There exist a different superspace called harmonic superspace (HSS) which has these invariant subspaces \cite{HSS}. Such an approach gives more adequate superfield description. The harmonic superspace is obtained from the superspace~(\ref{d4exsuperspace}) by extending it with bosonic coordinates which parametrize the R-symmetry group $\U(\N)$ space or a certain factor space of it.

The harmonic superspace approach is discussed in supersymmetric quantum mechanics in the next chapter.

\cleardoublepage
\chapter{Supersymmetry and harmonic superspace in quantum mechanics}\label{chap3}

\vfill

\begin{intro}

\askip
This chapter is devoted to basic introduction to supersymmetry in quantum mechanics. We consider the main properties of any supersymmetric system, explain what is superspace and how the superfield formalism can be used in order to construct genuine supersymmetric Lagrangians. A simplest example of a supersymmetric system is considered. It involves a particle with spin moving in a potential field in one-dimensional space.

\vskip 0.5cm
\askip
The second part of the chapter is devoted to harmonic superspace approach in supersymmetric quantum mechanics. The essential definitions and notations are introduced. They will be extensively used in the next chapter. In particular, the supermultiplets ${\bf (4,4,0)}$ and ${\bf (3,4,1)}$ are described. The former is relevant in the context of four-dimensional quantum mechanics, while the latter is used in construction of  three-dimensional systems.

\end{intro}
\vfill
\clearpage

\section{Supersymmetry in quantum mechanics}

A quantum-mechanical system with traditional commutation relations between coordinates and momenta and a traditional Hilbert space of states is described by its Hamiltonian function $H$. We introduce a set of complex operators $Q_i$ together with their Hermitian conjugated operators,
\begin{equation}
 \bar Q^i = \left(Q_i\right)^\dagger .
\end{equation}
The system with the Hamiltonian $H$ and {\em the supercharges} $Q_i$, $\bar Q^j$ is supersymmetric, by definition, if
\begin{eqnarray}
 \left\{Q_i,\bar Q^j\right\} &=& 2 \delta_i^j H  \label{c3e1}
\\[2mm] \label{c3e5}
 \big\{Q_i, Q_j\big\} &=& \left\{\bar Q^i, \bar Q^j\right\} = 0,
\end{eqnarray}
where, as usual, curly brackets denote the anticommutator.
In particular, note the important property $Q_i^2 = \left(\bar Q^i\right)^2 = 0$ for any $i$. Another important consequence is that the supercharges commute with the Hamiltonian:
\begin{equation}\label{c3e3}
 \big[H,\, Q_i\big] = \left[ H,\, \bar Q^i\right] = 0,
\end{equation}
which can be proven by direct computation. In this way, the supercharges $Q_i$ and $\bar Q^j$ are considered as conserved spinorial operators in the system.

Several comments are relevant here. The Latin indices $i$, $j$ denote supercharge numbers and vary in the following region:
\begin{equation}
 i,\, j = 1,\, 2, \, \dots, \, \N/2
\end{equation}
with number $\N$ being even integer, see below.

We distinguish the position of indices for the supercharge $Q_i$ and its Hermitian conjugated supercharge $\bar Q^i$. This is done because the $\SU(\frac{\N}2)$ subgroup of the R-symmetry group (in the $\N \ge 4$ case) acts on these indices differently, see Section~\ref{c3Rsym} for details.
This subgroup will be of special importance for us later, when we limit ourselves to $\N=4$ case and deploy harmonic superspace approach.
Spinor products and their complex conjugates are conveniently written in such notations. For instance, for any two {\em anticommuting} fields $\bar\psi^i$ and $\xi_j$
\begin{equation}
 \bar\psi\,\xi \equiv \bar\psi^i \,\xi_i,
\quad\quad\mbox{while}\quad\quad
 \left(\bar\psi\,\xi\right)^*=\bar\xi\,\psi \equiv  \bar\xi^i \,\psi_i,
\end{equation}
where $\bar\psi^i = \left(\psi_i\right)^*$ and $\bar\xi^i = \left(\xi_i\right)^*$.

Finally, one can always pass to a {\em real} basis in the vector space of supercharges, for instance, by using the definitions
\begin{equation}
 S_A =
\left\{
\begin{array}{l}
 Q_A+\bar Q^A, \quad\quad\quad\,\,\,\,\mbox{for}\quad A=1,\, 2,\, \dots,\, \N/2,
\\[2mm]
 i\left(Q_A-\bar Q^A\right),\quad\quad\mbox{for}\quad A = \N/2+1,\, \dots,\, \N
\end{array}
\right.
\end{equation}
which give commutation relations
\begin{equation}\label{c3SAB}
 \left\{S_A,\,S_B\right\}= 4\,\delta_{AB} \,H.
\end{equation}
The new indices $A$, $B$ vary from 1 to $\N$. Thereby, $\N$ counts the number of linearly independent real supercharges in the system at hand.
For instance, $\N=2$ corresponds to an ordinary supersymmetric quantum mechanics, while a system with $\N=4$ is endowed with extended supersymmetry.

Let us remark that there is also a different convention in quantum mechanics. According to it, $\N$ counts the number of linearly independent complex supercharges which is, of course, twice smaller than the number of real supercharges. Note that, in a four-dimensional field theory context, the supercharges represent complex Weyl doublets, and $\N$ counts the number of those doublets.

\section{Properties of supersymmetric quantum mechanics}\label{sec3.2}

We already saw in the previous chapter that a supersymmetric system is endowed with additional common properties, namely, vanishing of the vacuum energy, energy positiveness of any definite-energy state, equal number of bosons and fermions in a supermultiplet, their equal masses. Any supersymmetric quantum-mechanical system has similar features.

Let us describe them in detail. To this end, let us denote a normalized vacuum state as $\psi_{\rm vac}$,
\begin{equation}
 \left<{\psi_{\rm vac}}|{\psi_{\rm vac}}\right> =1,
\end{equation}
while any other similarly normalized state as $\psi$. For simplicity, all the states will be treated as normalizable. Also, we may denote a state either as $\psi$ or as $\ket \psi$ according to our convenience.

\subsection{Every eigenstate has non-negative energy}

This statement can be proven by sandwiching Eq.~(\ref{c3e1}) (with $i=j$ being fixed) with an  eigenstate $\psi$:
\begin{multline}\label{c3e2}
 2 E = \left<{\psi}\left| Q_i\left(Q_i\right)^\dagger\ \right|{\psi}\right>
  +\left<{\psi}\left| \left(Q_i\right)^\dagger Q_i\right| {\psi}\right>
\\[3mm]
=
 \left<\left(Q_i\right)^\dagger \psi\left| \left(Q_i\right)^\dagger \psi\right.\right>^*
  +\Big<Q_i \psi\Big| Q_i \psi\Big>^* .
\end{multline}
The second line in this equality is always non-negative. Thus,
\begin{equation}
 E\ge 0 .
\end{equation}

\subsection{Vanishing of the vacuum energy}

If supersymmetry is unbroken, the vacuum $\psi_{\rm vac}$ has exactly zero energy.
This statement straightforwardly follows from Eq.~(\ref{c3e2}) if one changes $E\rightarrow E_{\rm vac}$ and $\psi\rightarrow\psi_{\rm vac}$. Indeed, the minimum $E_{\rm vac}=0$ is achieved when
\begin{equation}\label{c3unbrokvac}
 Q_i\ket{\psi_{\rm vac}} = \bar Q^i\ket{ \psi_{\rm vac}} = 0,\quad\quad \mbox{for all $i$.}
\end{equation}
In fact, the conditions~(\ref{c3unbrokvac}) exactly correspond to the case of unbroken supersymmetry. The simplest example of supersymmetry breaking in quantum mechanics was constructed in \cite{witten81} (see \cite{409936} for a good pedagogical review).

\subsection{Supermultiplets}

The eigenstates of a supersymmetric Hamiltonian can be divided into {\em supermultiplets}. To this end, take a state $\psi$ with {\em positive} energy $E$. One can act on this state with supercharges $Q_i$ and $ \left(Q_j\right)^\dagger$ and obtain new states with exactly the same energy $E$. Indeed, using Eq.~(\ref{c3e3}), we conclude that
\begin{equation}
 H\ket{Q_i\,\psi} = Q_i\, H\ket{\psi} = E\ket{Q_i\psi},
\end{equation}
and similarly for $\left(Q_i\right)^\dagger$.

A supermultiplet is defined as a vector space of states of the same energy $E>0$ obtained by all possible actions of supercharges on a chosen reference state $\psi$. The basis in the supermultiplet can be explicitly constructed.
To do so, consider the following states:
\begin{equation}
 Q_{i_1}\,Q_{i_2}\,Q_{i_3}\,\dots\ket{\psi} .
\end{equation}
Due to the property $Q_i^2=0$ there  exists a  state
\begin{equation}
 \ket{\psi_{\rm low}} = Q_{k_1}\,Q_{k_2}\,\dots Q_{k_n}\ket{\psi},
\quad\quad
n\le \N/2
\end{equation}
which is not zero, $\left<\psi_{\rm low}|\psi_{\rm low}\right>\sim 1$, but all the supercharges $Q_i$ annihilate it:
\begin{equation}\label{c3e4}
 Q_i\ket{\psi_{\rm low}} = 0 \quad\quad \mbox{for all $i$}.
\end{equation}
The initial reference state $\psi$ can be restored from the state $\psi_{\rm low}$ in the following way:
\begin{equation}\label{c3e8}
 \left(Q_{k_1}\right)^\dagger \left(Q_{k_2}\right)^\dagger \dots \left(Q_{k_n}\right)^\dagger\ket{\psi_{\rm low}}
 = \left(2E\right)^n\ket{\psi},
\end{equation}
where Eqs.~(\ref{c3e1}) and~(\ref{c3e4}) were used.

Let us now show that the states
\begin{equation}\label{c3e6}
 \left(Q_{i_1}\right)^\dagger \left(Q_{i_2}\right)^\dagger \dots \left(Q_{i_f}\right)^\dagger\ket{\psi_{\rm low}},
\quad\quad i_1 < i_2 < \dots < i_f, \quad\quad f \le \N/2
\end{equation}
form the basis in the supermultiplet.

To this end, consider a state of type
\begin{equation}
 Q_{i_1}\left(Q_{j_1}\right)^\dagger Q_{i_2}\, Q_{i_3} \left(Q_{j_2}\right)^\dagger \dots \ket \psi
\end{equation}
and plug into it the expression for $\psi$ from Eq.~(\ref{c3e8}). One can use the relations~(\ref{c3e1}), (\ref{c3e5}) and order the supercharges in such a way that any $\left(Q_i\right)^\dagger$ stays to the left from any $Q_j$. After that, the identity (\ref{c3e4}) allows one to eliminate any terms which contain $Q_i$. For instance,
\begin{multline}
 Q_1 \left(Q_2\right)^\dagger \left(Q_1\right)^\dagger\ket{\psi_{\rm low}}
=
 \left(Q_2\right)^\dagger Q_1 \left(Q_1\right)^\dagger\ket{\psi_{\rm low}}
\\[2mm]
=
 -\left(Q_2\right)^\dagger  \left(Q_1\right)^\dagger Q_1\ket{\psi_{\rm low}}
 +\left(Q_2\right)^\dagger \left\{Q_1,\, \left(Q_1\right)^\dagger\right\}\ket{\psi_{\rm low}}
\\[2mm]
=
 2E\left(Q_2\right)^\dagger\ket{\psi_{\rm low}}.
\end{multline}
Thus, any state in the supermultiplet adds up to a linear combination of the states~(\ref{c3e6}).

The states~(\ref{c3e6}) are linearly independent, because their scalar products are zero. For instance, one has for the two states
$\left(Q_1\right)^\dagger\left(Q_2\right)^\dagger\ket{\psi_{\rm low}}$
and
$\left(Q_1\right)^\dagger\left(Q_3\right)^\dagger\ket{\psi_{\rm low}}$:
\begin{multline}\label{c3e7}
 \left<\left(Q_1\right)^\dagger \left(Q_2\right)^\dagger \psi_{\rm low}\left|\left(Q_1\right)^\dagger\left(Q_3\right)^\dagger \psi_{\rm low}\right.\right>
=
\left<\psi_{\rm low}\left| Q_2\, Q_1 \left(Q_1\right)^\dagger\left(Q_3\right)^\dagger \right|\psi_{\rm low}\right>^*
\\[2mm]
=
-\left<\psi_{\rm low}\left| Q_2 \left(Q_1\right)^\dagger Q_1\left(Q_3\right)^\dagger \right|\psi_{\rm low}\right>^*
+
\left<\psi_{\rm low}\left| Q_2 \left\{Q_1,\, \left(Q_1\right)^\dagger\right\}\left(Q_3\right)^\dagger \right|\psi_{\rm low}\right>^*
\\[2mm]
=2E \left<\psi_{\rm low}\left| \left\{Q_2,\, \left(Q_3\right)^\dagger\right\} \right|\psi_{\rm low}\right>^* = 0
\end{multline}
Evidently, the states~(\ref{c3e6}) themselves have positive normalization which follows from the calculations similar to above.

Let us remark that we assume that the energy $E$ is positive. If not, Eq.~(\ref{c3e8}) does not allow us to express $\psi$ through $\psi_{\rm low}$, and thus the considerations above cannot be applied.

As it follows from Eq.~(\ref{c3e6}), the dimension of the supermultiplet with $E>0$ is $2^{\N/2}$. In this way, in an $\N=2$ system each positive-energy state is doubly degenerate, while for a system with extended $\N=4$ supersymmetry each state with nonzero energy is four-times degenerate.

The states of zero energy -- {\em vacuums} -- are of special interest. Their number cannot be found from general considerations, and Eqs.~(\ref{c3unbrokvac}) have to be solved for that. If there are no vacuums in a theory, the supersymmetry is spontaneously broken (see e.g. Refs.~\cite{witten81,409936} for an example).

\subsection{Equal number of bosonic and fermionic states in a supermultiplet}\label{sec3.2.4}

In a field theory, supermultiplets involve bosonic and fermionic states. The same concerns supermultiplets in quantum mechanics.

Let us consider the states~(\ref{c3e6}). Each of them is characterized by the number of supercharges $f$ acting on $\psi_{\rm low}$.
We can introduce then the ``fermionic number operator'' $\hat N_F$ with eigenvalues $f$,
\begin{equation}
 \hat N_F\left(Q_{i_1}\right)^\dagger \left(Q_{i_2}\right)^\dagger \dots \left(Q_{i_m}\right)^\dagger\ket{\psi_{\rm low}}
=
 f\left(Q_{i_1}\right)^\dagger \left(Q_{i_2}\right)^\dagger \dots \left(Q_{i_m}\right)^\dagger\ket{\psi_{\rm low}}.
\end{equation}
By definition, the states for which $\hat N_F$ is even are called ``bosonic'', while the states for which $\hat N_F$ is odd -- ``fermionic''~\footnote{
In our general consideration this is a matter of mere convention: we could equally well call the states for which $\hat N_F$ is even fermionic.
For example, consider $\N=2$ supersymmetric quantum-mechanical system described by the Hamiltonian~(\ref{c3e20}). It describes a particle with spin 1/2. The corresponding supermultiplets involve two basis states of different spin directions, and each of them can be equally well called bosonic or fermionic.
}.
In particular, the state $\psi_{\rm low}$ is bosonic.

We have the following equalities for bosonic and fermionic states respectively:
\begin{equation}
\left(-1\right)^{\hat N_F} \ket{\psi_{\rm B}} = + \ket{\psi_{\rm B}},
\quad\quad
\left(-1\right)^{\hat N_F} \ket{\psi_{\rm F}} = - \ket{\psi_{\rm F}}.
\end{equation}
Using the definitions above, one can check that
\begin{equation}
 \left\{\left(-1\right)^{\hat N_F},\, Q_i\right\}
= \left\{\left(-1\right)^{\hat N_F},\, \bar Q^i\right\} = 0 .
\end{equation}
In this way, one obtains (the index $i$ is fixed)
\begin{multline}
{\rm Tr}\left(\left(-1\right)^{\hat N_F}\right)
=
\\[2mm]
=\frac 1{2E}{\rm Tr} \left( \left(-1\right)^{\hat N_F} \left\{ Q_i,\,\bar Q^i \right\}\right) =
\frac 1{2E}{\rm Tr} \left( - Q_i \left(-1\right)^{\hat N_F}\bar Q^i +\left(-1\right)^{\hat N_F}\bar Q^i\,  Q_i\right)
=0,
\end{multline}
where the cyclic property of the trace was used.
The trace is taken among the states of the (finite-dimensional) supermultiplet. For instance, it is the sum over the averages of the states~(\ref{c3e6}).

Summarizing all above, one concludes that any supermultiplet with positive energy has equal number of bosonic and fermionic degrees of freedom.

\section{R-symmetries}\label{c3Rsym}

Let us remark that the supercharges $Q_i$ and $\bar Q^j$ can be linearly transformed in such a way that the  form of Eqs.~(\ref{c3e1}), (\ref{c3e5})  does not change.
The corresponding group of automorphisms of the supersymmetry algebra is called {\em R-symmetry group}.
To deduce the most general form of such linear transformations, one takes the {\em ansatz}
\begin{equation}
 Q'_i =  U_{\!i}^{\,\,j} \, Q_j,
\quad\quad
 \bar Q'^i =  \bar Q^j \, (U^\dagger)_{\!j}^{\,\,i}
\end{equation}
involving matrix $U_{\!i}^{\,\,j}$ and its Hermitian-conjugated matrix $(U^\dagger)_{\!j}^{\,\,i}$
and substitutes it back into Eqs.~(\ref{c3e1}), (\ref{c3e5}). It appears that the matrix $U$ must be unitary, $U\,U^\dagger=1$. Thus, the R-symmetry group is $\U(\frac{\N}{2})$. It is clear now why the supercharge $Q_i$ carries the superscript index, while the supercharge $\bar Q^j$ carries the subscript index: they belong to complex-conjugated representations of the R-symmetry group.

In particular case of the $\N=2$ supersymmetry, the R-symmetry group is no more than just $\U(1)$ group of multiplications of the supercharges by a phase factor:
\begin{equation}
 Q'_i = {\rm e}^{i\vp} Q_i,
\quad\quad
 \bar Q'^j = {\rm e}^{-i\vp} \bar Q^j
\end{equation}
with $\vp$ being an arbitrary real number.

In the $\N \ge 4$ case, the R-symmetry group is $\U(\frac{\N}{2}) = \U(1)\times \SU(\frac{\N}{2})$, i.e. it contains the same $\U(1)$ subgroup of multiplications by phase factor and also the non-Abelian subgroup $\SU(\frac{\N}{2})$.

\section{Superspace and superfields}

This section is devoted to superfield approach in supersymmetric quantum mechanics. It is shown how to construct systems endowed with supersymmetry in a rather universal way. As we already know from the previous chapter, supersymmetry can be realized as a geometrical symmetry which acts on the coordinates on certain extended space. In quantum mechanics, this space includes the time coordinate $t$. The other space ``dimensions'' are of Grassmann nature, i.e. the corresponding coordinates {\em anticommute} among each other.

Superfield technique is very general. It allows one to construct genuine supersymmetric systems. It also serves as an efficient instrument for generalizing already known supersymmetric systems.

The extended space in quantum mechanics -- {\em superspace} -- is described by the following coordinates:
\begin{equation}\label{c3e9}
 \left\{t,\, \theta_i,\, \bar\theta^j\right\},
\quad\quad i,\, j = 1,\, 2, \, \dots, \, \N/2
\end{equation}
with the identification $\bar \theta^i=\left(\theta_i\right)^*$: the Grassmann coordinates $\theta_i$ and $\bar\theta^j$ are complex. They anticommute with each other:
\begin{equation}
 \big\{\theta_i,\, \theta_j\big\} =
\left\{\theta_i,\, \bar\theta^j\right\} =
\left\{\bar\theta^i,\, \bar\theta^j\right\} = 0.
\end{equation}
Let us remind that the upper index on the conjugated Grassmann variable reflects the way how it transforms under the action of the R-symmetry group.

Having thus defined the superspace, one introduces {\em superfields} depending on the superspace coordinates. The property $\theta^2=0$ for any Grassmann variable $\theta$ limits the number of terms in the expansion of a general superfield $\Phi(t,\, \theta_i,\,\bar\theta^j)$ as series in Grassmann variables.
Take, for instance, $\N=2$ case. Omitting the indices on $\theta_1$ and $\bar\theta^1$, one obtains
\begin{equation}
 \Phi(t,\,\theta,\,\bar\theta)
 = \phi(t) + \bar\psi (t)\theta + \xi (t) \bar\theta + D(t)\theta\bar\theta.
\end{equation}
Two complex fields $\phi(t)$, $D(t)$ and two complex Grassmann fields $\bar\psi(t)$, $\xi(t)$ depend only on the time variable and may represent physical degrees of freedom in supersymmetric quantum mechanics. However, usually this field content is excessive, i.e. it is possible to reduce the number of fields by additional constraints on the superfield $\Phi$. Such constraints must be covariant with respect to supersymmetry transformations. For instance, such a constraint can be reality condition for the superfield $\Phi$, see Section~\ref{c3realSF} for details.

\subsection{Supersymmetry transformations and differential operators on superspace}

Infinitesimal supersymmetry transformations in quantum mechanics are realized as shifts on the superspace~(\ref{c3e9}),
\begin{equation}\label{c3e11}
\begin{array}{l}
 t\rightarrow t +i\left(\epsilon_i \bar\theta^i + \bar\epsilon^i\theta_i\right),
\\[2mm]
 \theta_i\rightarrow\theta_i + \epsilon_i,
\\[2mm]
 \bar\theta^i\rightarrow \bar\theta^i + \bar\epsilon^i 
\end{array}
\end{equation}
with Grassmann parameter $\epsilon_i$, $\bar\epsilon^i=\left(\epsilon_i\right)^*$. Such transformations are induced by the operator
$i\left(\bar\epsilon^i Q_i - \epsilon_i \bar Q^i\right)$, where the supercharges
\begin{equation}\label{c3supchar}
Q_i=-i\pop{\bar\theta^i}+\theta_i \pop{t},
\quad\quad
\bar Q^i=i\pop{\theta_i}-\bar\theta^i\pop{t}
\end{equation}
satisfy the relations
\begin{eqnarray}
 \big\{Q_i,\, Q_j\big\} & = & \left\{\bar Q^i,\, \bar Q^j\right\} \,\,=\,\, 0,
\\[2mm] \label{c3e10}
\left\{Q_i,\, \bar Q^j\right\} &=& 2\delta^j_i\, i\partial_t 
\end{eqnarray}
with
\begin{equation}
 \partial_t =\frac{\partial}{\partial t}.
\end{equation}
The supercharges~(\ref{c3supchar}) together with the operator $H=i\partial_t$ realize a particular representation of the supersymmetry algebra~(\ref{c3e1}), (\ref{c3e5}) on the superspace~(\ref{c3e9}).

Let us also introduce {\em covariant superderivatives} $D^i$ and $\bar D_i$ defined as
\begin{equation}\label{c3superder}
D^i=\pop{\theta_i}-i\bar\theta^i\pop{t} , \vergule
\bar D_i=\pop{\bar\theta^i}-i\theta_i \pop{t}.
\end{equation}
These operators are of special interest, because they anticommute with the supercharges from above,
\begin{equation}
 \left\{D^i,\, Q_j\right\} =
\left\{D^i,\, \bar Q^j\right\} =
\left\{\bar D_i,\, Q_j\right\} =
\left\{\bar D_i,\, \bar Q^j\right\} = 0 ,
\end{equation}
meaning that they are covariant with respect to supertransformations~(\ref{c3e11}). Thereby, they can be used in covariant constraints on superfields to reduce their number of independent components. The anticommutation relations for the superderivatives are
\begin{equation}
\left\{D^i,\, D^j\right\}=
\left\{\bar D_i,\, \bar D_j\right\}= 0,
\quad\quad
\left\{D^i,\, \bar D_j\right\} = -2\delta^i_j\, i\partial_t .
\end{equation}

Note that, for later convenience, we intentionally use a different convention for the superderivatives~(\ref{c3superder}) as compared with the supercharges~(\ref{c3supchar}).

\subsection{The existence of analytical subspace in $\N=2$ SQM}

Like in supersymmetric field theory case, the $\N=2$ supersymmetric quantum-mechanical superspace $\left\{t,\,\theta,\,\bar\theta\right\}$ is endowed with two invariant subspaces --- {\em chiral},
\begin{equation}
 \left\{t_{\rm L},\,\theta\right\},
\quad\quad
 t_{\rm L} = t - i\theta\bar\theta,
\end{equation}
and {\em antichiral},
\begin{equation}
 \left\{t_{\rm R},\,\bar\theta\right\},
\quad\quad
 t_{\rm R} = t + i\theta\bar\theta.
\end{equation}
These two subspaces are invariant with respect to supersymmetry transformations~(\ref{c3e11}):
\begin{equation}\label{c3e12}
\begin{array}{ll}
 t_{\rm L} \rightarrow t_{\rm L} + 2i\bar\epsilon\,\theta,
\quad\quad &
 \theta\rightarrow\theta+\epsilon ,
\\[2mm]
 t_{\rm R} \rightarrow t_{\rm R} + 2i\epsilon\,\bar\theta,
\quad\quad &
 \bar\theta\rightarrow\bar\theta+\bar\epsilon .
\end{array}
\end{equation}
Practically this means that superfields which depend only on chiral (or antichiral) coordinates have twice as less Grassmann coordinates and thus their component expansion in Grassmann variables is shorter and contains smaller number of component fields. This feature of $\N=2$ superspace is of primary importance in construction of supersymmetric quantum-mechanical systems. Consider, for instance, a superfield $q\left(t_{\rm L},\,\theta\right)$ which depends only on the coordinates of the chiral subspace. It satisfies a covariant constraint
\begin{equation}
 \bar D\,q\left(t_{\rm L},\,\theta\right) =
\left(\pop{\bar\theta}-i\theta \pop{t}\right) q\left(t_{\rm L},\,\theta\right)
 =0 
\end{equation}
which is due to $\bar D\,t_{\rm L}=0$. Its component expansion
\begin{equation}
 q\left(t_{\rm L},\,\theta\right) = z(t) + \psi(t)\theta - \dot z(t)\,i\theta\bar\theta
\end{equation}
contains one complex variable $z(t)$ and its {\em superpartner} -- one complex Grassmann variable $\psi(t)$. This is indeed the minimum number of degrees of freedom one may have for such a superfield.

The supertransformations~(\ref{c3e12}) of the chiral superspace induce the supertransformations of the chiral superfield $q\left(t_{\rm L},\,\theta\right)$ and its components:
$q\left(t_{\rm L},\,\theta\right)\rightarrow q\left(t_{\rm L},\,\theta\right) + \delta q\left(t_{\rm L},\,\theta\right)$. Direct calculation yields:
\begin{equation}
\begin{array}{l}
 z(t)\rightarrow z(t) + \psi(t)\epsilon,
\\[2mm]
 \psi(t)\rightarrow \psi(t) + 2i\,\dot z(t)\bar\epsilon
\end{array}
\end{equation}
Note that the field $\psi$ transformation involves full time derivative.

\subsection{Real superfield in $\N=2$ case}\label{c3realSF}

Along with the (anti)chiral superfields, real superfields are widely used in the $\N=2$ supersymmetric quantum mechanics. Let us list their properties. We will use a real superfield below for the construction of an example of simple supersymmetric system.

The real superfield $v(t,\,\theta,\,\bar\theta)$ is defined by the reality condition,
\begin{equation}
 v=v^\dagger .
\end{equation}
Its component expansion reads:
\begin{equation}\label{c3e13}
 v(t,\,\theta,\,\bar\theta) = x(t) +\bar\psi(t)\theta +\bar\theta\psi(t) + D(t)\, \theta\bar\theta,
\end{equation}
where $\bar\psi=\psi^*$ and the fields $x(t)$ and $D(t)$ are real. Under the supertransformations~(\ref{c3e11}) the component fields transform as
\begin{equation}\label{c3e15}
\begin{array}{l}
 x\rightarrow x + \bar\psi\epsilon + \bar\epsilon\psi,
\\[2mm]
 \psi\rightarrow \psi  - i\dot x\, \epsilon,
\\[2mm]
\bar\psi\rightarrow \bar\psi  + i\dot x\, \bar\epsilon,
\\[2mm]
 D\rightarrow D - i\left(\dot{\bar\psi}\epsilon +\dot\psi\bar\epsilon\right)
\end{array}
\end{equation}
Note that the $D$-term transforms with full time derivative. It is this fact which allows one to build a supersymmetric system to which we now proceed.

\subsection{One-dimensional $\N=2$ supersymmetric quantum mechanics}\label{c3onedimech}

Let us illustrate the material above with an example of the simplest supersymmetric quantum-mechanical system of one real bosonic variable $x(t)$ and one complex Grassmann variable $\psi(t)$ \cite{witten81}. To this end, we take the real superfield~(\ref{c3e13}) and write the following action for it:
\begin{equation}\label{c3e14}
 S=\int dt\, d\bar\theta d\theta \left\{\frac 12\bar D v \, D v + \Lambda(v)\right\}
\end{equation}
which involves the covariant derivatives from Eq.~(\ref{c3superder}) and an arbitrary real function $\Lambda(v)$. The rules of Grassmann integration were discussed in Section~\ref{rugrin}. The integration over the Grassmann variables leaves only the $D$-term of the real superfield which stays under the integral in Eq.~(\ref{c3e14}). Consequently, the action~(\ref{c3e14}) is automatically supersymmetric. Indeed, under the supertransformations, the $D$-term transforms as a full time derivative, see Eq.~(\ref{c3e15}).

As one may guess, the first term in the action~(\ref{c3e14}) describes the kinetic term of the system, while the second term is the interaction term.
The integration over the Grassmann variables can be straightforwardly performed. Using the definition
$\int d\bar\theta d\theta\, \theta\bar\theta = 1$, one obtains
\begin{equation}\label{c3e17}
 S=\int dt\left\{\frac 12\dot x^2
  + \frac i2\left(\bar\psi\dot\psi-\dot{\bar\psi}\psi\right)
  + \frac 12 D^2
 - D\, W(x) - W'(x)\, \bar\psi\psi 
\right\},
\end{equation}
where we have introduced $W(x)=\Lambda'(x)$, and the prime denotes the derivative with respect to $x$. Only the last two terms come from the function $\Lambda(v)$ in the action. We now see that the variable $D(t)$ does not have kinetic term and, thus, it is not dynamical. One can integrate it out with its classical equations of motion,
\begin{equation}\label{c3e16}
 D=W(x).
\end{equation}
Putting Eq.~(\ref{c3e16}) back into the component action~(\ref{c3e17}), one finally obtains the final Lagrangian of the system:
\begin{equation}\label{c3e18}
 L = 
 \frac 12\dot x^2
 - \frac 12 W^2(x)
  + \frac i2\left(\bar\psi\dot\psi-\dot{\bar\psi}\psi\right)
- W'(x)\, \bar\psi\psi .
\end{equation}
This Lagrangian defines a traditional one-dimensional system with the usual kinetic term $\frac 12\dot x^2$ and the usual potential term $\frac 12 W^2(x)$.  The coordinate $x$, however, is coupled to a fermion $\psi$. Note that here one has one bosonic degree of freedom and two fermionic degrees  of freedom. Let us remark that the statement about the equality of the number of bosons and fermions concerns the number of states in a multiplet, not the number of fields in the system.

To understand better the nature of the fermionic degree of freedom, let us perform the Legendre transform of the Lagrangian to the Hamiltonian and quantize the system. The third term in the Lagrangian~(\ref{c3e18}) involves only one time derivative on $\psi$ and $\bar\psi$ and will not enter the Hamiltonian. Nevertheless, it does provide the anticommutation relations of $\psi$ and $\bar\psi$. Indeed, if one considers $\psi$ as a coordinate, then $\bar\psi$ would be the corresponding momentum and {\em vice versa}. Thus, the canonical anticommutation relations are
\begin{equation}\label{c3eee}
 \left\{\psi,\,\bar\psi\right\}=1 ,
\end{equation}
and the Hamiltonian reads:
\begin{equation}\label{c3e19}
 H_{\rm cl} =
 \frac 12 {p^2}
 + \frac 12 W^2(x)
+ W'(x)\, \bar\psi\psi ,
\end{equation}
where $p=\dot x$. This expression corresponds to the ``classical'' Hamiltonian, because the term with $\bar\psi\psi$ has ordering ambiguity. The recipe to solve this issue is known \cite{SMI}: one must take the classical supercharges (which can be obtained with N\"{o}ether theorem from the Lagrangian~(\ref{c3e18})) and order the problematic terms with $\bar\psi\psi$, if any, in certain way. After that, one must apply the commutation relations~(\ref{c3e1}) to obtain the ``quantum'' Hamiltonian which is indeed enjoys supersymmetry algebra. In fact, the supercharges do not have order ambiguity problem. Their expressions are as simple as
\begin{equation}
 Q= \frac{1}{\sqrt 2} \left(p+iW\right)\psi,
\quad\quad
 \bar Q= \frac{1}{\sqrt 2} \left(p-iW\right)\bar\psi .
\end{equation}
The anticommutator of supercharges shows the difference between the classical and the quantum Hamiltonian, namely, the quantum Hamiltonian is obtained from the classical Hamiltonian~(\ref{c3e19}) by the replacement
\begin{equation}\label{c3e21}
 \bar\psi\psi \rightarrow \frac{1}{2} \left(\bar\psi\psi - \psi\bar\psi\right) .
\end{equation}

One can realize the coordinate/momentum pair $\hat \psi$, $\hat {\bar\psi}$
as differential operators acting on the space of functions $f(\psi)$ of the argument $\psi$ in the following way~\footnote{We put ``hats'' on the operators for clearness.}:
\begin{equation}\label{c3psiop}
 \hat \psi = \psi,
\quad\quad
 \hat{\bar\psi} = \frac{\partial}{\partial \psi}.
\end{equation}
Alternatively, these operators can be realized as matrices. Indeed, the space of functions $f(\psi)$ is two-dimensional: $f(\psi)\equiv a + b\psi$. Let us introduce the basis functions
\begin{equation}
 1\equiv
\left(\begin{array}{c}
 1 \\ 0
\end{array}\right)
\quad\quad\mbox{and}\quad\quad
 \psi\equiv
\left(\begin{array}{c}
 0 \\ 1
\end{array}\right),
\end{equation}
so that
\begin{equation}
 f(\psi)=a+b\psi
 \equiv
\left(\begin{array}{c}
 a \\ b
\end{array}\right).
\end{equation}
Thus, the operators~(\ref{c3psiop}) have the following matrix form:
\begin{equation}
 \hat\psi =
\left(\begin{array}{cc}
 0 & 0\\ 1 &0
\end{array}\right)
\quad\quad\mbox{and}\quad\quad
 \hat{\bar\psi} =
\left(\begin{array}{cc}
 0 & 1\\ 0 &0
\end{array}\right).
\end{equation}
Thereby,
\begin{equation}\label{c3e22}
 \hat{\bar\psi}\hat\psi -\hat\psi\hat{\bar\psi}=
\left(\begin{array}{cc}
 1 & 0\\ 0 &-1
\end{array}\right) = \sigma_3 ,
\end{equation}
where $\sigma_3$ denotes the third Pauli matrix. Combining Eqs.~(\ref{c3e19}), (\ref{c3e21}) and~(\ref{c3e22}), one finally obtains the quantum Hamiltonian
\begin{equation}\label{c3e20}
 H =
 \frac 12 {p^2}
 + \frac 12 W^2(x)
+ \frac 12 W'(x)\sigma_3 ,
\end{equation}
which describes a particle with spin moving in one-dimensional space. This Hamiltonian is supersymmetric with supercharges introduced above. As a simple exercise, one can take $W(x) = \omega x$ which gives the system composed with non-interacting one-dimensional oscillator and one spin degree of freedom.

The reader is referred to Ref.~\cite{409936} for further details on this system, where the study of energy spectrum and quantum states is performed and where the illustration of spontaneous breaking of supersymmetry is presented.

\section{Harmonic superspace approach}

Let us emphasize that the existence of (anti)chiral subspace in the superspace~(\ref{c3e9}) which leads to the reduction of the component expansion of any (anti)chiral superfield is inherent only to the $\N=2$ case. The superspace~(\ref{c3e9}) in the $\N\ge 4$ quantum mechanics does not have this feature. One, however, may still use covariant constraints (e.g. reality condition or equations involving superderivatives) to reduce number of components in superfields. Still, this is sometimes not enough to reduce the number to the expected minimum. Moreover, the constraints may be rather sophisticated. Some successful examples of manipulation with superfields and their covariant constraints are described in Ref.~\cite{IKLecht}.

It appeared that the case of $\N=4$ is special: it also admits a superspace which has two invariant subspaces and allows one  to reduce the number of Grassmann variables in superfields by a factor of two. This is the so called harmonic superspace (HSS) approach \cite{HSS} invented by
Galperin, Ivanov, Ogievetsky and Sokatchev.
The key idea here is that the standard superspace~(\ref{c3e9}) should be supplemented with additional coordinates of bosonic nature.

Let us limit ourselves to the case of $\N=4$ supersymmetric quantum mechanics.

The harmonic superspace can be seen as one of the superspaces on which the supersymmetry group acts. The superspace~(\ref{c3e9}) is one of possible choices. All the conceivable spaces on which supersymmetry acts can be described as a factor
\begin{equation}
 \mbox{superspace} = \frac {\mbox{supersymmetry group}} {\mbox{one of its certain subgroups}}.
\end{equation}

We remind that $\N=4$ supersymmetry algebra in quantum mechanics is invariant under the $\SU(2)$ R-symmetry group (cf. Section~\ref{c3Rsym}). Till this moment, the latter was factored out from the considerations. However, while searching for conceivable superspaces, the R-symmetry group space can be added to the superspace~(\ref{c3e9}). It appears that the superfields on the extended superspace which deliver minimal component field content will be functions on a two-sphere ${\rm S}^2 = \SU(2)/\U(1)$ -- a factor of the R-symmetry group with respect to one of its $\U(1)$ subgroups.

\subsection{Notations}

The harmonic superspace (HSS) approach in quantum mechanics was developed in Ref.~\cite{IvLecht}.
The convention in this manuscript follows the convention of Ref.~\cite{KonSmi} and differs from
the convention of Ref.~\cite{IvLecht} by the change of time direction $t\rightarrow -t$.
With this, one reproduces the correct sign in the kinetic term for the spinor field in Eq.~(\ref{eq_41}).

From now and below, in $\N=4$ supersymmetry, we use a different notation for spinor indices: the indices from the beginning of the Greek alphabet
\begin{equation}
\alpha,\, \beta = 1, 2
\end{equation}
are used instead of the indices $i$, $j$. For instance, the  ordinary  $\ca N=4$ superspace is
\begin{equation}\label{c3orsusp}
\left\{t,\,\theta_\alpha,\,\bar\theta^\beta\right\},
\quad\quad
\bar \theta^\beta = (\theta_\beta)^* .
\end{equation}

For later references, let us repeat here the expression for the supercharges of Eq.~(\ref{c3supchar}):
\begin{equation}
Q_\alpha=-i\pop{\bar\theta^\alpha}+\theta_\alpha \pop{t} ,
\vergule
\bar Q^\alpha=i\pop{\theta_\alpha}-\bar\theta^\alpha\pop{t} ,
\end{equation}
and the superderivatives of Eq.~(\ref{c3superder}):
\begin{equation}\label{eq_superder}
D^\alpha=\pop{\theta_\alpha}-i\bar\theta^\alpha\pop{t} , \vergule
\bar D_\alpha=\pop{\bar\theta^\alpha}-i\theta_\alpha \pop{t}.
\end{equation}
The supersymmetry algebra~(\ref{c3e1}), (\ref{c3e5}) is
\begin{equation} \label{c3e5x}
\begin{array}{l}
 \left\{Q_\alpha,\bar Q^\beta\right\} = 2 \delta_\alpha^\beta H
\\[3mm]
 \big\{Q_\alpha, Q_\beta\big\} = \left\{\bar Q^\alpha, \bar Q^\beta\right\} = 0,
\end{array}
\end{equation}

\subsection{Raising and lowering spinor indices}

The $\SU(2)$ R-symmetry group admits the possibility of raising and lowering spinor indices with the invariant antisymmetric Levi-Civita tensors $\ve_{\alpha\beta}$ and $\ve^{\alpha\beta}$. By definition,
\begin{equation}
\begin{array}{ll}
 \ve_{\alpha\beta}=-\ve_{\beta\alpha}, \quad\quad &\ve_{12} = 1,
\\[2mm]
 \ve^{\alpha\beta}=-\ve^{\beta\alpha}, \quad\quad &\ve^{12} = -1,
\end{array}
\end{equation}
so that, for example, one has for a spinor $v_\alpha$:
\begin{equation}
 v^\alpha = \ve^{\alpha\beta}v_\beta,
\quad\quad
 v_\alpha = \ve_{\alpha\beta}v^\beta .
\end{equation}
Due to the invariance of the Levi-Civita tensors with respect to the action of $\SU(2)$ R-symmetry group, the equations involving them are also invariant.

Below, we will also introduce $\SU(2)$ Pauli-G\"{u}rsey group \cite{HSS} and dotted indices for it. Analogously, one can introduce Levi-Civita tensors with dotted indices:
\begin{equation}
\begin{array}{ll}
 \ve_{\dot\alpha\dot\beta}=-\ve_{\dot\beta\dot\alpha}, \quad\quad &\ve_{\dot 1 \dot 2} = 1,
\\[2mm]
 \ve^{\dot\alpha\dot\beta}=-\ve^{\dot\beta\dot\alpha}, \quad\quad &\ve^{\dot 1\dot 2} = -1.
\end{array}
\end{equation}

\subsection{Dealing with the sphere ${\rm S}^2$}

Before discussing the harmonic superspace as a whole it is instructive to study the coordinates on the $\SU(2)/\U(1)$ space which is a two-sphere. One could choose, for example, polar or stereographic coordinates on ${\rm S}^2$. However, it turns out much  more convenient to deal with the homogeneous coordinates on the $\SU(2)$ group space and constrain functions on it to live on the ${\rm S}^2$ space.

To elaborate more on this point, introduce the homogeneous complex coordinates $u^\pm_\alpha$, $\alpha=1,2$ called {\em harmonics} on a three-sphere $\SU(2)$. They satisfy the defining relations
\begin{equation}\label{c3e23}
u^{+\alpha} u^-_\alpha=1 ,
\quad\quad
u^-_\alpha = (u^{+\alpha})^* .
\end{equation}
(We are using raised indices in the above, see the previous subsection; for instance, $u^{+\alpha}=\ve^{\alpha\beta}u^+_\beta$.)
Note also an important identity
\begin{equation}\label{c3e24}
 u^+_\alpha u^-_\beta - u^-_\alpha u^+_\beta = \ve_{\alpha\beta}.
\end{equation}

We are interested in considering functions on the $\SU(2)/\U(1)$ space. Consider functions on a three-sphere which have definitive $\U(1)$ charge. For illustration, take a function $f^+(u)$ of charge +1. In can be expanded into series in harmonics $u^\pm_\alpha$:
\begin{equation}
 f^+(u) = f^\alpha u^+_\alpha + f^{\alpha\beta\gamma} u^+_\alpha u^+_\beta u^-_\gamma + \dots ,
\end{equation}
where the constants $f^\alpha$, $f^{\alpha\beta\gamma}$, $\dots$ can always be taken symmetric in their indices. Indeed, using the relation~(\ref{c3e24}), one can transform any product of harmonics to symmetric combinations plus products of harmonics of smaller orders. For example,
\begin{eqnarray}
 u^+_\alpha u^+_\beta u^-_\gamma &= &
  \frac 13\left(u^+_\alpha u^+_\beta u^-_\gamma+ u^-_\alpha u^+_\beta u^+_\gamma + u^+_\alpha u^-_\beta u^+_\gamma\right) +
\nonumber
\\[2mm]
 &&+\ \frac 13\left(u^+_\alpha u^+_\beta u^-_\gamma - u^-_\alpha u^+_\beta u^+_\gamma\right) +
\nonumber
\\[2mm]
 &&+\  \frac 13 \left(u^+_\alpha u^+_\beta u^-_\gamma - u^+_\alpha u^-_\beta u^+_\gamma\right) =
\nonumber
\\[2mm]
 &=& \frac 13\left(u^+_\alpha u^+_\beta u^-_\gamma+ u^-_\alpha u^+_\beta u^+_\gamma + u^+_\alpha u^-_\beta u^+_\gamma\right) +
  \frac 13 \ve_{\alpha\gamma}u^+_\beta + \frac 13 \ve_{\beta\gamma} u^+_\alpha .
\end{eqnarray}

With respect to the action of R-symmetry group, the function $f^+(u)$ undergo homogeneous $\U(1)$ phase transformations (according to its overall charge) and thus is well defined on a two-sphere $\SU(2)/\U(1)$. The same is also true for any function of harmonics with fixed $\U(1)$ charge.

In fact, the harmonics $u^\pm_\alpha$ are the fundamental spin 1/2 spherical harmonics familiar from quantum mechanics. This is why they are called harmonic variables.

The harmonics can be used for projection of spinor indices onto harmonic space. For instance, $f^+=u^+_\alpha f^\alpha$ and $f^-= u^-_\alpha f^\alpha$. The original spinor can be restored using Eq.~(\ref{c3e24}):
\begin{equation}
 f^\alpha =   u^{+\alpha}f^- -  u^{-\alpha}f^+.
\end{equation}

\subsection{Differential operators on $\SU(2)_{\rm R}$ group}

The differential operators
\begin{equation}\label{eq_Dpp}
D^{++}=u_\alpha^+\pop{u_\alpha^-},\quad\quad
D^{--}   =  u_\alpha^-\pop{u_\alpha^+}\ ,\quad\quad
D^0 = u_\alpha^+\pop{u_\alpha^+} - u_\alpha^-\pop{u_\alpha^-}
\end{equation}
are called {\it harmonic derivatives}. The operator $D^0$ plays a role of the ${\rm U}(1)$ charge operator. One has for a function $f^{+q}(u)$ of definite $\U(1)$ charge $+q$:
\begin{equation}
 D^0 f^{+q}(u) = q\, f^{+q}(u) .
\end{equation}
The coordinates $u^+_\alpha$ have charge 1, while the coordinates $u^-_\alpha$ have charge -1.

\subsection{$\N=4$ harmonic superspace}

The ${\cal N}=4$ harmonic superspace formalism in quantum mechanics was developed in Ref.~\cite{IvLecht}.
In this formalism, the superfields depend on time $t$ and on harmonics $u^{\pm \alpha}$ which
parametrize the R-symmetry group $\SU(2)$ of the ${\cal N} = 4$
superalgebra,  and on Grassmann variables $\theta_\alpha$, $\bar\theta^\beta$. The superspace is
\begin{equation}\label{c3e25}
\left\{t,\,\theta_\alpha,\,\bar\theta^\beta,\,u^\pm_\gamma\right\},
\quad\quad
\bar \theta^\beta = (\theta_\beta)^* .
\end{equation}
This is the so called {\em standard basis} in harmonic superspace.

Usually, instead of spinors $\theta_\alpha$ and $\bar\theta^\beta$ it is preferable to use the following harmonic projections:
\begin{equation}
 \theta^\pm = u_\alpha
^\pm \theta^\alpha ,
\quad\quad
\bar\theta^\pm = u_\alpha ^\pm
\bar\theta^\alpha .
\end{equation}
One can also define harmonic projections of superderivatives,
$D^\pm=u^\pm_\alpha D^\alpha$, $\bar D^\pm=u^\pm_\alpha \bar D^\alpha$.
One can check that
\begin{eqnarray}
 D^+ = \frac{\partial}{\partial \theta^-} - i\bar\theta^+ \frac{\partial}{\partial t},
&\quad\quad&
 D^- = -\frac{\partial}{\partial \theta^+} - i\bar\theta^- \frac{\partial}{\partial t},
\\[2mm]
 \bar D^+ = -\frac{\partial}{\partial \bar\theta^-} - i\theta^+ \frac{\partial}{\partial t},
&\quad\quad&
 \bar D^- = \frac{\partial}{\partial \bar\theta^+} - i\theta^- \frac{\partial}{\partial t},
\end{eqnarray}
in the standard basis~(\ref{c3e25}).

\subsection{Analytical basis in harmonic superspace}

The most striking feature of harmonic superspace is the
presence of an {\sl analytic subspace}
\begin{equation}\label{c3analsub}
\left\{t_{\rm A},\,
\theta^+,\, \bar\theta^+,\, u^{\pm \alpha}\right\} 
\end{equation}
in it (an analog of ${\cal N} = 2$ chiral superspace) involving the  ``analytic time''
\begin{equation}
t_A = t + i(\theta^+ \bar \theta^- + \theta^- \bar \theta^+)
\end{equation}
and containing twice as less fermionic coordinates.

Let us elaborate more on this point. It is convenient to go over to the  {\sl analytic basis} in harmonic superspace,
\begin{equation}\label{c3analbas}
\left\{t_{\rm A},\, \theta^\pm,\,\bar\theta^\pm,\,u^\pm_\alpha\right\}.
\end{equation}
In this basis, the covariant spinor derivatives $D^+,\ \bar D^+$ are as simple as
\begin{equation}\label{eq_Dana}
D^+=\pop{\theta^-},\vergule \bar D^+=-\pop{\bar\theta^-} .
\end{equation}
It is this fact which allows one to translate superfield constraints $D^+ f= \bar D^+ f$ for some superfield $f$ to be independent of $\theta^-$ and $\bar\theta^-$: $f=f(t_A,\,\theta^+,\,\bar\theta^+,\, u^\pm_\alpha)$.

One can check directly that the subspace~(\ref{c3analsub}) is invariant with respect to $\N=4$ supersymmetry transformations. Indeed,
using Eq.~(\ref{c3e11}), one obtains
\begin{equation}
\begin{array}{l}
 t_A\rightarrow t_A + 2i\left(\epsilon^- \bar\theta^+ - \bar\epsilon^-\theta^+ \right),
\\[2mm]
 \theta^\pm\rightarrow \theta^\pm + \epsilon^\pm,
\\[2mm]
 \bar\theta^\pm\rightarrow \bar\theta^\pm + \bar\epsilon^\pm,
\\[2mm]
 u^\pm_\alpha\rightarrow u^\pm_\alpha ,
\end{array}
\end{equation}
where
\begin{equation}
 \bar\epsilon^\alpha
= (\epsilon_\alpha)^*,
\quad\quad
 \epsilon^\pm = u^\pm_\alpha \epsilon^\alpha,
\quad\quad
 \bar\epsilon^\pm = u^\pm_\alpha \bar\epsilon^\alpha .
\end{equation}

Finally, let us also write the form of harmonic derivatives $D^{++}$ and $D^{--}$ from Eq.~(\ref{eq_Dpp}) in the analytic basis:
\begin{eqnarray}
  D^{++}=u^+_\alpha\pop{u^-_\alpha}+\theta^+\pop{\theta^-}+\bar\theta^+\pop{\bar\theta^-}
    +2i\theta^+\bar\theta^+ \pop{t_{\rm A}},
\\[2mm]
  D^{--}=u^-_\alpha\pop{u^+_\alpha}+\theta^-\pop{\theta^+}+\bar\theta^-\pop{\bar\theta^+}
    +2i\theta^-\bar\theta^- \pop{t_{\rm A}} .
\end{eqnarray}
Note that in the subspace~(\ref{c3analsub}) the second and the third terms vanish.

\subsection{Involution symmetry}\label{c3sectInvolution}

The superspace~(\ref{c3e25}) admits an {\em involution symmetry} which commutes with supersymmetry transformations \cite{IvLecht,HSS}. We denote the involution with a sign $\widetilde{\ }$, e.g. its action is $f\rightarrow \widetilde{f}$ for an arbitrary superfield $f(t_A,\,\theta^\pm,\, \bar\theta^\pm,\, u^\pm_\alpha)$.

By definition, the involution transformation acts just as the ordinary complex conjugation {\it except} its action on the harmonics $u^\pm_\alpha$ for which it is
\begin{equation}
  \label{harminv}
  \widetilde {u^\pm_\alpha}=u^{\pm \alpha},\vergule
  \widetilde {u^{\pm \alpha}}=-u^\pm_\alpha  .
\end{equation}
This gives
\begin{equation}
 \label{eq_involution}
  \widetilde{t_{\rm A}}=t_{\rm A},\vergule
  \widetilde {\theta^\pm}=\bar\theta^\pm,\vergule
  \widetilde {\bar\theta^\pm}=-\theta^\pm .
\end{equation}
The action of the involution transformation on harmonics can be seen as a composition of complex conjugation and point inversion on the sphere ${\rm S}^2$.
In general, the involution symmetry is very similar to the operation of complex conjugation, but is does not change $\U(1)$ charges of superfields. It allows one to put additional constraints on superfields (e.g. reality condition) and is used in construction of supersymmetric Lagrangians.

\subsection{Supermultiplets of different dimensions}

Let us now discuss possible minimal supermultiplets~\footnote{
In Section~\ref{sec3.2}, we discussed {\em supermultiplets of quantum states} (characterized by their wave functions), whereas here by a {\em supermultiplet} we mean a superfield with certain ``minimal'' superfield content.
In particular, the statement about the equality of the number of bosons and fermions (Section~\ref{sec3.2.4}) is not applicable to the number of physical bosonic and fermionic fields in a system. This property was already observed in the example of the simplest quantum mechanics, see Section~\ref{c3onedimech}.

Let us also remark that the above property is specific only to quantum mechanics.
} which are commonly involved in the superfield description. All such $\N=4$ supermultiplets are usually referred to with three numbers, $({\bf b, f, a})$, where
\begin{itemize}
 \item ${\bf b}$ is the number of physical bosonic degrees of freedom;
 \item ${\bf f}$ is the number of physical fermionic degrees of freedom;
 \item ${\bf a}$ is the number of auxiliary nondynamical bosonic degrees of freedom, which are integrated out of final Lagrangians.
\end{itemize}
The widely used supermultiplets are $({\bf 4, 4, 0})$ and $({\bf 3, 4, 1})$ \cite{GaNa,IvLecht,IKLecht,IvSmi} which usually describe four- and three-dimensional dynamics respectively. We discuss them in detail below. Also, the common way of obtaining $\N=8$ supersymmetry or higher dimensional theories with $\N=4$ supersymmetry is to take several superfields of such types.

Other supermultiplets include $({\bf 2, 4, 2})$ and $({\bf 1, 4, 3})$ \cite{IKLev,IKPash} which are usually used to build systems of many particles in two and one dimensions. These multiplets are not discussed in this manuscript.

In general, the number of physical fermions in all such multiplets is four, while the sum of physical and auxiliary bosonic degrees of freedom is also four. It is even possible to introduce the $({\bf 0, 4, 4})$ supermultiplet \cite{IvLecht}.

\subsection{Supermultiplet $({\bf 4,4,0})$}

  The derivative operators $D^+$, $\bar D^+$, $D^{++}$ (anti)commute with each other and with supercharges.
  Because of this, it is possible to consider a superfield $q^+$ with ${\rm U}(1)$ charge +1 satisfying
  \begin{equation}\label{eq_anal}
  D^+ q^+=0,\vergule
  \bar D^+ q^+=0,\vergule
  D^{++}q^+=0.
  \end{equation}
  In the analytic superspace coordinates, the first and the second equations mean
  that $q^+$ depend only on $\theta^+$ and $\bar\theta^+$, but not on $\theta^-$ and $\bar\theta^-$, see Eq.~(\ref{eq_Dana}).
  In this way, the first and the second equations form the so-called {\sl superfield analyticity conditions}.

When expanding the superfield $q^+(t_A,\, \theta^+,\, \bar\theta^+,\, u^\pm_\alpha)$ over spinor coordinates and the harmonics, one obtains an infinite set
  of physical fields. However,  imposing also the condition $D^{++} q^+ = 0$
  drastically reduces the number of such fields, making it finite. In the analytic basis~(\ref{c3analbas}),
  the solution of the constraints (\ref{eq_anal}) reads
  \begin{equation}\label{eq_q}
  q^+=x^\alpha(t_{\rm A})u^+_\alpha
  -2\theta^+\chi(t_{\rm A})-2\bar\theta^+\bar\chi'(t_{\rm A})
  -2i\theta^+\bar\theta^+\partial_{\rm A}x^\alpha(t_{\rm A}) u^-_\alpha
  \end{equation}
  with
\begin{equation}
  \partial_{\rm A}\equiv \frac{\partial}{\partial t_A}
\end{equation}
and the factors $-2$ introduced for convenience. Thus, the $({\bf 4,4,0})$ superfield $q^+$ involves two complex bosonic coordinates $x^\alpha$ and two complex fermions -- $\chi$, $\bar\chi'$.

The constraints $D^+ q^+ = \bar D^+ q^+ = 0$ are akin to the discussed previously chirality
constraints in ${\cal N} =1$ four-dimensional supersymmetric field theories. Such constraints appear
naturally in  the HSS formalism and are common also in four-dimensional field theories. A possibility to impose
the extra  constraint $D^{++} q^+ = 0$ is specific for quantum mechanics only,
where it has a pure kinematic nature. In $\N=2$ supersymmetric field theories, the relation $D^{++} q^+ = 0$ is not
a kinematic constraint, it is the equation of motion for the {\em free hypermultiplet}
derived from the action $S = \int d^4x \, du \, d^4\theta^+ \, \widetilde {q^+} D^{++} q^+$ \cite{HSS}.

  The constraints (\ref{eq_anal}) admit an  involution symmetry $q^+\rightarrow \widetilde {q^+}$ which commutes with supersymmetry transformations \cite{IvLecht,HSS}:
  \begin{equation}\label{eq_tildeq}
  \widetilde{q^+}=\left[x_\alpha(t_{\rm A})\right]^*u^+_\alpha
  -2\theta^+\bar\chi'^*(t_{\rm A})+2\bar\theta^+\chi^*(t_{\rm A})
  -2i \theta^+\bar\theta^+\partial_{\rm A}\left[x_\alpha(t_{\rm A})\right]^* u^-_\alpha .
  \end{equation}
  It is straightforward to see  that the field  $\widetilde {q^+}$ satisfies the same constraints (\ref{eq_anal}) as the field $q^+$.

As we will use the $({\bf 4,4,0})$ supermultiplet to construct supersymmetric quantum-mechanical system with four space dimensions, it will be more convenient for us to use the $q^+$ supermultiplet in different form. Namely, let us introduce the supermultiplet
\begin{equation}\label{c3e27}
 q^{+\dot\alpha} = \left\{q^+,\, \widetilde{q^+}\right\},
\quad\quad
\dot\alpha = 1, 2 .
\end{equation}

The involution symmetry can used  to impose the pseudoreality condition on the field $q^{+\dot\alpha}$,
\begin{equation}
\label{c3reality3}
  q^{+{\dot\alpha}}=\varepsilon^{{\dot\alpha}{\dot\beta}} \widetilde{\left(q^{+{\dot\beta}}\right)} ,
\end{equation}
which is in fact equivalent to Eq.~(\ref{c3e27}). In components,
\begin{equation}\label{c3eq_qdot4}
  q^{+\dot\alpha} = x^{\alpha\dot\alpha}(t_{\rm A})u^+_\alpha
  -2\theta^+\chi^{\dot\alpha}(t_{\rm A})-2\bar\theta^+\bar\chi^{\dot\alpha}(t_{\rm A})
  -2i\theta^+\bar\theta^+\partial_{\rm A}x^{\alpha\dot\alpha}(t_{\rm A}) u^-_\alpha  .
\end{equation}
The constraint~(\ref{c3reality3}) implies
\begin{equation}\label{c3xxx}
  x^{\alpha{\dot\alpha}}=-\left(x_{\alpha{\dot\alpha}}\right)^* ,\vergule
  \bar\chi^{\dot\alpha}=\left(\chi_{\dot\alpha}\right)^* .
\end{equation}

The form of the supermultiplet $q^{+\dot\alpha}$ suggests that one can associate the $\SU(2)$ group related to the dotted index $\dot\alpha$. This Pauli-G\"{u}rsey group \cite{HSS} is also realized on the $q^+$ supermultiplet, but not manifestly.

Consequently, the quantum-mechanical system which will be discussed in the next chapter, inherits the $\SO(4)=\SU(2)_{\rm R}\times \SU(2)_{\rm PG}$ group composed from the R-symmetry group and the Pauli-G\"{u}rsey group. In general, this rotational $\SO(4)$ group is completely broken by the presence of a four-dimensional gauge field, see next chapter.

\subsection{Supermultiplet $({\bf 3,4,1})$}

Instead of the coordinate superfield $q^{+\dot\alpha}$ one can deal with the analytic superfield $L^{++}$ of charge +2 which encompass the supermultiplet $(\bf{3,4,1})$ and is subjected to the constraints
\begin{equation}\label{a}
\begin{array}{c}
D^{+}L^{++} = \bar D^{+}L^{++} = 0,
\\[3mm]
D^{++}L^{++} = 0,
\quad\quad
 \widetilde{(L^{++})} = -L^{++},\blanc\blanc\,
\end{array}
\end{equation}

They restrict the analytic superfield $L^{++}$ to have the appropriate off-shell component field content, namely
$({\bf 3, 4, 1})$:
\begin{equation}\label{c3aaa}
L^{++} = \ell^{\alpha\beta}u^+_\alpha u^+_\beta
+ 2i\theta^+ \chi^\alpha u^+_\alpha
+2i\bar\theta^+ \bar\chi^\alpha u^+_\alpha +
\theta^+\bar\theta^+ [F - 2 i\dot{\ell}^{\alpha\beta} u^+_{\alpha} u^-_{\beta}]
\end{equation}
with
\begin{equation}\label{c3e3.5.41}
\left(\ell_{\alpha\beta}\right)^* = -\ell^{\alpha\beta},
\quad\quad
 (\chi^\alpha)^* = \bar\chi_\alpha .
\end{equation}
The multiplet $L^{++}$ involves the 3-dimensional target space coordinates $\ell^{\alpha\beta} = \ell^{\beta\alpha}$, their
fermionic partners $\chi^\alpha$, $\bar\chi^\alpha$ and a real auxiliary field $F$.

\subsection{Harmonic integrals}  
  
  The invariant actions involve the harmonic integral $\int du$. To find such integral of any function $f(u^\pm_\alpha)$, one should
  expand $f$ in the harmonic Taylor series and, for each term, do the integrals using the rules
  \begin{equation}
  \label{intharm}
  \int du\, 1=1,\vergule \int du\, u^+_{\{\alpha_1}\dots u^+_{\alpha_k}u^-_{\alpha_{k+1}}\dots u^-_{\alpha_{k+\ell}\}}=0 \ ,
  \end{equation}
  where the integrand in the right equation is  symmetrized over all indices.
  The values of the integrals of all other harmonic monoms (for example, $\int du \, u^+_\alpha u^-_\beta = \frac 12 \varepsilon_{\alpha\beta}$) follow from
  (\ref{intharm}) and the definitions (\ref{c3e23}), (\ref{c3e24}).

\subsection{Notations in four-dimensional mechanics}

We keep the notation of the previous chapter for the four-dimensional Euclidean space vector indices,
\begin{equation}
 \mu,\nu = 0,1,2,3.
\end{equation}
The Euclidean four-dimensional sigma-matrices, however, are different;
we use the following SO(4) notation (compare with Eq.~(\ref{sigma4d})):
\begin{equation}\label{sigmas}
(\sigma_\mu)_{\alpha \dot\alpha} = \left\{i, \vec\sigma \right\}_{\alpha\dot\alpha },
\quad\quad
\left(\sigma_\mu^\dagger\right)^{\dot\alpha\alpha}=\left\{-i,\vec\sigma\right\}^{\dot\alpha\alpha},
\end{equation}
where $\vec\sigma$ are ordinary Pauli matrices. (These are more or less the conventions of \cite{Wess} rotated to Euclidean space.)
The matrix $\sigma_\mu^\dagger$ is obtained from the matrix $\sigma_\mu$ by the operation of raising of indices:
\begin{equation}
\left(\sigma_\mu^\dagger\right)^{\dot\alpha\alpha}=
-\varepsilon^{\dot\alpha\dot\gamma}\varepsilon^{\alpha\gamma}(\sigma_\mu)_{\gamma\dot\gamma}\,.
\end{equation}

 The matrices $\sigma_\mu,\ \sigma^\dagger_\mu$ satisfy the identities
\begin{equation}\label{eq_sigma}
\begin{array}{c}
  \sigma_\mu\sigma^\dagger_\nu+\sigma_\nu\sigma^\dagger_\mu=
 \sigma^\dagger_\mu\sigma_\nu+\sigma^\dagger_\nu\sigma_\mu
=2\delta_{\mu\nu}, \\
  \sigma^\dagger_{\mu}\sigma_{\nu} - \sigma^\dagger_{\nu}\sigma_{\mu}
=2i\,\eta_{\mu\nu}^a \sigma_a, \\
  \sigma_{\mu}\sigma^\dagger_{\nu} - \sigma_{\nu}\sigma^\dagger_{\mu} =
2i\,\bar\eta_{\mu\nu}^a \sigma_a,
\end{array}
\end{equation}
where $\eta_{\mu\nu}^a$, $\bar\eta_{\mu\nu}^a$ are the  't~Hooft symbols,
\begin{equation}\label{tHooft2}
\eta^a_{ij} = \bar\eta^a_{ij} = \varepsilon_{aij},\ \ \eta^a_{i0} = - \eta^a_{0i} = \bar\eta^a_{0i} =  -\bar\eta^a_{i0} =  \delta_{ai}
\end{equation}
($\sigma_a$ -- Pauli matrices, indices $a$, $i$, $j$ run from 1 to 3).
They are self-dual and anti-self-dual respectively,
\begin{equation}\label{thooftselfdual}
  \eta_{\mu\nu}^a=\frac{1}{2}\varepsilon_{\mu\nu\rho\lambda}\eta_{\rho\lambda}^a,\vergule
  \bar\eta_{\mu\nu}^a=-\frac{1}{2}\varepsilon_{\mu\nu\rho\lambda}\bar\eta_{\rho\lambda}^a,
\end{equation}
with the convention
\begin{equation}
 \varepsilon_{0123} = -1.
\end{equation}
Another useful identity is
\begin{equation}\label{transpose}
\sigma_2 \sigma_\mu^T \sigma_2 = \ -\sigma^\dagger_\mu   .
\end{equation}

The $({\bf 4,4,0})$ supermultiplet~(\ref{c3eq_qdot4}) involves the bosonic field $x^{\alpha\dot\alpha}(t)$. In fact, such bosonic field is equivalent to a real four-vector in the Euclidean space. Let us describe the transformation between the spinor notation $v^{\alpha\dot\alpha}$ and the corresponding vector notation $v^\mu$ for an arbitrary field $v$:
\begin{equation} \label{spinor2vector}
\begin{array}{l}
v_{\alpha \dot \alpha} =  v_\mu (\sigma_\mu)_{\alpha \dot\alpha} ,
\\[2mm]
v_\mu = \frac 12 v_{\alpha \dot\alpha}
(\sigma_\mu^\dagger)^{\dot \alpha \alpha}\, =\, -\frac 12 v^{\alpha \dot\alpha}
(\sigma_\mu)_{\alpha \dot\alpha},
\\[2mm]
v^{\alpha \dot \alpha} = \varepsilon^{\alpha \beta}\varepsilon^{\dot\alpha\dot\beta } v_{\beta \dot\beta}
\, =\,  -v_\mu (\sigma_\mu^\dagger)^{\dot\alpha \alpha}.
\end{array}
\end{equation}
Particularly, it is straightforward to check that for the field $x^{\alpha\dot\alpha}$ with the constraint~(\ref{c3xxx}) the corresponding vector field $x^\mu$ is real.

\subsection{Relations for the $\eta$ symbols}

Here we give a list of relations for the 't~Hooft symbols $\eta^a_{\mu\nu}$ and
$\bar\eta^a_{\mu\nu}$, defined by Eqs.~(\ref{eq_sigma}), (\ref{tHooft2}):
\begin{equation}
\begin{array}{l}
  \eta^a_{\mu\nu}=-\eta^a_{\nu\mu},
\\[2mm]
  \eta^a_{\mu\nu}\eta^a_{\mu\lambda}=3\delta_{\nu\lambda},
\\[2mm]
    \eta^a_{\mu\nu}\eta^b_{\mu\nu}=4\delta^{ab},
\\[2mm]
  \eta^a_{\mu\nu}\eta^a_{\gamma\lambda}=\delta_{\mu\gamma}
    \delta_{\nu\lambda}-\delta_{\mu\lambda}\delta_{\nu\gamma}
    +\varepsilon_{\mu\nu\gamma\lambda},
\\[2mm]
  \varepsilon_{\mu\nu\lambda\sigma}\eta^a_{\gamma\sigma}=
    \delta_{\gamma\mu}\eta^a_{\nu\lambda}-\delta_{\gamma\nu}
      \eta^a_{\mu\lambda}+\delta_{\gamma\lambda}\eta^a_{\mu\nu},
\\[2mm]
  \eta^a_{\mu\nu}\eta^b_{\mu\lambda}=
    \delta^{ab}\delta_{\nu\lambda}
      +\varepsilon^{abc}\eta^c_{\nu\lambda},
\\[2mm]
  \varepsilon^{abc}\eta^b_{\mu\nu}\eta^c_{\gamma\lambda}=
    \delta_{\mu\gamma}\eta^a_{\nu\lambda}
    -\delta_{\mu\lambda}\eta^a_{\nu\gamma}
    -\delta_{\nu\gamma}\eta^a_{\mu\lambda}
    +\delta_{\nu\lambda}\eta^a_{\mu\gamma},
\\[2mm]
  \eta^a_{\mu\nu}\bar\eta^b_{\mu\nu}=0,
\\[2mm]
    \eta^a_{\gamma\mu}\bar\eta^b_{\gamma\lambda}=
      \eta^a_{\gamma\lambda}\bar\eta^b_{\gamma\mu}.
\end{array}
\end{equation}

To pass from the relations for $\eta^a_{\mu\nu}$, to those for
$\bar\eta^a_{\mu\nu}$ it is necessary to make the substitution
\begin{equation}
  \eta^a_{\mu\nu}\rightarrow\bar\eta^a_{\mu\nu},\quad
    \varepsilon_{\mu\nu\gamma\delta}\rightarrow
      -\varepsilon_{\mu\nu\gamma\delta}.
\end{equation}

  \cleardoublepage
\chapter{New supersymmetric models of quantum mechanics}\label{chap4}

\noindent\hspace*{\stretch{4}}

\vfill


\begin{intro}

\askip
This is the central chapter of the manuscript describing certain new SQM models discussed and studied in the papers~\cite{KonSmi,Ivanov:2009tw,IvKon}.
The explicit form of the corresponding superfield and component actions, as well as of the quantum Hamiltonians and supercharges is given. The brief summary of the results is the following.


\askip
It is shown that the Hamiltonian $H =\, /\!\!\!\!{ \D}^2$, where $\,/\!\!\!\!{     \D}$
is flat four-dimensional Dirac operator in an external {\em self-dual} gauge background, Abelian or non-Abelian, is supersymmetric with $\N=4$ supersymmetry.
A generalization of this Hamiltonian to the motion on a curved conformally flat four-dimensional manifold exists.
For an {\em Abelian} self-dual background, the corresponding Lagrangian can be derived from certain harmonic superspace expressions.

\askip
If the Hamiltonian involves a {\em non-Abelian} self-dual gauge field, one can construct the Lagrangian formulation by introducing auxiliary bosonic variables with Wess-Zumino type action.
For a special class of such Lagrangians when the gauge group is $\SU(2)$ and the gauge field is expressed in the `t~Hooft ansatz form, it is possible to give a superfield description using the harmonic superspace formalism. As a new explicit example, the ${\cal N}=4$ mechanics with {\em Yang monopole} in ${\mathbb R}^5$ (which coincides with an instanton on ${\rm S}^4$) is considered.

\askip
Independently, a similar system with $\N=4$ supersymmetry in {\em three dimensions} also admits the superfield description.
Although the three-dimensional system involves different superfields as compared with the four-dimensional case, its component Lagrangian and Hamiltonian appear to be the three-dimensional reduction of the mentioned four-dimensional system.
The off-shell $\N=4$ supersymmetry requires the gauge field to be a
static form of the 't Hooft ansatz for the four-dimensional self-dual SU(2)
gauge fields, that is a particular solution of Bogomolny equations
for {\em BPS monopoles}.

\end{intro}
\vfill
\clearpage

\section{Fermions in four-dimensional self-dual background}\label{sect2}

\subsection{Matrix description}

Consider the Dirac operator in flat four-dimensional  Euclidean space
\begin{equation}
\label{gamE}
/\!\!\! \!\D \ =\ \sum_{\mu=0,1,2,3} \D_\mu \gamma_\mu\ ,
\end{equation}
where $\D_\mu = \partial_\mu - i {\cal A}_\mu$ with $\A_\mu$ being a gauge field and $\gamma_\mu$ are Euclidean anti-Hermitian gamma--matrices,
\begin{equation}\label{eq_gamma4}
\gamma_\mu = \left( \begin{array}{cc} 0  & -\sigma^\dagger_\mu \\
                                      \sigma_\mu & 0
 \end{array} \right),\vergule
\left\{\gamma_\mu,\gamma_\nu\right\}=-2\delta_{\mu\nu}.
\end{equation}
The matrices $\sigma_\mu$ and $\sigma^\dagger_\mu$ were introduced in Eq.~(\ref{sigmas}).
The Hamiltonians
we are going to construct enjoy ${\rm SO}(4) = {\rm SU}(2) \times {\rm SU}(2)$ covariance such that the undotted spinor index refers
to the first ${\rm SU}(2)$ factor, while the dotted one to the second.

Consider the operator
\begin{equation}
\label{HD2}
H =  \frac 12\, /\!\!\!\!\D^2  \ =\ - \frac 12 \D^2  - \frac i4  {\cal F}_{\mu\nu} \gamma_{\mu} \gamma_{\nu}
 \ ,
\end{equation}
where ${\cal F}_{\mu\nu}=\partial_\mu{\cal A}_\nu-\partial_\nu {\cal A}_\mu-i\left[{\cal A}_\mu, {\cal A}_\nu\right]$ is the gauge field strength.
It is well known  that nonzero eigenvalues of the Euclidean Dirac operator come in pairs
$(-\lambda, \lambda)$ and hence the spectrum of the Hamiltonian $H$ is double-degenerate for all excited states.
This means that, for any external field ${\cal A}_\mu$, this
Hamiltonian is supersymmetric \cite{Gaume}  admitting two different anticommuting real supercharges: $/\!\!\!\!\D$ and $i/\!\!\!\! \D \gamma_5$
($\gamma_5 = \gamma_0 \gamma_1 \gamma_2 \gamma_3$).
Suppose now that the background field is self-dual,
\begin{equation}
\F_{\mu\nu} = \frac 12 \varepsilon_{\mu\nu\rho\delta} \F_{\rho\delta} \ \ \longleftrightarrow \ \ \F_{\mu\nu} = \eta^a_{\mu\nu} B_a\ ,
\end{equation}
where $\eta_{\mu\nu}^a$ are the 't~Hooft symbols defined in Eq.~(\ref{eq_sigma}).
One can be easily convinced that in this case the Hamiltonian admits {\it four} different   Hermitian square roots $S_A$ that satisfy
the extended supersymmetry algebra~(\ref{c3SAB}); we repeat these relations here:
\begin{equation}
\label{N4}
\{S_A, S_B\} = 4\delta_{AB} H.
\end{equation}
One of the choices is
\begin{equation}\label{matrS}
\begin{array}{l}
S_1  =   /\!\!\!\!\D =  \gamma_0 \D_0 + \gamma_1 \D_1 + \gamma_2 \D_2 + \gamma_3 \D_3,
   \\
S_2   = \gamma_0 \D_3  + \gamma_1 \D_2 - \gamma_2 \D_1 - \gamma_3 \D_0,     \\
S_3 = \gamma_0 \D_2 - \gamma_1 \D_3 - \gamma_2 \D_0 + \gamma_3 \D_1 ,    \\
S_4 = \gamma_0 \D_1 -\gamma_1 \D_0 + \gamma_2 \D_3 - \gamma_3 \D_2   .
\end{array}
\end{equation}
Introducing the complex supercharges
\begin{equation}
\label{QcherezS}
\begin{array}{c}
Q_1 = (S_1 - iS_2)/2,\vergule
Q_2 = (S_3 -  iS_4)/2, \\
\bar Q^1 = (S_1 + iS_2)/2,\vergule
\bar Q^2 = (S_3 + iS_4)/2,
\end{array}
\end{equation}
 we obtain the standard
 ${\cal N} = 4$ supersymmetry algebra~(\ref{c3e5x}).
Correspondingly, the excited spectrum of $H$ is four-fold degenerate, while the spectrum of $/\!\!\!\!\D$
consists of the quartets involving
two degenerate positive and two degenerate negative eigenvalues.
Note that, in contrast to $/\!\!\!\! \D$,  the operator $/\!\!\!\! \D \gamma_5$ is not expressed into a linear combination of $S_A$.
In other words, the ${\cal N}=2$ supersymmetry algebra with the operators $/\!\!\!\! \D (1 \pm \gamma_5)$ is not a subalgebra of the
${\cal N}=4$ algebra~(\ref{c3e5x}).

The algebra (\ref{N4}) with supercharges (\ref{matrS}) holds for any self-dual field,  irrespectively of whether
it is Abelian or non-Abelian. Thus,
the additional 2-fold degeneracy of the spectrum of the Dirac operator mentioned above should be there for a generic
self-dual field. One particular example of a non-Abelian self-dual field is the instanton solution, where this degeneracy
was observed back in \cite{JR} (see Eqs.~(4.15) there). The generalization of the Dirac operator and (anti)self-duality conditions to higher-dimensional manifolds was considered in Ref.~\cite{Kirchberg:2004za}.

\subsection{Covariant description}

To make contact with the Lagrangian (and, especially, superfield) description,
it is convenient to introduce holomorphic fermion variables $\psi_{\dot\alpha}$ and $\bar\psi^{\dot\alpha}$,
which satisfy the standard anticommutation relations
\begin{equation}
\label{antikompsi}
\{ \psi_{\dot\alpha}, \psi_{\dot\beta} \} \ =\ \{ \bar\psi^{\dot\alpha}, \bar\psi^{\dot\beta} \} \ =\ 0, \vergule
\{\bar \psi^{\dot\alpha}, \psi_{\dot\beta} \}
= \delta^{\dot\alpha}_{\dot\beta}
\end{equation}
and are realized as matrices in the following way:
\begin{eqnarray}
\label{psicherezgamma}
\psi_{\dot 1} = \frac {-\gamma_0 + i\gamma_3}2,\vergule
\bar\psi^{\dot 1} = \frac {\gamma_0 + i\gamma_3}2,  \nn\ \\
\psi_{\dot 2} = \frac {\gamma_2 + i\gamma_1}2,\vergule
\bar\psi^{\dot 2} = \frac {-\gamma_2 + i\gamma_1}2.\
\end{eqnarray}
Then two complex supercharges (\ref{QcherezS}) are expressed in a very simple way, namely
\begin{equation}
\label{eq_Q4}
\begin{array}{c}
  Q_\alpha= \left(\sigma_\mu  \bar\psi\right)_\alpha \left(\hat p_\mu-\ca A_\mu\right),
\\[3mm]
 \bar Q^\alpha= \left(\psi \sigma^\dagger_\mu  \right)^\alpha \left(\hat p_\mu-\ca A_\mu\right),
\end{array}
\end{equation}
with $\hat p_\mu = -i\partial_\mu$.
 The Hamiltonian (\ref{HD2}) is expressed in these terms as
\begin{equation}
\label{Hflatcherezpsi}
  H = \frac{1}{2} \left(\hat p_\mu-\ca A_\mu\right)^2
     +  \frac{i}{4}\, {\cal F}_{\mu\nu}\,\psi \sigma^\dagger_{\mu} \sigma_{\nu}\bar\psi\ .
\end{equation}
It is clear now why the spinor indices in Eqs.~(\ref{c3e5x}) are undotted, while in Eq.~(\ref{psicherezgamma}) they
are dotted:
the supercharges are rotated by the first ${\rm SU}(2)$ and the variables $\psi_{\dot\alpha}$ by the second~\footnote{Note that complex conjugation leaves the spinors in the same representation, the symmetry group here is ${\rm SO}(4)$ rather than ${\rm SO}(1,3)$.}.
A careful distinction between two different ${\rm SU}(2)$ factors allows one to understand better the reason
why the supercharges (\ref{eq_Q4}) satisfy the simple algebra~(\ref{c3e5x}) in a self-dual background.
The self-dual field density ${\cal F}$ carries in the spinor notation only dotted indices. Therefore any expression
involving ${\cal F}, \psi, \bar\psi$ is a scalar with respect to undotted ${\rm SU}(2)$. The only such scalar that can appear
in the right hand side of the anticommutators of the supercharges $\{Q_\alpha, \bar Q^\beta\}$ is the structure which is
proportional to $\delta_\alpha^\beta$, i.e. the Hamiltonian. No other operator is allowed.

In the {\it Abelian} case, the
supercharges (\ref{eq_Q4}) and the Hamiltonian (\ref{Hflatcherezpsi}) are scalar operators not carrying matrix
indices anymore. This allows one to derive  the Lagrangian,
\begin{equation}
  \label{Lflat}
  L =
    \frac{1}{2}\, \dot x_\mu\dot x_\mu
    +\ca A_\mu(x)\dot x_\mu
    +i{\bar\psi}^{\dot\alpha} \dot\psi_{\dot\alpha}
    -\frac{i}{4} {\cal F}_{\mu\nu}\,\psi\sigma^\dagger_{\mu}\sigma_{\nu}\bar\psi
  \ .
\end{equation}

In the non-Abelian case, the expressions (\ref{eq_Q4}) and (\ref{Hflatcherezpsi}) still keep their color matrix
structure, and one cannot derive the Lagrangian in a so straightforward way. One of the ways to handle the matrix structure is
 to introduce a set of color fermion variables (say, in the fundamental representation of the group)
and impose
the extra constraint considering only the sector with unit fermion charge \cite{Gaume}. An alternative non-Abelian construction of the Lagrangian
is presented in Section~\ref{sec4.3}. In this section, we limit ourselves only to Lagrangians for Abelian fields.

\subsection{The generalization to conformally flat metric}

As will be demonstrated explicitly in Section~\ref{sect4}, the component Lagrangian (\ref{Lflat}) follows from the
superfield action written earlier by Ivanov and Lechtenfeld  in the framework of harmonic superspace
approach \cite{IvLecht}.
We will see that one can naturally derive in this way a $\sigma$-model type generalization of the
Lagrangian (\ref{Lflat}) describing
the motion over the manifold with nontrivial conformally flat metric $ds^2 = \left\{f(x)\right\}^{-2} dx_\mu dx_\mu$.
It is written as follows:
\begin{multline}\label{eq_action}
     L \ =\
    \frac{1}{2}f^{-2}\, \dot x_\mu\dot x_\mu
     +i{\bar\psi}^{\dot\alpha} \dot{\psi}_{\dot\alpha}
    +\frac{1}{4} \left\{3\left(\partial_\mu f\right)^2- f\partial^2 f \right\} \psi^4
     +\frac i2 f^{-1}\partial_\mu f \,\dot x_\nu\,
      \psi  \sigma^\dagger_{[\mu}\sigma_{\nu]} \bar\psi
\\[3mm]
    +\ca A_\mu(x)\dot x^\mu
    - \frac{i}{4}f^2 {\cal F}_{\mu\nu}\,\psi\sigma^\dagger_{\mu}\sigma_{\nu}\bar\psi,
\end{multline}

The corresponding quantum N\"{o}ether supercharges and the Hamiltonian are
(see also Section~\ref{sec4.2.3} for details)
\begin{equation}\label{eq_Qf}
\begin{array}{l}
  Q_\alpha = f \left(\sigma_\mu \bar\psi\right)_\alpha \left(\hat p_\mu-\ca A_\mu\right)
-\psi_{\dot\gamma} \bar\psi^{\dot\gamma} \left(\sigma_\mu\bar\psi\right)_\alpha i\partial_\mu f,
\\[2mm]
  \bar Q^\alpha = \left(\psi\sigma^\dagger_\mu\right)^\alpha \left(\hat p_\mu-\ca A_\mu\right)f
+i\partial_\mu f \left(\psi\sigma^\dagger_\mu\right)^\alpha \cdot \psi_{\dot\gamma}\bar\psi^{\dot\gamma},
\end{array}
\end{equation}
\begin{multline}\label{eq_susyham}
  H=\frac{1}{2}f \left(\hat p_\mu-\ca A_\mu\right)^2 f
      +\frac{i}{4}f^2\, {\cal F}_{\mu\nu}\,\psi\sigma^\dagger_{\mu}\sigma_{\nu}\bar\psi
\\[2mm]
      - \frac 12 f i\partial_\mu f \,(\hat p_\nu-\ca A_\nu)\, \psi
   \sigma^\dagger_{[\mu} \sigma_{\nu]}\bar\psi
    + f\partial^2 f\left\{\psi_{\dot\gamma}\bar\psi^{\dot\gamma}-\frac{1}{2}\left(\psi_{\dot\gamma} \bar\psi^{\dot\gamma}\right)^2\right\}.
\end{multline}
On the other hand, one can explicitly calculate the anticommutators of the supercharges (\ref{eq_Qf})
 for any self-dual~\footnote{Anti-self-duality conditions are obtained when one interchanges
$\sigma_\mu$ and $\sigma^\dagger_\mu$ in all the formulas.
This is equivalent to the interchange of two spinor representations of \SO({4}).
}
field $\A_\mu(x)$, Abelian or non-Abelian, and verify that
the algebra (\ref{c3e5x}) holds.
While doing this, the use of the following Fierz identity
\begin{equation}
  \big(\bar\psi\sigma^\dagger_\mu\big)^\beta \big(\sigma_{\nu}\psi\big)_\alpha
  - \big(\sigma_{\mu}\bar\psi\big)_\alpha \big(\psi\sigma^\dagger_{\nu}\big)^\beta =
  \delta^\beta_\alpha \,\bar\psi \sigma^\dagger_\mu \sigma_\nu \psi \ ,
\end{equation}
which can be proven using (\ref{transpose}), is convenient.

Note that, with a nontrivial factor $f(x)$, the Dirac operator $\,/\!\!\!\!{     \D}$ in a conformally flat background can be expressed as a
linear combination of $Q_\alpha$ and $\bar Q^\alpha$ {\em only} if one also adds a certain {\em torsion} proportional to the derivatives of $f(x)$ \cite{arXiv:1107.1429}. The Hamiltonian (\ref{eq_susyham}) would also coincide with $\,/\!\!\!\!{     \D}^2/2$ in this case.
In fact, generalization of the system with the Hamiltonian~(\ref{Hflatcherezpsi}) to the conformally flat case, which preserves $\N=4$ supersymmetry, as in Eq.~(\ref{eq_susyham}) always involves the torsion field. The Hamiltonian~(\ref{Hflatcherezpsi}) also admits another type of $\N=4$ extension to a curved space endowed with Hyper-K\"{a}hler metric and without a torsion \cite{Kirchberg:2004za}. In its most generality, the Hamiltonian involves weak HKT geometry -- the generalization of Hyper-K\"{a}hler metric obtained by introduction of conformal factor and torsion~\cite{arXiv:1107.1429}.

The model (\ref{eq_action}), (\ref{eq_Qf}), (\ref{eq_susyham}) is a close relative to the model constructed in
Ref.~\cite{Smilga:1986rb} (see Eqs.~(30) and~(31) there), which describes the motion on a {\it three}-dimensional conformally
flat manifold in external magnetic field and a scalar potential. In fact, the latter model
can be obtained from the former, if assuming that the metric
and the vector potential ${\cal A}_\mu \equiv (\Phi, \vec{\A})$ depend only on three spatial coordinates $x_i$.
 If assuming further that the metric is flat, one is led to the Hamiltonian \cite{de Crombrugghe:1982un}
\begin{equation}\label{eq_3dham}
  H \ = \ \frac{1}{2}\left(\hat {\vec p}-\vec{\ca A}\right)^2+\frac{1}{2}U^2+\vec\nabla U \,\psi\vec\sigma\bar\psi,
\end{equation}
which is supersymmetric under the condition $ {\cal F}_{ij}  = \varepsilon_{ijk}
 \partial_k U $ (the 3-dimensional reduction of the four-dimensional self-duality condition). It was noticed in Ref.~\cite{Smilga:1986rb}
that the effective Hamiltonian of a chiral supersymmetric electrodynamics
in finite spatial volume belongs to this class with $U \propto 1/|\vec{A}|$.
The vector potential $\vec{\cal A}(\vec{A})$ describes
in this case a Dirac magnetic
monopole such that the Berry phase appears. The three dynamical variables $\vec{A}$ (do not confuse with curly $\vec{\cal A}$ !)
have in this case the meaning of
the zero Fourier harmonic of the vector
potential in the original field theory. In the leading order, the metric is flat.
When higher loop corrections are included, a (conformally flat !) metric on the moduli space $\{\vec{A}\}$ appears.

Performing the Hamiltonian reduction of Eq.~(\ref{eq_susyham}) with non-Abelian $\A_\mu$, a non-Abelian
generalization of Eq.~(\ref{eq_3dham}) can easily be derived. It keeps the gauge structure of
Eq.~(\ref{eq_3dham}) with matrix-valued $\vec \A$ and $U$ satisfying the condition
 $\F_{ij}=\varepsilon_{ijk}\D_k U$. Note that such Hamiltonian does {\sl not} coincide
with the non-Abelian 3-dimensional Hamiltonian derived in Ref.~\cite{BKS}.

\subsection{Example of constant gauge field}\label{sect3}

As an illustration, consider the system described by  the Hamiltonian (\ref{Hflatcherezpsi})
 in a constant self-dual Abelian background. The constant self-dual field strength $\F_{\mu\nu}
= \eta_{\mu\nu}^a B_a $ is parametrized by three independent components. Let us direct $B^a$ along the third axis,
$B_a=(0,0,B)$, and  choose the gauge
\begin{equation}\label{eq_constfieldA}
   \A_0=Bx_3,\vergule
   \A_2=Bx_1,\vergule
   \A_1=\A_3=0.
\end{equation}
 The Hamiltonian (\ref{Hflatcherezpsi}) acquires the form
\begin{multline}\label{eq_constfield}
 H=\left\{\frac 12 \left(\hat p_0-Bx_3\right)^2+\frac 12 \hat p_3^2
              +B \left(\chi_1\bar\chi^1-\frac 12\right)\right\}
\\
   +\left\{\frac 12 \left(\hat p_2-Bx_1\right)^2+\frac 12 \hat p_1^2
              +B \left(\chi_2\bar\chi^2-\frac 12\right)\right\}.
\end{multline}
For convenience, we have introduced notations
 $\chi_1=\bar\psi^{\dot 1}$, $\bar\chi^1=\psi_{\dot 1}$, $\chi_2=\psi_{\dot 2}$, $\bar\chi^2=\bar\psi^{\dot 2}$.
The  Hamiltonian is thus reduced to the sum $H_1 + H_2$ of two independent (acting in different Hilbert spaces)
supersymmetric Hamiltonians, each describing the 2-dimensional  motion of an electron in homogeneous orthogonal to the plane
magnetic field $\vec{B}$.
The bosonic sector of each such Hamiltonian corresponds to the spin projection $\vec{s} \vec{B}/|\vec{B}| = -1/2$, and the fermionic
sector to the spin projection $\vec{s} \vec{B}/|\vec{B}| = 1/2$.
This is the first and the simplest supersymmetric quantum problem ever
considered \cite{Landau}.
 The energy levels for each Hamiltonian are
$\varepsilon_i=B\left(n_i+\frac 12+s_i\right)$, $n_i\ge 0$ -- integers, $s_i=\pm \frac 12$. Each level of $H_i$ is doubly degenerate. Besides, there is an
infinite degeneracy associated with the positions of the center of the orbit along the axes 1 and 3 that
are  proportional to the integrals of motion $p_2$ and $p_0$. The full spectrum
\begin{equation}\label{eq_25}
 E=B\left(n_1+n_2+1+s_1+s_2\right)
\end{equation}
is thus 4-fold degenerate at each level (except for the singlet state with $E=0$) for given $p_0$, $p_2$.

It might be instructive to explicitly associate this degeneracy with the action of supercharges (\ref{eq_Q4}).
Let us assume for definiteness $B > 0$. One can represent $Q_\alpha$ as
\begin{equation}
 Q_1=\sqrt{2B}\left(b\chi_1+a^\dagger\bar\chi^2\right),
\hskip 1cm
 Q_2=\sqrt{2B}\left(a\chi_1-b^\dagger\bar\chi^2\right)\ ,
\end{equation}
where $a^\dagger$, $b^\dagger$ and  $a$, $b$ are the creation and annihilation operators,
\begin{equation}
 a=\frac{1}{\sqrt{2B}}\left(\hat p_1 - iBx_1  \ + i p_2 \right),
\vergule
 b=\frac{1}{\sqrt{2B}}\left(\hat p_3 - iBx_3 \ + i p_0\right)\ ,
\end{equation}
\begin{equation}
 \left[a,a^\dagger\right]=1,\vergule
 \left[b,b^\dagger\right]=1.
\end{equation}
In these notations, the Hamiltonian (\ref{eq_constfield}) takes a very simple form
\begin{equation}
 H=
    B\left\{
       a^\dagger a+b^\dagger b
       +\chi_1\bar\chi^1+\chi_2\bar\chi^2
    \right\}.
\end{equation}

Obviously, the energy levels of the Hamiltonian (\ref{eq_constfield}) are defined by two integrals of motion
$p_{2,0}$,
two oscillator excitation numbers $n_{1,2}$ and two spins $s_{1,2}$, as in Eq.~(\ref{eq_25}).
For each  $p_2, p_0$, there is a unique ground zero energy state $|0 \rangle$
annihilated by all supercharges. A quartet of excited states
can be  represented as
 \begin{equation}
 \ket{n_1,n_2},\quad
 Q_1^\dagger\ket{n_1,n_2},\quad
 Q_2^\dagger\ket{n_1,n_2},\quad
 Q_1^\dagger Q_2^\dagger\ket{n_1,n_2} \ ,
\end{equation}
where the state
$$ \ket{n_1,n_2}\equiv\chi_1\cdot\left(a^\dagger\right)^{n_1}\left(b^\dagger\right)^{n_2}\ket{0}$$
of energy  $E=B(n_1+n_2+1)$ is annihilated by both $Q_1$ and $Q_2$.

For each $p_2, p_0$, there are $N$ such quartets at the energy level $E = BN$.

\section{Harmonic superspace description in the Abelian case}\label{sect4}

In this section, we derive  the Hamiltonian (\ref{eq_susyham}) from the harmonic superspace approach.
The introduction to the harmonic superspace and its salient features and definitions in application to quantum
mechanical problems were already discussed in the previous chapter.
 The relevant superfield action was written in \cite{IvLecht}. Let us show here that the corresponding component Lagrangian
coincides with (\ref{eq_action}). The corresponding supercharges (\ref{eq_Qf}) and the Hamiltonian (\ref{eq_susyham}) involve
an Abelian self-dual gauge field ${\cal A}_\mu (x)$. The non-Abelian field case is discussed later in this chapter.

\subsection{Superfield content}

Let us introduce a doublet of superfields $q^{+{\dot\alpha}}$ with charge +1 ($D^0 q^{+\dot\alpha} = q^{+\dot\alpha}$)
satisfying the constraints~(\ref{eq_anal}), (\ref{c3reality3}).
The index ${\dot\alpha}$ is the fundamental representation index of an additional external (Pauli-G\"{u}rsey) group ${\rm SU}(2)$.
The solution for these constraints in the analytical basis was written in Eqs.~(\ref{c3eq_qdot4}), (\ref{c3xxx}).
It can be presented in the central basis~(\ref{c3e25}) as $q^{+{\dot\alpha}}=u^+_\alpha q^{\alpha{\dot\alpha}}$, where
$q^{\alpha{\dot\alpha}}$ does not depend on $u^\pm_\alpha$ (the latter follows from the constraint $D^{++} q^{+{\dot\alpha}} = 0$
and the definition $D^{++} = u^+_\alpha \pop{u^-_\alpha}$).
It is convenient to go over to the four-dimensional vector  notation~(\ref{spinor2vector}), introducing
\begin{equation}\label{eq_qqp}
  q_\mu= -\frac 12\left(\sigma_\mu\right)_{\alpha{\dot\alpha}} \, q^{\alpha{\dot\alpha}},\vergule
  q^{+{\dot\alpha}}=-q_\mu\left(\sigma^\dagger_\mu\right)^{{\dot\alpha} \alpha}u_\alpha^+.
\end{equation}
 Now, $q_\mu$ is a vector with respect to
the group ${\rm SO}(4)={\rm SU_{\rm R}}(2)\times{\rm SU_{\rm PG}}(2)$, with the first factor representing the
 $\ca N=4$ R-symmetry group and the second one being the Pauli-G\"{u}rsey global ${\rm SU}(2)$ group which rotates the dotted ``flavor'' indices.

The pseudoreality condition (\ref{c3reality3}) implies that the superfield $q_\mu$ is { real}. The latter
 is expressed in components as follows:
  \begin{equation}\label{eq_qmu}
  q_\mu=x_\mu
+\theta\sigma_\mu\chi+\bar\theta\sigma_\mu\bar\chi
-\frac{i}{2}\dot x_\nu\, \bar\theta \sigma_{[\mu}\sigma^\dagger_{\nu]}\theta
    +\frac{i}{2}\bar\theta\sigma_\mu\dot \chi \,\theta^2
    -\frac{i}{2}\theta\sigma_\mu\dot{\bar\chi} \,\bar\theta^2
  -\frac{1}{4}\ddot x_\mu \,\theta^4 \ ,
  \end{equation}
where $\theta^2 \equiv \theta^\alpha\theta_\alpha$, $\bar\theta^2 \equiv \bar\theta^\alpha\bar\theta_\alpha$,
$\theta^4 \equiv \theta^2\bar\theta^2$.
Moreover, the first equality in Eq.~(\ref{c3xxx}) implies that
\begin{equation}\label{c4Xreal}
x_\mu = -\frac 12x^{\alpha \dot\alpha} (\sigma_\mu)_{\alpha \dot\alpha} 
\end{equation}
is also real, and we are left with four dynamic bosonic variables.

\subsection{Superfield action}

The classical $\ca N=2$ supersymmetric action for the superfield $q_\mu$ can now be written. It consists of two parts, $S=S_{\rm kin}+S_{\rm int}$.
The kinetic part,
\begin{equation}
\label{Lkin}
  S_{\rm kin}=\int  dt\, d^4\theta du\, R'_{\rm kin}(q^{+{\dot\alpha}}, q^{-{\dot\beta}}, u^\pm_\gamma)=\int dt\,d^4\theta\, R_{\rm kin}(q_\mu),
\end{equation}
depends on  an arbitrary function $R_{\rm kin}(q_\mu)$. Note that one can forget the harmonic superspace coordinates here and work in an ordinary superspace. In other words, the kinetic term $S_{\rm kin}$ does not require the additional coordinates $u^\pm_\alpha$ of the harmonic superspace.
Plugging  (\ref{eq_qmu}) into (\ref{Lkin}) and adding/subtracting proper total derivatives, one obtains
\begin{multline}\label{eq_41}
  S_{\rm kin}=\int dt\left\{
    \frac{1}{2}g(x)\, \dot x_\mu\dot x_\mu
    +\frac{i}{2}g(x)\left(\bar\chi^{\dot\alpha}\dot\chi_{\dot\alpha}-\dot{\bar\chi}^{\dot\alpha}\chi_{\dot\alpha}\right)
  \right.
\\[3mm]
  \left.
    +\frac{1}{8}\partial^2 g(x) \,\chi^4
    -\frac{i}{4}\partial_\mu g(x) \,\dot x_\nu\,
      \chi\sigma^\dagger_{[\mu}\sigma_{\nu]}\bar\chi
  \right\},
\end{multline}
where $g(x)=\frac{1}{2}\partial^2_x R_{\rm kin}(x)$ and
$\chi^4=\chi^{\dot\alpha}\chi_{\dot\alpha}\,\bar\chi^{\dot\beta}\bar\chi_{\dot\beta}$.

To couple $x_\mu$ to an external gauge field, one should add  the interaction term $S_{\rm int}$ that represents an integral over
{\it analytic} superspace in the harmonic superspace,
\begin{equation}\label{eq_38}
  S_{\rm int}=\int dt\, du\, d\bar\theta^+ d\theta^+ R_{\rm int}^{++}\left(q^{+{\dot\alpha}}(t_{\rm A},\theta^+,\bar\theta^+),u^\pm_\gamma\right).
\end{equation}
 We choose $R_{\rm int}^{++}$ (it carries the charge +2, $D^0 R_{\rm int}^{++}=2R_{\rm int}^{++}$)
satisfying the condition $\widetilde {R_{\rm int}^{++}} = - R_{\rm int}^{++}$ (the involution operation $\widetilde X$
was defined in Section~\ref{c3sectInvolution}) such that the action (\ref{eq_38}) is real. In contrast to the kinetic term, the interaction term involves the dependence on harmonics $u^\pm_\gamma$ and thus cannot be written in terms of superfields of ordinary superspace~(\ref{c3orsusp}).

To do the integral over $\theta^+$ and $\bar\theta^+$, we substitute Eq.~(\ref{c3eq_qdot4}) into~(\ref{eq_38}) and expand the latter in Taylor series over $\theta^+$, $\bar\theta^+$, keeping only terms $\sim\theta^+\bar\theta^+$:
\begin{multline}
 R_{\rm int}^{++}(q^{+\dot\alpha}, u^\pm_\gamma)
=
\partial_{+\dot\alpha} R_{\rm int}^{++}\cdot
  \left(-2i\theta^+\bar\theta^+ u^-_\alpha\dot x^{\alpha\dot\alpha}\right)
\\[2mm]
 +2\partial_{+\dot\alpha}\partial_{+\dot\beta}R_{\rm int}^{++}
  \cdot\theta^+\bar\theta^+
  \left(\chi^{\dot\alpha}\bar\chi^{\dot\beta}+\chi^{\dot\beta}\bar\chi^{\dot\alpha}\right)
+\dots
\end{multline}
(ellipsis denote terms not proportional to $\theta^+\bar\theta^+$) with
\begin{equation}
  \partial_{+\dot\alpha} R_{\rm int}^{++} (x,u)\equiv
  \frac{\partial R_{\rm int}^{++} (x^{+{\dot\gamma}}, u^\pm_\gamma)}{\partial x^{+{\dot\alpha}}}
\end{equation}
and similarly for $\partial_{+\dot\alpha}\partial_{+\dot\beta}R_{\rm int}^{++}$.
Let us also pass to vector notation $x_\mu$ for the coordinates $x^{\alpha\dalpha}$, see Eq.~(\ref{c4Xreal}). Consequently,
\begin{equation}
x^{+{\dot\alpha}}
\equiv x^{\alpha\dot\alpha}u^+_\alpha
=-x_\mu \left(\sigma^\dagger_\mu\right)^{{\dot\alpha} \alpha}u^+_\alpha.
\end{equation}
Then
\begin{equation}
\label{Lint}
  S_{\rm int}=
    \int dt\,du\left\{2i\left(\sigma^\dagger_\mu\right)^{{\dot\alpha} \alpha} \partial_{+\dot\alpha} R_{\rm int}^{++}\, u_\alpha^- \cdot \dot x_\mu
    -4\chi^{\dot\alpha}\bar\chi^{\dot\beta}\,\partial_{+\dot\alpha}\partial_{+\dot\beta} R_{\rm int}^{++}
    \right\}.
\end{equation}
Now, define the gauge field,
\begin{equation}
\label{defA}
  \ca A_\mu(x) \equiv \int du\left\{2i\left(\sigma^\dagger_\mu\right)^{{\dot\alpha} \alpha} \partial_{+\dot\alpha} R_{\rm int}^{++}\, u_\alpha^-\right\}.
\end{equation}
As the action (\ref{Lint}) is real, the field $\ca A_\mu(x)$ is also real. It automatically has zero divergence,
\begin{equation}
\partial_\mu \A_\mu=0.
\end{equation}

The field strength is expressed as
\begin{equation}
\label{defF}
  \F_{\mu\nu} = \partial_\mu \ca A_\nu-\partial_\nu \ca A_\mu
  = -2 \eta_{\mu\nu}^a  \int du\, \partial_{+\dot\alpha}\partial_{+\dot\beta} R_{\rm int}^{++}\,
\varepsilon^{{\dot\alpha}{\dot\gamma}}
    \left( \sigma_a  \right)^{\!{\dot\beta}}_{\,\,{\dot\gamma}}
\end{equation}
(the identities (\ref{eq_sigma}) were used). It is obviously self-dual because the 't~Hooft symbols are self-dual, see Eq.~(\ref{thooftselfdual}).
 With the definitions (\ref{defA}) and (\ref{defF}) in hand, one can represent the interaction term~(\ref{Lint}) simply as
\begin{equation}
\label{43}
  S_{\rm int}=\int dt\left\{\ca A_\mu(x)\dot x_\mu
  -\frac{i}{4}\F_{\mu\nu}\,\chi\sigma^\dagger_{\mu}\sigma_{\nu}\bar\chi\right\}.
\end{equation}

Finally, one can get rid of the factor $g(x)$ in the fermion kinetic term~(\ref{eq_41}) by introducing canonically conjugated
\begin{equation}\label{eq_canon}
 \psi_{\dot\alpha}=f^{-1}(x)\chi_{\dot\alpha},
\quad\quad
 \bar\psi^{\dot\alpha} = f^{-1}(x) \bar\chi^{\dot\alpha}
\end{equation}  
with
\begin{equation}\label{eq_fg}
f(x) = g^{-1/2}(x)\equiv \left[\frac{1}{2}\,\partial^2_\mu R_{\rm kin}(x)\right]^{-1/2}.
\end{equation}
Adding the kinetic term in~(\ref{eq_41}) to the interaction term~(\ref{43}),
one can explicitly check that the Lagrangian $L = L_{\rm kin}+L_{\rm int}$ coincides, up to a total derivative, with (\ref{eq_action}).

As was noticed, the field $A_\mu$ naturally obtained in the HSS framework satisfies the constraint  $\partial_\mu \A_\mu=0$
\cite{IvLecht}.
This does not really impose a restriction, however, because gauge transformations of $A_\mu$ that shift it by
the gradient of an arbitrary
 function amount to adding a total derivative in the Lagrangian (\ref{43}).

\subsection{Supertransformations, quantization and Weyl ordering}\label{sec4.2.3}

By construction, the action with the Lagrangian~(\ref{eq_action}) is invariant under the following supersymmetry transformations:
\begin{equation}\label{c4susytrans} 
\begin{array}{c}
  x_\mu\rightarrow x_\mu
  +f\epsilon\sigma_\mu\psi
  +f\bar\epsilon\sigma_\mu\bar\psi,
\\[3mm]
  f\psi_{\dot\alpha}\rightarrow f\psi_{\dot\alpha}
  +i\dot x_\mu \left(\bar\epsilon\sigma_\mu\right)_{{\dot\alpha}},
\\[3mm]
  f\bar\psi^{\dot\alpha}\rightarrow f\bar\psi^{\dot\alpha}
  -i\dot x_\mu \left(\sigma^\dagger_\mu\epsilon\right)^{{\dot\alpha}}.
\end{array}
\end{equation}
 The N\"{o}ether classical supercharges expressed in terms of  $\psi_{\dot\alpha}$ and $\bar\psi^{\dot\alpha}$, $x_\mu$ and their canonical momenta,
\begin{equation}
\label{pmu}
p_\mu \ =\ f^{-2}\dot x_\mu + {\cal A}_\mu - \frac i2 f^{-1} \partial_\nu f\,\psi \sigma^\dagger_{[\mu} \sigma_{\nu]} \bar \psi\ ,
\end{equation}
are
\begin{equation}\label{c4clQ}
\begin{array}{ccl}
 Q_\alpha &=& f \left(\sigma_\mu \bar\psi\right)_\alpha \left( p_\mu-\ca A_\mu\right)
-i  \partial_\mu f \psi_{\dot\gamma} \bar\psi^{\dot\gamma} \left(\sigma_\mu\bar\psi\right)_\alpha,
\\[2mm]
 \bar Q^\alpha &=& \mbox{[complex conjugate]}  .
\end{array}
\end{equation}

The quantization procedure of the corresponding classical Hamiltonian has order ambiguity problem for the bosonic operators $\hat p_\mu = -i\partial_\mu$ and $x^\mu$ as well as for the fermionic operators $\hat \psi_\dalpha$ and $\hat{\bar\psi}^\dalpha$ with anticommutation relations~(\ref{antikompsi}). (We temporary restore ``hats'' on fermionic operators in order to distinguish them from the corresponding classical anticommuting variables $\psi_\dalpha$ and $\bar\psi^\dalpha$.) One must thus define an ordering procedure in such a way that the supersymmetry algebra~(\ref{c3e5x}) would hold.
The solution to this problem is known \cite{SMI} and it prescribes to order the supercharges in a certain way, while the quantum Hamiltonian should be obtained from the anticommutator $\left\{Q_\alpha,\,\bar Q^\alpha\right\}$.

It is prescribed by Ref.~\cite{SMI} to order the operators in the classical supercharges~(\ref{c4clQ}) with the so called {\em Weyl ordering} procedure: any product of operators must be substituted with its totally symmetrized expression, taking into account the commuting/anticommuting nature of the operators. For instance, the expression $x^1 x^2 p_3$, upon quantization, becomes
\begin{equation}
 x^1 x^2p_3 \quad\longrightarrow\quad \frac 16\left(x^1 x^2 \hat p_3
  +x^2 x^1 \hat p_3 + x^1 \hat p_3 x^2 + x^2 \hat p_3 x^1
  + \hat p_3 x^1 x^2 + \hat p_3 x^2 x^1\right),
\end{equation}
while the expression $\psi_\dgamma\bar\psi^\dgamma\, \bar\psi^\dbeta$ becomes
\begin{multline}\label{c4.2.20}
 \psi_\dgamma\bar\psi^\dgamma\, \bar\psi^\dbeta
\quad\longrightarrow\quad
 \frac 16 \left(
  \hat\psi_\dgamma \hat{\bar\psi}^\dgamma \hat{\bar\psi}^\dbeta
  + (-1)\hat\psi_\dgamma\hat{\bar\psi}^\dbeta\hat{\bar\psi}^\dgamma
  +(-1)\hat{\bar\psi}^\dgamma\hat\psi_\dgamma \hat{\bar\psi}^\dbeta
\right.
\\[2mm]
\left.
  +\hat{\bar\psi}^\dbeta\hat\psi_\dgamma\hat{\bar\psi}^\dgamma
  +\hat{\bar\psi}^\dgamma\hat{\bar\psi}^\dbeta\hat\psi_\dgamma
  +(-1)\hat{\bar\psi}^\dbeta\hat{\bar\psi}^\dgamma\hat\psi_\dgamma
  \right).
\end{multline}

Let us elaborate more on the ordering of the expression $f(x) p_\mu$. For this, consider  Weyl ordering of the expression $x^n p$, where $x$ is any of the coordinates $x^\mu$, while $p$ is the corresponding conjugated momentum which, upon quantization, becomes $\hat p =-i\partial/\partial x$. One has:
\begin{multline}
 x^n p \quad\longrightarrow\quad
\frac{1}{n+1}
\left(x^n \hat p + x^{n-1} \hat p x + x^{n-2} \hat p x^2
  +\dots + \hat p x^n\right)
\\[2mm]
=
 x^n\hat p +\frac{1}{n+1}x^{n-1} \left(1+2+\dots+n\right)\cdot
  \left[\hat p,\,x\right]
\\[2mm]
=
 x^n \hat p +\frac 12 n x^{n-1}
  \left[\hat p,\,\hat x\right]
\end{multline}
Thus, in fact, Weyl ordering of the product $f(x)p_\mu$ gives a simple formula:
\begin{equation}\label{c4.2.22}
 f(x)p_\mu \quad\longrightarrow\quad \frac 12 \big[f(x)\hat p_\mu + \hat p_\mu \, f(x)\big]
  =f(x)\hat p_\mu - \frac 12 i\,\partial_\mu f(x).
\end{equation}
In the same way, the expression in Eq.~(\ref{c4.2.20}) can be simplified which gives
\begin{equation}\label{c4.2.23}
 \psi_\dgamma\bar\psi^\dgamma\, \bar\psi^\dbeta
\quad\longrightarrow\quad
 \hat \psi_\dgamma \hat{\bar\psi}^\dgamma \,\hat {\bar\psi}^\dbeta
  -\frac 12 \hat{\bar\psi}^\dbeta .
\end{equation}
Finally,  combining Eqs.~(\ref{c4.2.22}), (\ref{c4.2.23}), one obtains:
\begin{equation}
 fp_\mu \bar\psi^\dbeta - i\partial_\mu f\, \psi_\dgamma\bar\psi^\dgamma \,\bar\psi^\dbeta
 \quad\longrightarrow\quad
 f\hat p_\mu \hat{\bar\psi}^\dbeta -i\partial_\mu f \,\hat \psi_\dgamma \hat{\bar\psi}^\dgamma \,\hat{\bar\psi}^\dbeta .
\end{equation}
This gives the quantum supercharges~(\ref{eq_Qf}). One can check that the anticommutator $\{Q_\alpha, \bar Q^\alpha\}$ gives the quantum Hamiltonian~(\ref{eq_susyham}).

\section{The component Lagrangian in the non-Abelian case}\label{sec4.3}

For a matrix-valued non-Abelian self-dual field  ${\cal A}_\mu$, the (scalar) Lagrangian cannot be straightforwardly derived from the Hamiltonian
(\ref{eq_susyham}) by a Legendre transformation as it was done in the Abelian case, Eq.~(\ref{eq_action}). Nevertheless, this can be done in the case of $\SU(N)$ gauge group by introducing extra ``semi-dynamical'' fields $\varphi_i$
in the fundamental representation of $\SU(N)$ and the auxiliary ${\rm U}(1)$ gauge field $B(t)$.
The second line in (\ref{eq_action}) is then generalized to
\begin{equation}\label{eq_lagrNA}
 L_{\rm int}^{\SU(N)}=i\bar\varphi^i\left(\dot\varphi_i+iB\varphi_i\right)
  +kB
  +\A_\mu^a T^a \,\dot x_\mu-\frac{i}{4}f^2\F_{\mu\nu}^a T^a\,\psi\sigma_\mu^\dagger\sigma_\nu\bar\psi
\end{equation}
with integer $k$ and
\begin{equation}\label{eq31}
T^a=\bar\varphi^i\left(t^a\right)^{\,\, j}_{\! i} \varphi_j,
\end{equation}
$t^a$ being standard $\SU(N)$ algebra generators.
The interaction Lagrangian~(\ref{eq_lagrNA}) possesses \mbox{$\N=4$} supersymmetry.
The corresponding supersymmetry transformations are written below in Eqs.~(\ref{susytran}).
It is not difficult to check that it is invariant with respect to the non-Abelian gauge transformations of the target space:
\begin{equation}\label{gauge}
\begin{array}{c}
 \A_\mu^a\, t^a\rightarrow U^\dagger \A_\mu^a\, t^a U+iU^\dagger\partial_\mu U
\\[2mm]
 \varphi_i\rightarrow\left(U^\dagger\varphi\right)_i,\,\,\,\,
 \bar\varphi^i\rightarrow \left(\bar\varphi U\right)^i ,
\end{array} 
\end{equation}
where $U(x)\in\SU(N)$.
In addition, the expression~(\ref{eq_lagrNA}) is also invariant with respect to the following gauge transformations of auxiliary fields $B(t)$ and $\vp_i$:
 \begin{equation}\label{gaugeM}
B(t) \to \ B(t) + \frac {d \alpha(t)}{dt}, \quad\quad
\varphi_i(t) \to e^{-i\alpha(t)} \varphi_i (t)  .
 \end{equation}

It is not immediately clear how to extend the Abelian superfield
description to a general non-Abelian case, i.e. to the gauge group
${\rm SU}(N)\,$.
We succeeded in constructing such a description for the particular case
of $\SU(2)$ self-dual or anti-self-dual gauge fields expressed in the form
 \begin{equation}
\label{Hooft} {\cal A}^a_\mu \ =\ -\bar \eta^a_{\mu\nu}
\partial_\nu \ln h(x) \quad {\rm or}\quad
{\cal A}^a_\mu \ =\ -\eta^a_{\mu\nu} \partial_\nu \ln h(x)
 \end{equation}
respectively, with harmonic function $h(x)$,
\begin{equation}
 \partial_\mu^2 h(x) = {\rm a\ finite\ sum\ of \ delta\ functions}.
\end{equation}
This is the so called 't Hooft ansatz for a multi-instanton $\SU(2)$ solution \cite{tHooft} with the 't~Hooft symbols $\eta^a_{\mu\nu}$ defined previously in Eq.~(\ref{eq_sigma}).
If one takes the function $h(x)$ to be vanishing at $|x|\rightarrow \infty$, then this function can be presented as the following sum over instantons:
\begin{equation}
 h(x)=1+\sum\limits_I \frac{c_I}{\left(x^\mu-b^{\mu}_I\right)^2}.
\end{equation}
It involves particular instanton positions $b_I^\mu$ as well as the numbers $c_I$ associated with each instanton.

\subsection{Quantization of auxiliary fields in the $\SU(2)$ gauge group case}

Let us understand how the interaction Lagrangian~(\ref{eq_lagrNA}) gives rise to the matrix Hamiltonian~(\ref{eq_susyham}). This is achieved upon quantization of the auxiliary variables $\vp_\alpha$ and $\bar\vp^\beta$. We consider only the particular case of $\SU(2)$ gauge group when the indices $i$, $j$ for the auxiliary variables take only two values, 1 and 2, and are denoted as $\alpha$, $\beta$. See Section~\ref{c4.4.5} for a general discussion of $\SU(N)$ gauge group.

Observe that the variables $\varphi_\alpha$ enter the Lagrangian with only one time derivative. Thus, they are
not full-fledged dynamic variables (like $x_\mu$) and not auxiliary fields (like $B(t)$ field or $\omega_{1,2}$ fields in Eq.~(\ref{vp}), see below). They have a kind
of intermediate nature. In the context of ${\cal N}=4$ SQM models, such variables (together
with their analytic superfield carriers $v^+, \widetilde{v^+}$, see Eqs.~(\ref{vp}), (\ref{vptild}) below) were introduced in \cite{FIL0,FIL}.
See also \cite{BKS} for a recent application.
To understand  better the nature of the auxiliary fields, perform the quantization.

The canonical commutation relations following from the action (\ref{eq_lagrNA}) through the standard Dirac prescription  are
\begin{equation}
\label{phicomm}
 [ \varphi_\alpha, \bar \varphi^\beta] = \delta_\alpha^\beta \ , \ \ \ \ \ \ \ \ \ \ \ \
[\varphi_\alpha, \varphi_\beta] = [\bar \varphi^\alpha, \bar \varphi^\beta] = 0  .
 \end{equation}

The fact that $k$ must be integer leads to the finite representations of the operator algebra $\varphi_\alpha$, $\bar\varphi^\alpha$.
Indeed, consider the constraint
\begin{equation}
\bar\varphi^\alpha\varphi_\alpha = k\,, \lb{classConstr}
\end{equation}
which follows from \p{eq_lagrNA} by varying with respect to $B$. All real positive values of $k$ are classically allowed.
As we will shortly see, in the quantum theory, $k$ must be integer, but not necessary positive.

Consider the case of positive values of the integer $k$. In quantum theory, one can choose $\varphi_\alpha \equiv \partial/\partial \bar\varphi^\alpha$ and impose \p{classConstr} on the wave functions:
 \begin{equation}
 \label{svjazk}
 \bar \varphi^\alpha \varphi_\alpha \Psi =  \bar\varphi^\alpha \frac \partial {\partial \bar\varphi^\alpha} \Psi  = k\Psi\ .
 \end{equation}
In other words, the wave functions represent homogeneous polynomials of $\bar\varphi^\alpha$ of (an integer) degree $k\,$. In the case $k<0$
the algebra (\ref{phicomm}) is the same, but one must
choose $\bar\varphi^\alpha= - \partial/\partial\varphi_\alpha$ and consider polynomials of $\varphi_\alpha$ of degree $|k|$.
The number of such (linearly independent) polynomials is $|k|+1$.
 Moreover, it is also easy to see that the operators (\ref{eq31}) (which enter the interaction Lagrangian~(\ref{eq_lagrNA})) satisfy the following algebra:
 \begin{equation}
\label{algTa}
[T^a, T^b] \ =\ i\varepsilon^{abc} T^c .
 \end{equation}
In addition, assuming $k>0$ and taking into account (\ref{svjazk}), one derives
 \begin{equation}
\label{Kasimir}
T^a T^a \ =\ \frac 14 \left[ (\bar\varphi^\alpha \varphi_\alpha)^2 + 2 (\bar\varphi^\alpha \varphi_\alpha) \right] \
=\  \frac k2 \left(\frac k2 +1\right) .
 \end{equation}
In other words, $T^a$ can be treated as the generators of $\SU(2)$ in the representation
of spin $k/2$.

This way of quantizing semi-dynamical variables $\varphi_\alpha, \bar\varphi^\alpha$
was employed in Ref.~\cite{FIL}. Alternatively, one could interpret
$\varphi_\alpha, \bar\varphi^\alpha$ with the constraint \p{classConstr} as a kind of the target harmonic
variables representing a sphere ${\rm S}^2$, solve \p{classConstr} in terms of the stereographic projection coordinates
and quantize the system (see, for example, Ref.~\cite{cpnm}).

A nice feature is that this gauge $\SU(2)$ group is in fact the R-symmetry group of $\N=4$ supersymmetry algebra.

The crucial role of the constraint \p{svjazk} is to restrict the space of quantum states of the considered model
to the {\it finite} set of irreducible $\SU(2)$ multiplets of fixed spins (e.g., of the spin $k/2$ in the bosonic sector).
This is an essential difference of this approach from that employed, e.g.,  in \cite{Bala} (and later in \cite{BKS,KLS})
where no analog of the constraints
\p{classConstr} and \p{svjazk} was imposed, thus allowing for the space of states to involve an {\it infinite} number of $\SU(2)$ multiplets
of all spins. The quantization scheme which we follow here was earlier used in the SQM context in \cite{FIL0,FIL} and can be traced back to the work \cite{Polych}.

\subsection{Why the number $k$ must be integer}\label{ss4.3.2}

 Let us restrict ourselves by the first two terms in the interaction Lagrangian~(\ref{eq_lagrNA}).
The action
 \begin{equation}
 \label{SphiB}
 S \ =\ \int dt\, \Big[ i \bar \varphi^i (\dot{\varphi}_i + iB \varphi_i) \ + \ kB \Big]
 \end{equation}
much resembles the three-dimensional Chern-Simons action,
 \begin{equation}
\label{CS}
S_{\CS} = \kappa \int \left( A \wedge dA - \frac {2i}3 A\wedge A \wedge A \right) .
 \end{equation}
In both systems, the canonical Hamiltonian is zero, the canonical momenta are algebraically expressed through coordinates,
 and the quantization consists in imposing certain second class constraints (for a nice review of the classical and quantum aspects of
the Chern-Simons theory, see \cite{Dunne}).
Another well-known feature of CS theory is the quantization of the coupling, $k_{\CS} = 4\pi \kappa$ = integer. This follows from the requirement
for the Euclidean path integral to be invariant with respect to large (topologically nontrivial) gauge transformations. As was mentioned above, in
our case the coefficient $k$ is also quantized. This can be derived following a similar reasoning.

Consider the Euclidean version of the action~(\ref{SphiB}), where one changes the time $t$ to the Euclidean time $\tau$ by
\begin{equation}
 t=-i\tau 
\end{equation}
and regularize it in the infrared by putting it on a finite Euclidean interval $\tau \in (0,\beta)$ and imposing the periodic boundary conditions. This is of course equivalent to do calculations at finite temperature $T = 1/\beta$.

Notice first that the action (\ref{SphiB}) is invariant with respect to gauge transformations~(\ref{gaugeM}) which, in the Euclidean version of the theory, become
 \begin{equation}
\label{gaugeE}
B(\tau) \to \ B(\tau) + i \frac {d\alpha(\tau)}{d\tau} ,
\quad\quad
\varphi_i(\tau) \to e^{-i\alpha(\tau)} \varphi_i (\tau)  .
 \end{equation}

Let us discover topologically nontrivial gauge transformation in the Euclidean version
of this theory with the periodic boundary conditions
\begin{equation}
 B(\beta) = B(0), \quad \quad\varphi_i(\beta) = \varphi_i(0).
\end{equation}
The only admissible gauge transformations (\ref{gaugeE}) are those which do not break these periodicity conditions.
We see that the  transformation  with $\alpha(\tau) = 2\pi \tau/\beta$ is topologically nontrivial: it cannot be reduced to a chain
of infinitesimal transformations. This transformation shifts the Euclidean version of the last term in the action~(\ref{SphiB}) by an {\it imaginary} constant,
$\Delta S_{\rm FI} = -2\pi i k$. The requirement that the Euclidean path integrals (involving the factor $e^{-S_{\rm FI}}$) are not changed
leads \cite{Polych} to the quantization condition
 \begin{equation}
\label{kvantovanie}
k \ =\ {\rm integer}
 \end{equation}
Thus, a benign quantum theory can only be defined if this requirement is fulfilled.

\section{Harmonic superspace Lagrangian with non-Abelian gauge fields}\label{sect4.4}

\subsection{Superfield content}

To construct the action involving non-Abelian gauge fields, introduce, as earlier, a
 doublet of superfields $q^{+{\dot\alpha}}$ with charge +1
satisfying the constraints~(\ref{eq_anal}), (\ref{c3reality3}).
{\it On top of that}, we introduce an analytic gauge superfield $V^{++}$  of charge +2 satisfying the constraints
\begin{equation}
\label{constrV}
 D^+ V^{++}  = \bar D^+ V^{++} = 0 \ ,\blanc  V^{++} = \widetilde {V^{++}}
\end{equation}
and the ``matter'' superfield $v^+$ of charge +1.  The  constraints it satisfies,
  \begin{equation}\label{covarconstr}
  D^+ v^{+} =0,\blanc
  \bar D^+ v^{+} = 0,\blanc
  (D^{++} +i V^{++} ) v^{ +} =0 \, ,
\end{equation}
 differ from~(\ref{eq_anal}) by the presence of the covariant harmonic derivative ${\cal D}^{++} = D^{++} + iV^{++}$
\cite{HSS}.
The constraint ${\cal D}^{++} v^+ = 0$ is covariant with respect to gauge transformations
\begin{equation}
\label{gaugeharm}
V^{++} \ \to \ V^{++} + D^{++} \Lambda,\blanc
 v^+ \ \to\ e^{-i\Lambda} v^+ ,
\blanc
D^+\Lambda=\bar D^+\Lambda=0\ .
\end{equation}
We can use this gauge freedom to eliminate almost all components from $V^{++}$ and to present it as
\begin{equation}
\label{VPP}
V^{++} \ =\ 2i\, \theta^+ \bar \theta^+ B,
\end{equation}
where the gauge field $B(t)$ is real.
This is a one-dimensional counterpart of the familiar Wess-Zumino gauge in four-dimensional theories.
Observe also that Eq.~(\ref{gaugeM}) is a remnant of  gauge transformations (\ref{gaugeharm}), which survives
in the Wess-Zumino gauge (\ref{VPP}).

Then the superfield $v^+$ is expressed in the analytical basis as
\begin{equation}
\label{vp}
  v^+=\phi^\alpha u^+_\alpha
  -2\theta^+\omega_1-2\bar\theta^+\bar\omega_2
  -2i\theta^+\bar\theta^+  (\dot \phi^\alpha + i B\phi^\alpha) u^-_\alpha\ ,
\end{equation}
from which it follows that
\begin{equation}
\label{vptild}
 \widetilde  {v^+}= \bar \phi^\alpha u^+_\alpha
  -2\theta^+\omega_2  + 2\bar\theta^+ \bar\omega_1
  -2i\theta^+\bar\theta^+ ({\dot {\bar \phi}^\alpha} - iB \bar \phi^\alpha) u^-_\alpha\, \
\end{equation}
with $\bar \phi^\alpha = (\phi_\alpha)^*$.
Thus, the fields $\phi_\alpha$ and $\bar\phi^\alpha$ carry nonzero opposite $\U(1)$ charges associated with the auxiliary gauge field $B$.

\subsection{Superfield action}

The  $\ca N=4$ supersymmetry-invariant action consists of three parts, $S=S_{\rm kin}+S_{\rm int} + S_{\rm FI}$.
The kinetic part is more convenient to express in the central basis
$\{t,\, \theta_\alpha,\, \bar\theta^\beta\}$.
It has  the same form as in Eq.~(\ref{Lkin}) and its component expansion coincides with the first line in Eq.~(\ref{eq_action}), where the same change of variables~(\ref{eq_canon}) and~(\ref{eq_fg}) is performed as in the Abelian case.

The interaction part is taken as
\begin{equation}
\label{Sint}
 S_{\rm int}= -\frac 12 \int \, dt \, du\, d\bar \theta^+d\theta^+\,
  K\left(q^{+\dot{\alpha}}, u^\pm_\beta\right)  v^+ \widetilde{v^+}  ,
\end{equation}
where the condition $\widetilde K=K$ is imposed to ensure the action to be real.
Finally, we add the Fayet-Iliopoulos term
\begin{equation}
\label{FI}
S_{\rm FI} \ =\ -\frac {i k}2  \int \, dt \, du \, d\bar \theta^+ d\theta^+ \, V^{++}\
=\ k\int dt\,  B   ,
\end{equation}
which is invariant under gauge transformations (\ref{gaugeharm}).

Let us concentrate on the interaction part. It is convenient to introduce new variables
 \begin{equation}
\label{varphi}
\varphi_\alpha \ =\ \phi_\alpha \sqrt {h(x) }\ ,
 \end{equation}
 where
\begin{equation}\label{h}
h(x) = \int du\, K(x^{+\dot{\alpha}}, u^\pm_\beta )
\quad\quad\quad
 \big(x^{+\dot\alpha}=x^{\alpha\dot\alpha}u_\alpha^+\big),
\end{equation}
is a harmonic function~\footnote{We assume here that $h(x)>0$. The case $h(x)<0$ is treated similarly, if one redefines $h(x)\rightarrow -h(x).$}.   Indeed,
\begin{equation}
\partial_\mu^2 h(x) = 4\,\varepsilon^{\dot\alpha\dot\beta}\int du\,
\partial_{+\dot\alpha}\partial_{-\dot\beta}K(x^{+\dot\gamma}, u^\pm_\beta) = 0\,. 
\end{equation}

Substituting (\ref{c3eq_qdot4}), (\ref{vp}) and (\ref{vptild}) into (\ref{Sint}) and eliminating the auxiliary  fermionic degrees
of freedom $\omega_{1,2}$, $\bar \omega_{1,2}$ by their algebraic equations of motion, we derive after some algebra
\begin{equation}
\label{Lint2}
 L_{\rm int} \ =\
 i \bar \varphi^\alpha ({\dot \varphi}_\alpha + iB \varphi_\alpha)
 -  \frac 12 \bar \varphi^\beta \varphi_\gamma
\left({\cal A}_{\alpha{\dot\alpha}} \right)_{\!\beta}^{\,\,\gamma} {\dot x}^{\alpha {\dot \alpha}}
  - \frac i4 \left( {\cal F}_{{\dot\alpha}{\dot\beta}}\right)^{\,\,\gamma}_{\!\beta}
 \chi^{\dot\alpha}
\bar \chi^{\dot\beta} \bar\varphi^\beta \varphi_\gamma
\,. \end{equation}
Here
\begin{equation}
\label{Adef}
\left({\cal A}_{\alpha{\dot\alpha}} \right)_{\!\beta}^{\,\,\gamma}  = -\frac {2i}{\int du \, K} \int du \, \partial_{+\dot\alpha}
K \left( u^{+\gamma}\varepsilon_{\alpha\beta}
  - \frac 12
u^+_\alpha \delta^\gamma_\beta \right)
=\frac{i}{h}
\left(\varepsilon_{\alpha\beta} \, \partial^\gamma_{\ \dot\alpha} h - \frac 12
 \delta^\gamma_\beta \,\partial_{\alpha{\dot\alpha}}h  \right)
\end{equation}
($\partial_{\alpha\dot\alpha}\equiv \left(\sigma_\mu\right)_{\alpha\dot\alpha}\partial_\mu =-2\partial/\partial x^{\alpha\dot\alpha}$)
is a Hermitian traceless matrix -- the gauge field, and
\begin{equation}
\label{Fdef}
\left( {\cal F}_{{\dot\alpha}{\dot\beta}} \right)^{\,\,\gamma}_{\!\beta}
= \left(\F_{\mu\nu}\right)^{\,\,\gamma}_{\!\beta}
    \left(\sigma_\mu^\dagger\sigma_\nu\right)_{\dot\alpha\dot\beta}
  =\partial_{\delta\dot\alpha}
( {\cal A}^{\delta}_{\ {\dot\beta}} )^{\,\,\gamma}_{\!\beta}
-i ( {\cal A}_{\delta {\dot\alpha} } )^{\,\,\lambda}_{\!\beta}
  ( {\cal A}^{\delta}_{\ {\dot\beta} } )^{\,\,\gamma}_{\!\lambda} \   +
({\dot\alpha} \leftrightarrow {\dot\beta} )
 \end{equation}
is its self-dual part. It is easy to check explicitly, that the anti-self-dual part of the gauge field $\A_\mu$ vanishes,
\begin{equation}\label{Fanti}
 \left(\F_{\alpha\beta}\right)_{\!\gamma}^{\,\,\delta}
  =\left(\F_{\mu\nu}\right)_{\!\gamma}^{\,\,\delta} \left(\sigma_\mu\sigma_\nu^\dagger\right)_{\alpha\beta}
    =-\partial_{\alpha\dot\alpha}(\A_\beta^{\ \dot\alpha})^{\,\,\delta}_{\!\gamma}
    +i(\A_{\alpha\dot\alpha})^{\,\,\lambda}_{\!\gamma}
      (\A_\beta^{\ \dot\alpha})^{\,\,\delta}_{\!\lambda}
        +\left(\alpha\leftrightarrow\beta\right)
    =0.
\end{equation}
Thus, the field strength ${\cal F}_{\mu\nu}^a$ is self-dual and belongs
to the representation $(0,1)$ of ${\rm SO}(4)=\SU(2)\times\SU(2)$. Passing to ${\cal A}_\mu^a$ such that
$\left(\A_\mu\right)^{\,\,\gamma}_{\!\beta}=\A_\mu^a \left(\sigma_a\right)^{\,\,\gamma}_{\!\beta}\!\!/2$,
we find that the representation \p{Adef} precisely amounts to the self-dual 't~Hooft ansatz~(\ref{Hooft}), left equation.
The anti-self-dual expression from the right equation arises if one interchanges dotted and undotted indices,
i.e. effectively interchanges $\sigma_\mu$ and $\sigma^\dagger_\mu$. This also implies passing to the harmonics
$u^\pm_{\dot\alpha}$ and in fact to another ${\cal N}=4$ supersymmetry, with the second SU(2) (acting on dotted indices)
as the R-symmetry group.

Finally, substituting
$\bar\varphi^\beta\varphi_\gamma=T^a\left(\sigma_a\right)_{\!\gamma}^{\,\,\beta}$
and $\chi_{\dot\alpha}=f\psi_{\dot\alpha}$ into (\ref{Lint2}), where $T^a$ is defined in (\ref{eq31}) with $t^a=\frac 12\sigma_a$,
one convinces himself that the interaction term together with the FI term \p{FI} yields just (\ref{eq_lagrNA}) for the SU(2) gauge group case, and the quantum Hamiltonian derived from the Lagrangian $L_{\rm kin} + L_{\rm int} + L_{\rm FI}$ has the form (\ref{eq_susyham})
with ${\cal A}_\mu \equiv {\cal A}_\mu^a T^a$ and ${\cal F}_{\mu\nu} \equiv   {\cal F}_{\mu\nu}^a T^a $.

\subsection{Supersymmetry transformations}

The full Lagrangian in the non-Abelian case representing the sum of Eq.~(\ref{eq_lagrNA}) and the first line of Eq.~(\ref{eq_action}) is invariant, up to a total derivative, with respect
to ${\cal N}=4$ supersymmetry transformations (in the infinitesimal form) of Eqs.~(\ref{c4susytrans}) supplemented with transformations for the auxiliary fields:
\begin{equation}\label{susytran}
\begin{array}{c}
  \varphi_i\rightarrow \varphi_i
   +i f\left(t^a\varphi \right)_i \A_\mu^a
    \left(\epsilon\sigma_\mu\psi+\bar\epsilon\sigma_\mu\bar\psi\right),
\\[3mm]
  \bar\varphi^i\rightarrow \bar\varphi^i
   -i f\left(\bar\varphi t^a\right)^i \A_\mu^a
    \left(\epsilon\sigma_\mu\psi+\bar\epsilon\sigma_\mu\bar\psi\right).
\end{array}
\end{equation}
Note that the above formulas are written for the case of the gauge group $\SU(N)$ when the semi-dynamical fields $\vp_i$, $\bar\vp^j$ belong to a fundamental representation of $\SU(N)$.

\subsection{$\N=4$ supersymmetry with Yang monopole}

We have constructed the superfield action for the ${\cal N}=4$ supersymmetric quantum mechanics
corresponding to the Hamiltonian (\ref{eq_susyham}) with a non-Abelian $\SU(2)$ gauge field ${\cal A}_\mu$ which
lives on a conformally flat 4-manifold and is representable in the 't Hooft ansatz form (\ref{Hooft}).

As an example of such a field, let us quote the instanton solution on ${\rm S}^4$. Generically, it depends on the radius $R$ of the sphere
and the instanton size $\rho$. The configurations of maximal size, $\rho = R$, present a particular interest.
In the stereographic coordinates on ${\rm S}^4$,
 \begin{equation}
\label{metricS4}
 ds^2 \ =\ \frac {4R^4 dx_\mu^2}{(x^2 + R^2)^2}\ ,
 \end{equation}
they are expressed by the same formulas as the flat instantons in the singular gauge,
  \begin{equation}
   \label{inst}
  {\cal A}_\mu^a \ =\ \frac{2R^2 \bar\eta^a_{\mu\nu} x_\nu }{x^2(x^2+ R^2)} \quad \mbox{or} \quad
  ({\cal A}_{\alpha\dot\alpha})^{\,\,\gamma}_{\!\beta} = - \frac{2i \,R^2}{x^2(x^2 + R^2)}\left(\varepsilon_{\alpha\beta}x^\gamma_{\dot\alpha}
  - \frac{1}{2}\,\delta^\gamma_\beta \,x_{\alpha\dot\alpha} \right),
 \end{equation}
and
\begin{equation}
({\cal F}_{\dot\alpha\dot\beta})^{\,\,\gamma}_{\!\beta} = \frac{8i \,R^2}{x^2(x^2 + R^2)^2}\left(x^\gamma_{\dot\beta}x_{\beta\dot\alpha}
+ x^\gamma_{\dot\alpha}x_{\beta\dot\beta}\right). \lb{FstSing}
\end{equation}
The corresponding functions in Eq.~(\ref{h}) are taken in the form
\begin{equation}\label{Kh}
 K(x^{+\dot\alpha},u^\pm_\beta)
  =1+\frac{1}{\left(c^-_{\dot\alpha}x^{+\dot\alpha}\right)^2}\ ,
\blanc
 h(x)\equiv \int du\, K(x^{+\dot\alpha},u^\pm_\beta)=1+\frac{R^2}{ x_\mu^2},
\end{equation}
where $c^-_{\dot\alpha}=c_{\ \dot\alpha}^\alpha u^-_\alpha$, $c^{\alpha\dot\alpha}$ --
constant vector and $R^2=1/c^2_\mu$. The integral on the right hand side of Eq.~(\ref{Kh})
can be calculated as the power series in
$c^-_{\dot\alpha}c^{+\dot\alpha}=-c_\mu^2$ or directly after noting that the form of this integral is ${\rm SO}(4)$
invariant and putting $c_\mu=(c,0,0,0)$, $x_\mu=(x_1,x_2,0,0)$~\footnote{
Let us describe the latter possibility in more detail. If $c_\mu$ and $x_\mu$ are chosen as mentioned above, this gives
\begin{equation}
 c_\dalpha^- x^{+\dalpha} = -c\, x^0 + i c\, x^1 \left(u_1^+ u_1^- - u_2^+ u_2^-\right).
\end{equation}
To calculate the integral in Eq.~(\ref{Kh}), one realizes the harmonics $u^\pm_\alpha$ in the familiar stereographic parametrization \cite{HSS}:
\begin{equation}
\left(\begin{array}{ll}
 u_1^+ & u_1^-
\\
 u_2^+ & u_2^-
\end{array}\right)
=
\frac{1}{\sqrt{1+t \bar t}}
\left(
\begin{array}{ll}
 e^{i\psi} & -\bar t\, e^{-i\psi}
\\
 t\, e^{i\psi} & e^{-i\psi}
\end{array}
\right),
\end{equation}
where $t\in {\mathbb C}$, $0 \le \psi < 2\pi$. It is also necessary to define the measure of integration,
\begin{equation}
 \int du 
\quad\longrightarrow\quad
 \frac{i}{4\pi^2}\int_0^{2\pi} d\psi
  \int \frac{dt\, d\bar t}{(1+t\bar t)^2} .
\end{equation}
After this, the computation of the integral
\begin{equation}
 \frac {i}{2\pi}\int \frac{dt\,d\bar t}{(1+t\bar t)^2}
\left[1 + \frac{(1+t\bar t)^2}{\left(c\,x^0 (1+t\bar t)+ic\, x^1(t+\bar t)\right)^2}\right]
\end{equation}
gives~(\ref{Kh}).
}.

The field $\A_\mu^a$ can be brought to the nonsingular gauge
\begin{equation}\label{non-singular}
 \A_\mu^a=\frac{2\eta_{\mu\nu}^a x_\nu }{x^2+R^2}\,, \qquad {\cal F}^a_{\mu\nu}
\ =\ - \frac {4R^2\eta^a_{\mu\nu}}{(x^2 + R^2)^2}\,,
\end{equation}
by the gauge transformations (\ref{gauge}) with $U(x)=-i\sigma_\mu
x_\mu/\sqrt{x^2}$ (this particular $U(x)$ form is prompted by the form of the field
strength \p{FstSing}). The action density $\sim \F_{\mu\nu}
\F^{\mu\nu}$ is the same in this case at all points of ${\rm S}^4$. It is
worth noting that the singular gauge transformation converts the
undotted gauge group indices into the dotted ones: the self-dual
gauge potential and the field strength in the spinorial notation
become
\begin{equation}
({\cal A}_{\alpha\dot\alpha})^{\!\dot\gamma}_{\,\,\dot\beta} = \frac{2i}{x^2 + R^2}\left(\varepsilon_{\dot\alpha\dot\beta}x^{\dot\gamma}_{\alpha}
  - \frac{1}{2}\,\delta^{\dot\gamma}_{\dot\beta} \,x_{\alpha\dot\alpha} \right),\quad
({\cal F}_{\dot\alpha\dot\beta})^{\!\dot\gamma}_{\,\,\dot\delta} = -\frac{8i \,R^2}{(x^2 + R^2)^2}
\left(\varepsilon_{\dot\alpha\dot\delta}\delta^{\dot\gamma}_{\dot\beta} + \varepsilon_{\dot\beta\dot\delta}\delta^{\dot\gamma}_{\dot\alpha}\right)
\end{equation}
and, also, $\varphi_\alpha \rightarrow \varphi^{\dot\alpha} = -i\varphi_\alpha\,
x^{\alpha\dot\alpha}/\sqrt{x^2}$,
$\bar\varphi^\alpha \rightarrow \bar\varphi_{\dot\alpha} = -i\bar\varphi^\alpha\,
x_{\alpha\dot\alpha}/\sqrt{x^2} \,$.

Note that, the field (\ref{inst}), (\ref{non-singular}) describes the Yang monopole living
in ${\mathbb R}^5$ \cite{Yang}. The potential \p{non-singular} has a nice group-theoretical meaning as one of the two SU(2)
connections on the coset manifold $\SO(5)/[\SU(2)\times \SU(2)] \sim {\rm S}^4$ (see e.g. \cite{gp}). It coincides with the flat self-dual instanton
only in the conformally flat parametrization of ${\rm S}^4$ as in \p{metricS4}. When coupled to the world-line through
our semi-dynamical variables $\varphi_\alpha, \bar\varphi^\alpha$, the 5-dimensional Yang monopole is reduced to this SU(2)
connection defined on ${\rm S}^4$.

Let us elaborate on this point in more detail, choosing, without loss of generality, $R =1$ in the above formulas.
Consider the following  $d=1$ bosonic Lagrangian with
the ${\mathbb R}^5$ target space and an additional coupling to Yang monopole
\begin{equation}
L_{\,{\mathbb R}^5} = \frac 12\left(\dot{y}_5 \dot{y}_5 +  \dot{y}_\mu \dot{y}_\mu\right) + \B_\mu^{\, a} (y) T^a \,\dot y_\mu \,.\lb{Yang1}
\end{equation}
Here, $\B_\mu^{\,a} $ is the standard form of the Yang monopole in the ${\mathbb R}^5$ coordinates,
\begin{equation}
\B_\mu^{\,a} = \frac{\eta_{\mu\nu}^a y_\nu }{r(r + y_5)}\,, \quad r = \sqrt{y_5^2 + y^2_\mu}\,,
\end{equation}
$T^a$ are defined as in \p{eq31} with $t^a = {\textstyle \frac{1}{2}}\sigma^a\,$, and  the action for the semi-dynamical
variables $\varphi_\alpha, \bar\varphi^{\alpha}$ is omitted. Now one passes to the polar decomposition of ${\mathbb R}^5$ into a radius $r$ and the angular part ${\rm S}^4\,$, $(y_5, y_\mu) \;\rightarrow \; (r, \tilde{y}_5, \tilde{y}_\mu)\,, \; \tilde{y}_5 = \sqrt{1 - \tilde{y}^2_\mu}\,$,
and rewrites \p{Yang1} as
\begin{equation}
L_{\,{\mathbb R}^5} =
\frac 12\dot{r}{}^2 + \frac 12 r^2\left(\dot{\tilde{y}}_5 \dot{\tilde{y}}_5 + \dot{\tilde{y}}_\mu \dot{\tilde{y}}_\mu\right)
+ \frac{\eta_{\mu\nu}^a \tilde{y}_\nu \dot{\tilde{y}}_\mu\,T^a }{1 + \sqrt{1 - \tilde{y}^2_\mu}} \,.\lb{Yang2}
\end{equation}
The coordinates $\tilde{y}_\mu$ give a particular parametrization of ${\rm S}^4$. Passing to the stereographic coordinates is accomplished by
the redefinition
$$
\tilde{y}_\mu = 2\, \frac{x_\mu}{ 1 + x^2}\,,
$$
which casts \p{Yang2} into the form
\begin{equation}
L_{\,{\mathbb R}^5} =
\frac 12\left\{\dot{r}{}^2 + 4 r^2\frac{\dot{x}_\mu\dot{x}_\mu}{(1 + x^2)^2}
\right\}
+ \frac{2 \eta_{\mu\nu}^a x_\nu \dot{x}_\mu\,T^a }{1 + x^2} \,.\lb{Yang3}
\end{equation}
One sees that the ${\rm S}^4$ metric \p{metricS4} (with $R =1$) and the instanton
vector potential \p{non-singular} appear.

Thus, current approach, as a by-product, provides a solution
to the long-standing problem of constructing ${\cal N}=4$ supersymmetric quantum mechanics with Yang monopole (see e.g. \cite{gknty} and references therein).
Obviously, the component Lagrangian (\ref{eq_action}) (with the relevant function $f(x)$) is just
the ${\rm S}^4$ part of the Lagrangian \p{Yang3} with the ``frozen'' radial variable $r = 1$. Presumably, one can restore the
full 5-dimensional kinetic part in \p{Yang3} by adding a coupling to the appropriately constrained scalar ${\cal N}=4$
zero-charge superfield $X(t,\theta, \bar\theta)$ which describes an off-shell multiplet $({\bf 1, 4, 3})$ with
one physical bosonic field \cite{IKLev}, such that $X|_{\theta = \bar\theta = 0} = r\,$.

\subsection{Some remarks in the non-Abelian case}\label{c4.4.5}

The problem of finding a superfield formulation for a generic $\SU(N)$ self-dual field is more complicated and is still an open question.
However, by introducing extra variables $\varphi_i$, it is always possible to write a {\it component} Lagrangian
(\ref{eq_lagrNA}) (together with the first line in (\ref{eq_action})) corresponding
to the matrix Hamiltonian (\ref{eq_susyham}).

This observation has actually nothing to do with supersymmetry. It boils
down to the following. Consider the eigenvalue problem for a usual Hermitian matrix $H_{jk}$. It can be treated as a Schr\"odinger problem
$\hat H \Psi(\varphi_j) = \lambda \Psi(\varphi_j)$ with the constraint $\hat G \Psi = 0$, where
 \begin{equation}
\label{HGphi}
 \hat H \ =\ \varphi_j H_{jk} \frac {\partial}{\partial \varphi_k} \ , \ \ \ \ \ \ \ \ \ \ \hat G = \varphi_j
\frac {\partial}{\partial \varphi_j} - 1 .
 \end{equation}
The corresponding Lagrangian
 is
\begin{equation}
\label{Lphi}
L \ =\ i\bar \varphi_j \dot \varphi_j - B(\bar \varphi_j \varphi_j - 1) -  \bar\varphi_j H_{jk} \varphi_k ,
\end{equation}
where $\bar\varphi_j=(\varphi_j)^*$.
This easily generalizes to the case where $H$ is an operator depending on a set of canonically conjugated variables
$\{p_\mu, x_\mu \}$. The only difference is that $-H_{jk}$ is now replaced  by the matrix $L_{jk}$ obtained from $H_{jk}$ by the appropriate
Legendre transformation~\footnote{This elementary observation should be well known, for example, in matrix models.
 Surprisingly, it is not found it in such a ``chemically pure'' form in the
literature, but similar constructions were discussed, e.g., in Refs.~\cite{Polych,topkvant}.}.

The initial goal was to find a Lagrangian representation for the Hamiltonian (\ref{eq_susyham}) with matrix-valued $\A_\mu$, $\F_{\mu\nu}$.
The construction just described, with $\varphi_i$ in the fundamental representation of $\SU(N)$, leads
to the $N\times N$ matrix Hamiltonian. The Lagrangian (\ref{Lphi}) coincides in this case with the Lagrangian (\ref{eq_lagrNA})
with the choice $k=1\,$, to which the first line from Eq.~(\ref{eq_action}) is also added.

Obviously, one can describe the Hamiltonians in higher representations of $\SU(N)$ in a similar way, by choosing the number of components $\varphi_i$
equal to the dimension of the representation. We have seen, however, that in the $\SU(2)$ case one can be more economic, introducing
only a couple of dynamic variables $\varphi_\alpha$ and multiplying the term proportional to $B$ in the Lagrangian by an arbitrary
integer $k$. This leads to the Hamiltonian in the representation of spin $|k|/2$. Certain $\SU(N)$ representations
(namely, the symmetric products of $|k|$ fundamental or $|k|$ antifundamental representations) can also be attained in this way.

One can also construct in this way a ${\cal N} =2 $ supersymmetric Lagrangian for the Hamiltonian (\ref{eq_susyham})
with generic (not necessarily self-dual) ${\cal A}_\mu$. A similar construction (but with extra fermionic rather than bosonic variables)
was in fact discussed in Ref.~\cite{Gaume}.
A beauty of the harmonic superspace approach explored here is, however, that such extra variables and the constraint
(\ref{svjazk}) are not introduced by hand, but
arise naturally from the manifestly off-shell supersymmetric superfield actions.

\section{Three-dimensional SQM in non-Abelian monopole background}

The Hamiltonian~(\ref{eq_susyham}) presents the generalization of the Hamiltonian~(\ref{Hflatcherezpsi}) to the conformally flat metric case in {\em four dimensions}. We succeeded in the construction of this generalization using the superfield formalism. In this section, we employ similar construction and generalize the {\em three-dimensional} system described by with the Hamiltonian~(\ref{eq_3dham}) to the conformally flat case.
Although the resulting Hamiltonian, the supercharges and the component Lagrangian appear to be just the three-dimensional reduction of the four-dimensional counterpart, the superfield formalism in the three-dimensional case involves a different superfield for the space coordinates $x^i$ and is thus implemented differently.

\subsection{Superfield content and superfield action}

In this section, instead of the coordinate superfield $q^{+\dot\alpha}$ one deals with the analytic superfield $L^{++}$ which encompass the multiplet $(\bf{3,4,1})$ and is subjected to the constraints~(\ref{a}).
The superfield $V^{++}$ and the auxiliary superfields $v^+$ and $\widetilde{v^+}$ are defined by the same Eqs.~(\ref{constrV}), (\ref{covarconstr}). The superfield $V^{++}$ in the Wess-Zumino gauge (\ref{VPP}) is expressed through one independent component $B(t)$.
We remind that $B(t)$ is a real one-dimensional ``gauge field'' which transforms as $B \to  B + \dot\lambda\,$, with $\lambda(t)$ being the parameter of the residual gauge U(1) symmetry.

The explicit expressions for the superfields $q^{+\dot\alpha}$, $v^+$, $\widetilde{v^+}$ and $V^{++}$ are written in Eqs.~(\ref{c3eq_qdot4}), (\ref{vp}), (\ref{vptild}) and (\ref{VPP}) respectively.
The component expansion of the analytic superfield $L^{++}$ can be found in Eqs.~(\ref{c3aaa}), (\ref{c3e3.5.41}).
The multiplet $L^{++}$ involves the three-dimensional target space coordinates $\ell^{\alpha\beta} = \ell^{\beta\alpha}$, their fermionic partners $\chi^\alpha$, $\bar\chi^\alpha$ and a real auxiliary field $F$.
Let us remark that the three-dimensional case involves only one $\SU(2)$ (R-symmetry) group and thus no dotted indices present in the description.

Note also that the constraint $\bar\chi_\alpha = \left(\chi^\alpha\right)^*$ involves different position of spinor indices compared to Eq.~(\ref{c3xxx}) in the four-dimensional case (see Section~\ref{sec4.5.7} below). The transition from the spinor notation $\ell^{\alpha\beta}$ to the vector notation $\ell^i$,
\begin{equation}\label{e4.5.1}
 \ell_\alpha^\beta=\ell_i\left(\sigma_i\right)_{\!\alpha}^{\,\,\beta},\blanc
  \ell_i=\frac 12 \ell^\alpha_\beta\left(\sigma_i\right)_{\!\alpha}^{\,\,\beta},
  \blanc
 i=1,2,3
\end{equation}
($\sigma_i$ are Pauli matrices and, as usual, spinor indices are raised and lowered with antisymmetric Levi-Civita tensors $\ve_{\alpha\beta}$ and $\ve^{\alpha\beta}$), for the three-dimensional coordinates is considered in Section~\ref{sec4.5.3}.
The condition~(\ref{c3e3.5.41}) ensures that the coordinates $\ell_i$ are real.

The full Lagrangian ${\cal L}$ entering the  $\N=4$ invariant  off-shell action $S= \int dt {\cal L}$ consists of the three pieces~\footnote{The first
superfield formulation of general $({\bf 3, 4, 1})$ SQM without background gauge field couplings was given in \cite{IvSmi}.}
\begin{multline}
{\cal L} = {\cal L}_{\rm kin} + {\cal L}_{\rm int} + {\cal L}_{\rm FI}
  = \int du\, d^4\theta\, R_{\rm kin}(L^{++}, L^{+-}, L^{--}, u)
\\[3mm]
- \frac{1}{2} \int du\, d\bar\theta^+ d\theta^+\, K(L^{++}, u) v^+\widetilde{v^+}
-\frac {i k}2  \int  \, du \, d\bar \theta^+ d\theta^+ \, V^{++} , \lb{ACT}
\end{multline}
where $L^{+-}=\frac 12D^{--}L^{++}$ and $L^{--}=D^{--}L^{+-}$. The superfield functions $R_{\rm kin}$ and $K$
bear an arbitrary dependence on their arguments~\footnote{
The superfield Lagrangian~(\ref{ACT}) is written in the non-Abelian case. In the Abelian case, the superfield Lagrangian is simpler as it does not involve the auxiliary superfields $V^{++}$, $v^+$, $\widetilde{v^+}$:
\begin{equation}
{\cal L} = {\cal L}_{\rm kin} + {\cal L}_{\rm int}
  = \int du\, d^4\theta\, R_{\rm kin}(L^{++}, L^{+-}, L^{--}, u)
+\int du\, d\bar\theta^+ d\theta^+\, K^{++}(L^{++}, u).
\end{equation}
Here the interaction term is defined by a function $K^{++}(L^{++},\,u)$ of charge +2.
One can check that although the corresponding component Lagrangian as well as the expression for the Abelian gauge field differ from that in the non-Abelian case, the quantum Hamiltonian and the supercharges in the Abelian case have exactly the same form as in the non-Abelian case.
}.
The meaning of three terms in \p{ACT} is explained below.

\subsection{From harmonic superspace to components}

The first, sigma-model-type term in Eq.~\p{ACT}, after integrating over Grassmann and harmonic variables, yields the generalized
kinetic terms for $\ell^{\alpha\beta}, \chi^\alpha, \bar\chi_\alpha$:
\begin{multline}\label{kin_term}
{\cal L}_{\rm kin}
  =\frac 18f^{-2}\left(-2\dot \ell_{\alpha\beta}\dot\ell^{\alpha\beta}+F^2\right)
  +\frac i2 f^{-2}\left(\bar\chi_\alpha\dot\chi^\alpha-{\dot{\bar\chi}}_\alpha\chi^\alpha\right)
  +\frac 1{4}\left(\partial_{\alpha\beta}\partial^{\alpha\beta}f^{-2}\right)\chi^4
\\[2mm]
  +\frac{i}{f^3}\dot \ell^{\alpha\beta}\big\{
    \partial_{\alpha\gamma}f\chi_\beta\bar\chi^\gamma+\partial_{\beta\gamma}f\chi^\gamma\bar\chi_\alpha
  \big\}
  -\frac{1}{f^3}F\chi^\alpha\bar\chi^\beta\partial_{\alpha\beta}f,
\end{multline}
where $\chi^4=\chi^\alpha\chi_\alpha\, \bar\chi^\beta\bar\chi_\beta$,
$\partial_{\alpha\beta} \equiv \frac{\partial}{\partial \ell^{\alpha\beta}}$
and $f(\ell)$ is a conformal factor.
The calculations are most easily performed in the central basis, where
$L^{++}=u^+_\alpha u^+_\beta L^{\alpha\beta}\left(t,\theta_\gamma,\bar\theta^\delta\right)$. Then
\begin{equation*}
f^{-2}(\ell)=-\partial_{\alpha\beta}\partial^{\alpha\beta}
  \int R_{\rm kin}\left(
    \ell^{\alpha\beta}u^+_\alpha u^+_\beta,
    \ell^{\alpha\beta}u^+_\alpha u^-_\beta,
    \ell^{\alpha\beta}u^-_\alpha u^-_\beta
  \right)\,du.
\end{equation*}

The fermionic kinetic term can be brought to the canonical form by the change of variables
\begin{equation}\label{changetopsi}
 \chi^\alpha=f \psi^\alpha,\blanc
 \bar\chi_\alpha=f\bar\psi_\alpha.
\end{equation}
It is worth pointing out that the R-symmetry $\SU(2)$ group amounts to the rotational ${\rm SO(3)}$ group in the $\mathbb{R}^3$ target space
parametrized by $\ell^{i}$ from Eq.~(\ref{e4.5.1}). The conformal factor $f(\ell)$ can bear an arbitrary dependence on $\ell^{\alpha\beta}$,
so this ${\rm SO(3)}$ can be totally broken in the Lagrangian \p{kin_term}.

The second piece in Eq.~(\ref{ACT}) describes the coupling to an external non-Abelian gauge field.
Performing the integration over $\theta^+$, $\bar\theta^+$ and $u^\pm_\alpha$, eliminating the auxiliary fermionic fields $\omega_{1,2}$
and, finally, rescaling the bosonic doublet variables as $\varphi_\alpha \ =\ \phi_\alpha \sqrt {h(\ell) }$, where
\begin{equation}
 h(\ell)=\int du\, K\left(\ell^{\alpha\beta}u^+_\alpha u^+_\beta,u^\pm_\gamma\right), \lb{hdef}
\end{equation}
after some algebra one obtains
\begin{equation}\label{spinor_lagr}
{\cal L}_{\rm int}=i\bar\vp^\alpha\left(\dot\vp_\alpha+iB\vp_\alpha\right)
  +\bar\vp^\gamma\vp^\delta \frac 12
    \left(\A_{\alpha\beta}\right)_{\gamma\delta} \dot \ell^{\alpha\beta}
  -\frac 12 F\,\bar\vp^\gamma\vp^\delta \,U_{\gamma\delta}
    + \chi^\alpha\bar\chi^\beta\bar\vp^\gamma\vp^\delta
      \nabla_{\alpha\beta}U_{\gamma\delta}.
\end{equation}
Here the non-Abelian background gauge field and the scalar (matrix) potential are fully specified by the function $h$:
\begin{equation}
\left(\A_{\alpha\beta}\right)_{\gamma\delta}
  =\frac i{2h} \Big\{
    \ve_{\gamma\beta}\partial_{\alpha\delta}h
    +\ve_{\gamma\alpha}\partial_{\beta\delta}h
    +\ve_{\delta\beta}\partial_{\alpha\gamma}h
    +\ve_{\delta\alpha}\partial_{\beta\gamma}h
  \Big\},\blanc
 U_{\gamma\delta}=\frac{1}{h}\partial_{\gamma\delta}h\,. \lb{GpotMpot}
\end{equation}
By its definition, the function $h$ obeys the 3-dimensional Laplace equation,
\begin{equation}
\partial^{\alpha\beta}\partial_{\alpha\beta}\,h = 0\,. \lb{Lapl}
\end{equation}

Using the explicit expressions \p{GpotMpot}, it is straightforward to check the relation
\begin{equation}\label{samo}
 \left(\F_{\alpha\beta}\right)_{\gamma\delta}
   =2i\nabla_{\alpha\beta}U_{\gamma\delta},
\end{equation}
where
\begin{equation}\label{fij}
 \left(\F_{\alpha\beta}\right)_{\gamma\delta}
 =-2\partial_{\alpha}^{\;\lambda}\left(\A_{\lambda\beta}\right)_{\gamma\delta}
  +i\left(\A_{\alpha}^{\,\,\,\lambda}\right)_{\!\gamma\sigma}
    \left(\A_{\lambda\beta}\right)^{\,\,\sigma}_{\delta}
  +\left(\alpha\leftrightarrow \beta\right),
\end{equation}
\begin{equation}\label{nabla}
 \nabla_{\alpha\beta}U_{\gamma\delta}=-2\partial_{\alpha\beta}U_{\gamma\delta}
  +i\left(\A_{\alpha\beta}\right)_{\gamma\lambda}U^{\,\,\lambda}_{\!\delta}
  +i\left(\A_{\alpha\beta}\right)_{\delta\lambda}U^{\,\,\lambda}_{\!\gamma},
\end{equation}
and $\left(\F_{\alpha\beta}\right)_{\gamma\delta}$ is related to the standard gauge field strength in the vector notation, see below.
As we shall see soon, the condition \p{samo} is none other than the static
form of the general self-duality condition for the $\SU(2)$ Yang-Mills field on $\mathbb{R}^4\,$ (see Eq.~(\ref{samoV})), i.e.
the Bogomolny equations for BPS monopoles \cite{Bog},
while \p{GpotMpot} provides a particular solution to these equations, being a static form
of the 't~Hooft ansatz \cite{tHooft}.

Note that the relation \p{samo} is covariant and the Lagrangian \p{spinor_lagr} is form-invariant under the
``target space'' $\SU(2)$ gauge transformations written in Eq.~(\ref{gauge}).
This is not a genuine symmetry; rather, it is a reparametrization of the Lagrangian which allows one to cast
the background potentials \p{GpotMpot}  in some different equivalent forms.
It is worth noting that the gauge group indices coincide with those of the R-symmetry group, like in the
four-dimensional case. Nevertheless, the ``gauge'' reparametrizations \p{gauge} do not affect
the doublet indices of the target space coordinates $\ell^{\alpha\beta}$ and their superpartners
present in the superfield $L^{++}$. They act only on the semi-dynamical spin variables
$\varphi_\alpha, \bar\varphi^\alpha$ and gauge and scalar potentials \p{GpotMpot}.

Finally, the last piece in Eq.~(\ref{ACT}) yields the Fayet-Iliopoulos term,
\begin{equation}\label{FI2}
 {\cal L}_{\rm FI}=k B\,.
\end{equation}
In the quantum case, the coefficient $k$ is quantized, $k \in \mathbb{Z}\,$, on the same grounds as in the 4-dimensional case, see Section~\ref{ss4.3.2}.

\subsection{Vector notations in three dimensions}\label{sec4.5.3}

It is instructive to rewrite the above relations and expressions, including the full Lagrangian (\ref{ACT}),
in the vector notations. To this end, let us pass from $\ell^{\alpha\beta}$ to $\ell_i$ as in Eq.~(\ref{e4.5.1}) and associate the vector $\A_i$ to the gauge field $\A_{\alpha\beta}$ (with additional matrix indices which are omitted here for simplicity) by the rule
\begin{equation}
 \A_\alpha^\beta=\A_i\left(\sigma_i\right)_{\!\alpha}^{\,\,\beta},\blanc
  \A_i=\frac 12 \A^\alpha_\beta\left(\sigma_i\right)_{\!\alpha}^{\,\,\beta},
  \blanc
 i=1,2,3. \lb{sp_vec}
\end{equation}
One can check that the coordinates $\ell_i$ are real while the matrix $\left(\A_i\right)_{\!\gamma}^{\,\,\delta}$ is Hermitian. Note also the relation between the partial derivatives $\partial_{\alpha\beta} = \partial/\partial \ell^{\alpha\beta}$ and
$\partial_i =  \partial/\partial \ell_i$:
\begin{equation}
\partial_{\alpha\beta}= -\frac{1}{2}\left(\sigma_i\right)_{\!\alpha\beta}\partial_i,\blanc
  \partial_i= -\left(\sigma_i\right)_{\!\alpha}^{\,\,\beta}\partial^\alpha_\beta\,.
\end{equation}
Let us also make a similar conversion of the gauge group indices,
\begin{equation}
 M_{\!\gamma}^{\,\,\delta}=\frac{1}{2} M^a \left(\sigma_a\right)_{\!\gamma}^{\,\,\delta},
  \blanc
 M^a=M_{\!\delta}^{\,\,\gamma}\left(\sigma_a\right)_{\!\gamma}^{\,\,\delta},
  \blanc
 a=1,2,3\,,
\end{equation}
for any Hermitian traceless $2\times 2$ matrix $M$.

In the new notations, the total Lagrangian \p{ACT} takes the following form:
\begin{multline}\label{vectLagr}
{\cal L}=
  \frac 12 f^{-2}\dot \ell_i^2
  +\A_i^a T^a\dot \ell_i
  +i\bar\vp^\alpha\left(\dot\vp_\alpha+iB\vp_\alpha\right)+kB
  +i\bar\psi_\alpha\dot\psi^\alpha
  +f^2\nabla_i U^a T^a\,\psi\sigma_i\bar\psi
\\[2mm]
  +\frac{1}{4}\left\{f\partial_i^2 f-3\left(\partial_i f\right)^2\right\}\psi^4
  +2 f^{-1}\varepsilon_{ijk}\partial_i f\,\dot\ell_j\, \psi\sigma_k\bar\psi
\\[2mm]
 +\frac 18f^{-2}F^2
  +\frac 12 F\left( U^a T^a
- f^{-1}\partial_i f\,\psi\sigma_i\bar\psi\right).
\end{multline}
where $T^a$ defined in Eq.~(\ref{eq31}).
Here
\begin{equation}
\nabla_i U^a=\partial_i U^a +\varepsilon^{abc}\A_i^b U^c
\end{equation}
and the Bogomolny equations \p{samo} relating $\A_i^a$ and $U^a$ are equivalently
rewritten in the more familiar form,
\begin{equation}
{\cal F}_{ij}^a = \varepsilon_{ijk}\nabla_k U^a\,, \lb{samoV}
\end{equation}
where ${\cal F}_{ij}^a = \partial_i\A_j^a - \partial_j\A_i^a
+\varepsilon^{abc}\A_i^b\A_j^c$.
Finally, the gauge field and the matrix potential defined in \p{GpotMpot} are rewritten as
\begin{equation}\label{3dsamo}
 \A_i^a
  =-\varepsilon_{ija} \partial_j \ln h,\blanc
  U^a=-\partial_a \ln h\,, \qquad \Delta\,h = 0\,.
\end{equation}

After eliminating the auxiliary field $F$ by its equation of motion,
\begin{equation}
F = 2 f^2\left( f^{-1}\partial_i f \,\psi\sigma_i \bar\psi - U^aT^a\right)\,, \lb{EqF}
\end{equation}
the Lagrangian \p{vectLagr} takes the form
\begin{multline}
{\cal L} =
  \frac 12 f^{-2}\dot \ell_i^2
  +\A_i^a T^a\dot \ell_i
  +i\bar\vp^\alpha\left(\dot\vp_\alpha+iB\vp_\alpha\right)+kB
  +i\bar\psi_\alpha\dot\psi^\alpha
  +f^2\psi\sigma_i\bar\psi\left(\nabla_i +f^{-1}\partial_i f\right) U^aT^a
\\[3mm]
+\,\frac{1}{4}\left\{f\partial_i^2 f- 4\left(\partial_i f\right)^2\right\}\psi^4
  +2 f^{-1}\varepsilon_{ijk}\partial_i f\,\dot{\ell}_j\, \psi\sigma_k\bar\psi
 -\frac 12 f^2 ( U^a T^a)^2\,.\label{vectLagr1}
\end{multline}
We see that this Lagrangian involves three physical bosonic fields $\ell_i$ and
four physical fermionic fields $\psi_\alpha\,$. It is fully specified by two independent functions: the metric conformal factor $f(\ell)$ which
can bear an arbitrary dependence on $\ell_i$ and the function $h(\ell)$ which satisfies the 3-dimensional Laplace equation
and determines the background non-Abelian gauge and scalar potentials. The representation \p{hdef} for $h$ in terms
of the analytic function $K(\ell^{++}, u)$ yields in fact a general solution of the 3-dimensional Laplace equation \cite{WhitWat}.
If one takes the function $h(\ell)$ to be vanishing at $|\vec \ell|\rightarrow \infty$, then this function can be presented as the following sum over monopoles:
\begin{equation}
 h(\ell)=1+\sum\limits_M \frac{c_M}{\left|\vec \ell-\vec b_M\right|}.
\end{equation}
It involves particular monopole positions $\vec b_M$ as well as the numbers $c_M$ associated with each monopole.

The Lagrangian \p{vectLagr1} also contains the ``semi-dynamical'' spin variables $\varphi_\alpha, \bar\varphi^\alpha\,$,
the role of which is the same as in the four-dimensional case: after quantization they ensure that $T^a$ defined in~(\ref{eq31})
become matrix SU(2) generators corresponding to the spin $|k|/2$ representation.

\subsection{Supertransformations of component fields}

The component action corresponding to the Lagrangian \p{vectLagr} is partly on shell
since we have already eliminated the fermionic fields of the auxiliary $v^+$
multiplet by their algebraic equations of motion. The fields of the coordinate multiplet $L^{++}$
are still off shell.
The ${\cal N}=4$ transformations leaving invariant the action $S = \int dt \,{\cal L}$ look most transparent being expressed in terms of the
component fields $\ell_i, F, \chi^\alpha$, $\bar\chi^\alpha$, $\phi^\beta$, $\bar\phi^\beta$:
\begin{equation}
\begin{array}{l}
\ell_i\rightarrow \ell_i  +i\epsilon\sigma_i\chi + i\bar\epsilon\sigma_i\bar\chi,
\\[2mm]
F\rightarrow F - 2\epsilon^\alpha\dot{\chi}_\alpha - 2\bar\epsilon^\alpha\dot{\bar\chi}_\alpha,
\\[2mm]
\chi^\alpha\rightarrow \chi^\alpha - \frac 12 iF\bar\epsilon^\alpha - (\bar\epsilon\sigma_i)^\alpha\,\dot{\ell}_i,
\\[2mm]
\bar\chi^\alpha\rightarrow \bar\chi^\alpha  +\frac 12 iF\epsilon^\alpha + (\epsilon\sigma_i)^\alpha\,\dot{\ell}_i,
\\[2mm]
\phi^\alpha\rightarrow \phi^\alpha - {i}\big(\epsilon^\alpha \chi\sigma_i\phi
+ \bar\epsilon^\alpha \bar\chi\sigma_i\phi\big)\partial_i\ln h,
\\[2mm]
\bar\phi^\alpha\rightarrow \bar\phi^\alpha - {i}\left(\epsilon^\alpha \chi\sigma_i\bar\phi
+ \bar\epsilon^\alpha \bar\chi\sigma_i\bar\phi\right)\partial_i\ln h.
\end{array}
\lb{Transf}
\end{equation}
These transformations can be deduced from the analytic subspace realization of ${\cal N}=4$ supersymmetry~(\ref{c3e11}),
with taking into account the compensating $\U(1)$ gauge transformations of the superfields $v^+, \widetilde{v}^+$ and $V^{++}$
needed to preserve the WZ gauge \p{VPP}. Note that $\delta B = 0$ under ${\cal N}=4$ supersymmetry. This transformation law matches with the ${\cal N}=4$ superalgebra in WZ gauge, taking into account that the translation of $B$ looks
as a particular U(1) gauge transformation of the latter.

The Lagrangian~(\ref{vectLagr1}) is invariant, modulo a total time derivative,  under the transformations \p{Transf} in which $F$ is expressed from \p{EqF}.

\subsection{Hamiltonian and supercharges}
The Lagrangian \p{vectLagr1} is the point of departure for setting up the Hamiltonian formulation of the model under consideration and
quantizing the latter.
After substitution
of $\SU(2)$ spin-$k/2$ generators instead of $T^a$, the quantum Hamiltonian of this system takes the form
\begin{multline}\label{decrom}
  H=\frac{1}{2}f \left(\hat p_i-\ca A_i\right)^2 f
+\frac 12 f^2 U^2
-f^2\nabla_i U\psi\sigma_i\bar\psi
\\[2mm]
+\Big\{
\ve_{ijk}f\partial_i f\left(\hat p_j-\A_j\right)
-f\partial_k f U
\Big\}\psi\sigma_k\bar\psi
    + f\partial^2 f\left\{\psi^{\gamma}\bar\psi_{\gamma}-\frac{1}{2}\left(\psi^{\gamma} \bar\psi_{\gamma}\right)^2\right\},
\end{multline}
which is just a static 3-dimensional reduction of the 4-dimensional Hamiltonian given by Eq.~(\ref{eq_susyham}).
In this expression, the gauge field $\A_i = \A_i^aT^a$ and the scalar potential $U = U^aT^a$ are $\SU(2)$ matrices subjected
to the constraint (\ref{samoV}). It is also easy to find the supercharges $Q_\alpha, \bar Q^\beta$:
\begin{equation}\label{QQ}
\begin{array}{l}
  Q_\alpha = f \left(\sigma_i \bar\psi\right)_\alpha \left(\hat p_i-\ca A_i\right)
-\psi^{\gamma} \bar\psi_{\gamma} \left(\sigma_i\bar\psi\right)_\alpha i\partial_i f
-ifU\bar\psi_\alpha,
\\[3mm]
  \bar Q^\alpha = \left(\psi\sigma_i\right)^\alpha \left(\hat p_i-\ca A_i\right)f
+i\partial_i f \left(\psi\sigma_i\right)^\alpha \psi^{\gamma}\bar\psi_{\gamma}
+ifU\psi^\alpha,
\end{array}
\end{equation}
The ordering ambiguity arising in the case of the general conformal factor $f(\ell)$ can be fixed, as usual, by Weyl ordering procedure~\cite{SMI}, see Section~\ref{sec4.2.3} for details.

Let us emphasize that the only condition required from the background matrix fields ${\cal A}_i$ and $U$ for the generators
$Q_\alpha$ and $\bar Q^\beta$ to form ${\cal N}=4$ superalgebra \p{c3e5x} is that these fields satisfy the Bogomolny equations \p{samoV}.
Thus the expressions \p{decrom} and \p{QQ} define the ${\cal N}=4$ SQM model in the field of {\it arbitrary} BPS monopole,
not necessarily restricted to the ansatz \p{3dsamo}. Also, one can extend the gauge group $\SU(2)$ to $\SU(N)$ in \p{decrom} and \p{QQ}.

Let us remark that the three-dimensional Hamiltonian~(\ref{decrom}) and the supercharges~(\ref{QQ}) were considered for the first time in Ref.~\cite{Smilga:1986rb} (in the Abelian case).

\subsection{$\N=4$ supersymmetry with Wu-Yang monopole}

Finally, as a simple example of the monopole background consistent with the off- and on-shell ${\cal N}=4$ supersymmetry,
let us consider a particular 3-dimensional spherically symmetric case. It corresponds
to the most general $\SO(3)$ invariant solution of the Laplace equation for the function $h$,
\begin{equation}
h_{{\rm so}(3)}(\ell) = c_0 + c_1\,\frac{1}{\sqrt{\ell^2}}\,. \lb{so3}
\end{equation}
The corresponding potentials calculated according to Eqs.~\p{3dsamo} read
\begin{equation}
{\cal A}^a_i= \varepsilon_{ija}\frac{\ell_j}{\ell^2} \,\frac{c_1}{c_1 + c_0\sqrt{\ell^2}}, \quad\quad U^a =
\frac{\ell_a}{\ell^2} \,\frac{c_1}{c_1 + c_0\sqrt{\ell^2}}. \lb{WYa}
\end{equation}
This configuration becomes the Wu-Yang monopole \cite{WYa} for the choice $c_0 = 0\,$.
It is easy to find the analytic function $K(\ell^{++},u)$
which generates the solution \p{so3} (see \cite{IvLecht}):
\begin{multline}
  h_{{\rm so}(3)}(\ell) = \int du\, K_{{\rm so}(3)}(\ell^{++}, u)\,, \quad K_{so(3)}(\ell^{++}, u)
= c_0 + c_1 \left(1 + a^{--}\hat{\ell}^{++}\right)^{-\frac{3}{2}}\,, \lb{Kso3}
\\[3mm]
 \ell^{++} \equiv \hat{\ell}^{++} + a^{++}\,, \quad
a^{\pm\pm} = a^{\alpha\beta}u^\pm_\alpha u^\pm_\beta\,, \quad a^{\alpha}_\beta a^\beta_\alpha = 2\,.
\end{multline}

One could equally choose as $h(\ell)$, e.g., the well-known multi-center solution to the Laplace equation, with the broken $\SO(3)$.
Note that the ${\cal N}=4$ mechanics with
coupling to Wu-Yang monopole was recently constructed in \cite{BKS}, proceeding from a different approach, with
the built-in $\SO(3)$ invariance and the treatment of spin variables in the spirit of Ref.~\cite{Bala}.
Our general consideration shows, in particular, that the demand of $\SO(3)$ symmetry
is not necessary for the existence of ${\cal N}=4$ SQM models with non-Abelian
monopole backgrounds.

\subsection{Relation to four-dimensional ${\cal N}=4$ SQM system}\label{sec4.5.7}

It is instructive to show that \p{3dsamo} can indeed be viewed as a 3-dimensional reduction of the 't~Hooft ansatz for the solution
of general self-duality equation in $\mathbb{R}^4$ for the gauge group $\SU(2)$, with the identification $U^a =  \A_0^a\,$,
while the condition \p{samoV} is the 3-dimensional reduction of this equation.

To establish this relationship,
we use the following rules of correspondence between the $\SO(4)= \SU(2)\times \SU(2)$ spinor formalism
and its $\SU(2)$ reduction:
\begin{equation} \lb{dict}
\begin{array}{l}
\left(\sigma_\mu\right)_{\alpha\dot\beta}\rightarrow
  \left\{i\delta_\alpha^{\beta},
   \left(\sigma_i\right)_{\!\alpha}^{\,\,\beta}\right\},
\\[2mm]
 \varepsilon^{\dot\alpha\dot\beta} \rightarrow - \varepsilon_{\alpha\beta},
\quad\quad
   \varepsilon_{\dot\alpha\dot\beta} \rightarrow - \varepsilon^{\alpha\beta},
\\[2mm]
  x_{\alpha\dot\beta}\rightarrow \ell_\alpha^{\beta},
\quad\quad\quad
x^{\alpha\dot\beta}\rightarrow -\ell^\alpha_{\beta},
\\[2mm]
  \psi_{\dot\alpha}\rightarrow\psi^{\alpha}.
\end{array}
\end{equation}
This reflects the fact that the R-symmetry $\SU(2)$ in the $({\bf 3, 4, 1})$ models can be treated as a diagonal subgroup in the
symmetry group $\SO(4)=  \SU(2)\times \SU(2)$ of the $({\bf 4, 4, 0})$ models, with the $\SU(2)$ factors acting, respectively,
on the undotted and dotted indices.

The self-dual $\mathbb{R}^4$ SU(2) gauge field in the 't Hooft ansatz is written in the spinor notation in Eq.~(\ref{Adef}).
Then, using the rules \p{dict}, one performs the reduction $\mathbb{R}^4 \rightarrow \mathbb{R}^3$ as
\begin{equation}\label{Ared}
\begin{array}{l}
({\cal A}_{\alpha\dot\beta})_{\!\gamma}^{\,\,\delta}
\; \rightarrow \; iU_{\!\gamma}^{\,\,\delta} \delta_\alpha^\beta + ({\cal A}_{\alpha}^{\beta})_{\!\gamma}^{\,\,\delta},\quad\quad
({\cal A}_{\alpha}^{\alpha})_{\!\gamma}^{\,\,\delta} = 0,
\\[3mm]
h(x)\; \rightarrow \;h(\ell), \quad\quad
\partial_\beta^\alpha\partial^\beta_\alpha\,h = 0.
\end{array}
\end{equation}

Upon this reduction, the four-dimensional ansatz \p{Adef} yields precisely \p{GpotMpot}, while the general self-duality condition~(\ref{Fanti})
goes over into the Bogomolny equations \p{samo}. Of course, the same reduction can be performed in the vector notation,
with
$\F_{\mu\nu}\rightarrow\left\{\F_{ij}, \F_{0k}=\nabla_k U\right\}$,
and Eqs.~\p{samoV}, \p{3dsamo} as an output.

Thus, the general gauge field background prescribed by the off-shell ${\cal N}=4$
supersymmetry in this  $({\bf 3,4,1})$ system is a static form of the 't Hooft ansatz
for the self-dual $\SU(2)$ gauge field in $\mathbb{R}^4\,$.
This suggests that the above bosonic target space reduction has its superfield counterpart relating the four-dimensional system described in Section~\ref{sect4.4}
to the one considered here.

Indeed, the superfield $({\bf 3,4,1})$ action \p{ACT} can be obtained from
the $({\bf 4,4,0})$ multiplet action composed from Eqs.~(\ref{Lkin}), (\ref{Sint}), (\ref{FI}) via the ``automorphic duality'' \cite{GR} by considering a restricted
class of the $({\bf 4,4,0})$ actions
with $\U(1)$ isometry and performing a superfield gauging of this isometry by an extra gauge superfield $V^{++}{}'$ along the general
line of Ref.~\cite{DI}. Actually, the bosonic target space reduction we have just described corresponds to the shift isometry
of the analytic superfield $q^{+\dot\alpha}$ accommodating the $({\bf 4,4,0})$ multiplet, namely,
to $q^{+\dot\alpha} \rightarrow q^{+\dot\alpha} + \omega u^{+ \dot\alpha}\,$. It is the invariant projection
$q^{+\dot\alpha}u^+_{\dot\alpha}$ which is going to become the $({\bf 3,4,1})$ superfield $L^{++}$ upon gauging
this isometry and choosing the appropriate manifestly ${\cal N}=4$ supersymmetric gauge.

An important impact of this superfield reduction on the structure of the component action is the appearance
of the new induced potential bilinear in the gauge group generators $\sim U^2 = U^a U^b T^a T^b\,$. It comes out
as a result of eliminating the auxiliary field $F$ in the off-shell $({\bf 3, 4, 1})$ multiplet, and so
is necessarily prescribed by ${\cal N}=4$ supersymmetry. It is interesting that analogous potential
terms were introduced in \cite{H2} at the bosonic level for ensuring the existence of some hidden
symmetries in the models of 3-dimensional particle in a non-Abelian monopole background.

The same reduction ${\mathbb R}^4\rightarrow {\mathbb R}^3$ can be performed
at the level of Hamiltonian and supercharges. In particular, the reduction
of the Hamiltonian of the four-dimensional system of Eq.~(\ref{eq_susyham}) yields the 3-dimensional Hamiltonian~(\ref{decrom}).



\cleardoublepage
\chapitresimple{Conclusion}

\bigskip

We studied some rather general off-shell ${\cal N}=4$ supersymmetric coupling of the $d=1$ coordinate supermultiplets ${\bf (4,4,0)}$ and ${\bf (3,4,1)}$ to an external self-dual (or anti-self-dual) Abelian gauge field and discussed the ${\bf (4,4,0)}$ case in details.
Our main framework was the harmonic superspace approach.

The use of an analytic ``semi-dynamical'' multiplet $({\bf 4,4,0})$
with the Wess-Zumino type action allowed us to make coupling of the coordinate multiplets ${\bf (4,4,0)}$ and ${\bf (3,4,1)}$ to an
external $\SU(2)$ gauge field. This auxiliary multiplet incorporates  $\SU(2)$ doublet of bosonic spin variables which are crucial for arranging couplings to non-Abelian gauge fields.
In the four-dimensional case,
the off-shell ${\cal N}=4$ supersymmetry restricts the non-Abelian gauge field to be self-dual (or anti-self-dual) and in a form of the 't~Hooft ansatz for $\SU(2)$ gauge field.
In the three-dimensional case, the non-Abelian gauge field is
a three-dimensional reduction of this 't~Hooft ansatz, i.e. a particular solution of the Bogomolny monopole equations.
Additionally, in three dimensions, at the component level, the coupling to a gauge field is necessarily accompanied by an induced potential which is bilinear in the $\SU(2)$ generators and arises as a result of eliminating the auxiliary field in the coordinate ${\bf (3,4,1)}$ multiplet.

The explicit form of the Hamiltonians and the supercharges were presented. The corresponding expressions respect $\N=4$ {\em on-shell} supersymmetry for any self-dual or anti-self-dual, Abelian or non-Abelian gauge field in four-dimensions, not necessarily in the 't~Hooft ansatz form. In three dimensions, an arbitrary BPS monopole background can be used in the non-Abelian case.

It is worthwhile to note that similar constraints (Bogomolny equations) on the external non-Abelian three-dimensional gauge field were
found in \cite{Berry}, while considering an ${\cal N}=4$ extension of Berry phase in quantum mechanics. However, no invariant actions
and/or the explicit expressions for the Hamiltonian and ${\cal N}=4$ supercharges were presented there.

The nonlinear counterpart of $q^{+\dalpha}$ multiplet is discussed in \cite{arXiv:1107.1429}. In this case, the bosonic target geometry is more general as compared to the conformally-flat geometry associated with the linear $({\bf 4, 4, 0})$  multiplet.

Among the possible directions of further study, we mention the construction of higher ${\cal N}$ SQM models
with non-Abelian gauge field backgrounds, e.g. ${\cal N}=8$ ones,
as well as studying various supersymmetry-preserving reductions of these models to lower-dimensional target bosonic manifolds
by the gauging procedure of \cite{DI}. Actually, the method of the auxiliary ``semi-dynamical'' $({\bf 4,4,0})$
multiplet with the Wess-Zumino type action, which was successfully applied in our construction here, could work with
the equal efficiency for constructing a Lagrangian description of other supersymmetric quantum-mechanics problems involving
the coupling to an external non-Abelian gauge field. Besides the obvious examples
of quantum Hall effect (or Landau problem) in higher dimensions (see e.g. the discussion in \cite{gknty}), let us also mention
supersymmetric Wilson loop functionals which can be interpreted in terms of a non-Abelian version of Chern-Simons
(super)quantum mechanics \cite{HT}, with the parameter along the loop as an evolution parameter. We hope that
the quantized semi-dynamical variables could provide a new efficient tool to study this class
of problems.

The 't Hooft type ansatz \p{Hooft} and the choice of $\SU(2)$ as the gauge group are required for the existence
of the {\it off-shell} superfield formulation of the discussed SQM systems. It is not known whether the most general system can be derived from some
off-shell superfield formalism, with general instanton/monopole backgrounds obtained from the ADHM construction \cite{ADHM} or its three-dimensional reduction.
Additionally, there remains a problem of extending the models to a generic $\SU(N)$
gauge group.
Possibly, the above issues are related to the generalization of the interaction term~(\ref{Sint})  to
\begin{equation*}
 S_{\rm int}= \int \, dt \, du\, d\bar \theta^+d\theta^+\,
  K^{++}\left(q^{+\dot{\alpha}},\, u^\pm_\beta,\,  v^+ \widetilde{v^+} \right) .
\end{equation*}
It would be also interesting to study SQM models with nonlinear counterpart of the semi-dynamical multiplet $({\bf 4, 4, 0})$ \cite{DI}.



\cleardoublepage
\addcontentsline{toc}{chapter}{\numberline{}Bibliography}

\renewcommand\bibname{Bibliography}






\end{document}